\documentclass[aps,prapplied,reprint,superscriptaddress,nofootinbib,longbibliography]{revtex4-1}
\usepackage[utf8]{inputenc}
\usepackage[normalem]{ulem}
\usepackage{graphicx}
\usepackage[version=4]{mhchem}
\usepackage{mathtools}
\usepackage{physics}
\usepackage{amsmath}
\usepackage{siunitx}
\usepackage{gensymb}
\usepackage{textcomp}
\usepackage{newtxmath}
\usepackage{multirow}
\usepackage{makecell}
\usepackage{booktabs}
\usepackage{xcolor}
\usepackage[utf8]{inputenc}

\usepackage[colorlinks=true,linktocpage=true,breaklinks=true]{hyperref}

\usepackage[a-2b,mathxmp]{pdfx}[2018/12/22]

\emergencystretch=\maxdimen
\begin{document}
\newcommand{\Hom}{Fe$_{50}$Ni$_{50}$}
\newcommand{\FeBot}{Fe$_{100-x}$Ni$_{x}$}
\newcommand{\NiBot}{Fe$_{x}$Ni$_{100-x}$}
\newcommand{\alphaEff}{\alpha_\mathrm{eff}}
\newcommand{\alphaInt}{\alpha_\mathrm{int}}
\newcommand{\alphaExt}{\alpha_\mathrm{TMS}}
\newcommand{\DampSOT}{\theta_\mathrm{AD}} 

\title{Vertically Graded FeNi Alloys with Low Damping and a Sizeable Spin-Orbit Torque}

\author{Rachel E. Maizel}\email{maizer@vt.edu}
\affiliation{Department of Physics, Virginia Tech, Blacksburg, VA, USA}
\author{Shuang Wu}
\affiliation{Department of Physics, Virginia Tech, Blacksburg, VA, USA}

\author{Purnima P. Balakrishnan}
\affiliation{NIST Center for Neutron Research, National Institute of Standards and Technology, Gaithersburg, MD, USA} 
\author{Alexander J. Grutter}
\affiliation{NIST Center for Neutron Research, National Institute of Standards and Technology, Gaithersburg, MD, USA} 

\author{Christy J. Kinane}
\affiliation{ISIS-Neutron and Muon Source, STFC Rutherford Appleton Laboratory, Didcot, United Kingdom}
\author{Andrew J. Caruana}
\affiliation{ISIS-Neutron and Muon Source, STFC Rutherford Appleton Laboratory, Didcot, United Kingdom}

\author{Prabandha Nakarmi}
\affiliation{Department of Physics and Astronomy, University of Alabama, Tuscaloosa, AL, USA}
\author{Bhuwan Nepal}
\affiliation{Department of Physics and Astronomy, University of Alabama, Tuscaloosa, AL, USA}
\author{David A. Smith}
\affiliation{Department of Physics, Virginia Tech, Blacksburg, VA, USA}
\author{Youngmin Lim}
\affiliation{Department of Physics, Virginia Tech, Blacksburg, VA, USA}
\author{Julia L. Jones}
\affiliation{Department of Physics, Virginia Tech, Blacksburg, VA, USA}
\author{Wyatt C. Thomas}
\affiliation{Department of Physics, Virginia Tech, Blacksburg, VA, USA}

\author{Jing Zhao}
\affiliation{Department of Geosciences, Virginia Tech, Blacksburg, VA, USA}
\author{F. Marc Michel}
\affiliation{Department of Geosciences, Virginia Tech, Blacksburg, VA, USA}
\affiliation{Academy of Integrated Science, Virginia Tech, Blacksburg, VA, USA}
\author{Tim Mewes}
\affiliation{Department of Physics and Astronomy, University of Alabama, Tuscaloosa, AL, USA}
\author{Satoru Emori}\email{semori@vt.edu}
\affiliation{Department of Physics, Virginia Tech, Blacksburg, VA, USA}
\affiliation{Academy of Integrated Science, Virginia Tech, Blacksburg, VA, USA}
\date{August 14, 2024}

\begin{abstract}
Energy-efficient spintronic devices require a large spin-orbit torque (SOT) and low damping to excite magnetic precession. In conventional devices with heavy-metal/ferromagnet bilayers, reducing the ferromagnet thickness to $\sim$1 nm enhances the SOT but dramatically increases damping. Here, we investigate an alternative approach based on a 10 nm thick single-layer ferromagnet to attain both low damping \emph{and} a sizable SOT. Instead of relying on a single interface, we continuously break the bulk inversion symmetry with a vertical compositional gradient of two ferromagnetic elements: Fe with low intrinsic damping and Ni with sizable spin-orbit coupling.  We find low effective damping parameters of $\alphaEff < 5\times10^{-3}$ in the FeNi alloy films, despite the steep compositional gradients. Moreover, we reveal a sizable anti-damping SOT efficiency of $|\DampSOT| \approx 0.05$, even \emph{without} an intentional compositional gradient. Through depth-resolved x-ray diffraction, we identify a lattice strain gradient as crucial symmetry breaking that underpins the SOT. Our findings provide fresh insights into damping and SOTs in single-layer ferromagnets for power-efficient spintronic devices. 
\end{abstract}

\maketitle

\section{Introduction} \label{sec:intro}
Spin-orbit torques (SOT) can control the magnetic states of memories and magnetization dynamics
in oscillators driven by electric current \cite{Hirohata2020, Shao2021, Manchon2019}. So-called ``Type-Y'' SOT devices\footnote{The ``Y'' in ``Type-Y'' defines the magnetization axis with respect to the drive electric current, which flows along the ``X'' axis~\cite{Fukami2016b}. The magnetization axis of a Type-Y SOT device is in-plane and orthogonal to the current. There are also Type-X devices (magnetization axis along the current) and Type-Z devices (magnetization axis out-of-plane), but they generally require higher power consumption than Type-Y devices.} can attain especially low power consumption by fulfilling two criteria~\cite{Fukami2016b}: (1) low damping to reduce loss in precessional magnetization dynamics and (2) a strong anti-damping SOT to enable free magnetic precession. 

\begin{figure}[t]
 \centering
    \includegraphics[width = 3.3in, height = 2.3in]{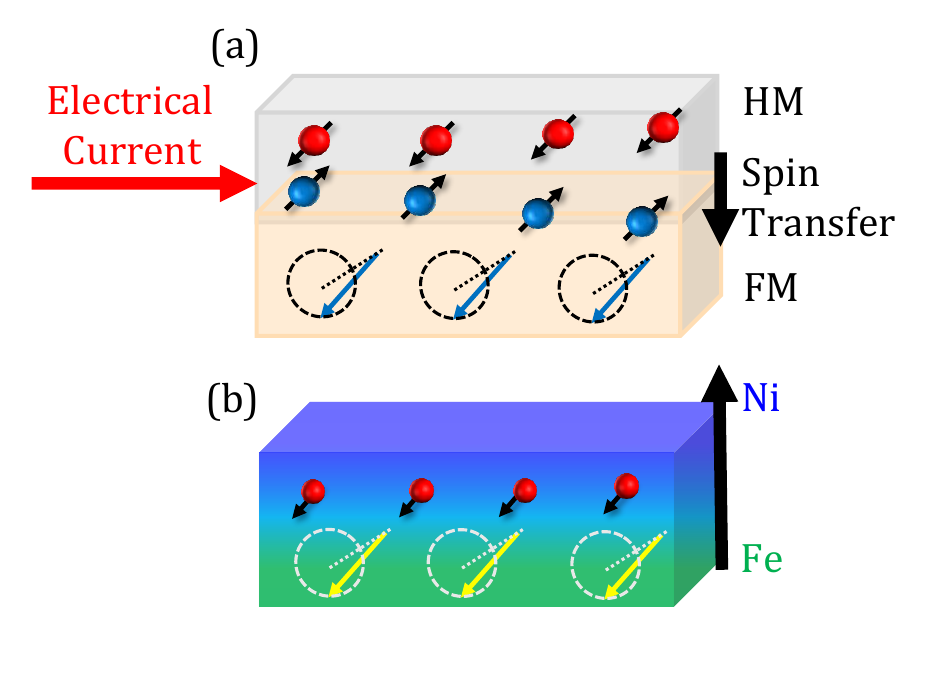}
    \caption{(a) HM/FM bilayer: the transfer of non-equilibrium spin accumulation across the HM/FM interface generates SOTs, driving precession of the magnetization. (b) Vertically graded alloy: the intentional asymmetry in Fe:Ni composition permits a net non-equilibrium spin accumulation within the bulk, generating SOTs.}
    \label{fig:SHE}
\end{figure}

SOTs require symmetry breaking to enable an uncompensated spin accumulation acting on the magnetization~\cite{Shao2021, Manchon2019, Davidson2020a}. Conventional SOT devices achieve this by heavy-metal/ferromagnet (HM/FM) bilayers [Fig. \ref{fig:SHE}(a)], where the HM/FM interface provides the requisite symmetry breaking. 
Passing an electric current through the HM, e.g., with a strong spin Hall effect~\cite{Sinova2015}, causes the conduction electron spins with opposite polarizations to deflect toward opposite surfaces of the HM layer. The HM/FM interface develops a non-equilibrium spin accumulation, which is transferred to the FM, exerting SOTs on the magnetization. In addition to the spin Hall effect in the HM [Fig.~\ref{fig:SHE}(a)], various coexisting mechanisms may yield SOTs in HM/FM bilayers such as the Rashba-Edelstein effect, spin-filtering and precession due to the interfacial spin-orbit field, and self-generated torques with a HM sink~\cite{Manchon2019, Davidson2020a, Kim2020}. Moreover, some bilayers incorporate 3$d$ transition metals instead of HMs and still exhibit substantial SOTs~\cite{Taniguchi2015, Baek2018, Chuang2019, Montoya2024}. 

Nevertheless, all of these reported mechanisms require a transfer of angular momentum across the bilayer interface. Thus, regardless of the mechanism, the net spin accumulation is greatest at the interface and decreases sharply within the FM thickness~\cite{Kim2020, Haney2013a, Fan2014}. Consequently, conventional bilayers tend to have a thin FM with a thickness of $t_\mathrm{FM} \sim 1$ nm to maximize the torque on the magnetization. 

However, while decreasing $t_\mathrm{FM}$ increases the SOTs, it has the undesirable effect of increasing the effective damping. For example, spin-pumping damping scales as ${t_\mathrm{FM}}^{-1}$~\cite{Tserkovnyak2005}, and two-magnon scattering\footnote{Two-magnon scattering is the decay of uniformly precessing magnetic moments ($k = 0$ spin wave mode) into a finite-wavelength ($k\neq 0$) spin-wave mode~\cite{McMichael2004}. Two-magnon scattering by itself does not involve energy dissipation from magnetization dynamics to the lattice, so it is not ``damping'' in the strictest sense. Yet, here, we consider two-magnon scattering to be part of the ``effective damping'' as it contributes to the broadening of the resonance linewidth (i.e., deteriorates the quality factor of the magnetic precession) and may adversely impact the efficiency and stability of Type-Y SOT devices.}
at the interface scales as ${t_\mathrm{FM}}^{-2}$ \cite{Zhu2019b}.
In other words, there exists a fundamental trade-off in conventional bilayer devices: SOTs are enhanced at the expense of higher damping. 
Attaining low damping \emph{and} strong SOTs remains an outstanding challenge for developing more power-efficient spintronic devices. 

Here, we present an encouraging route toward low damping and a sizable anti-damping SOT in \emph{single-layer} FMs. Unlike bilayers that break symmetry at the interface, our present approach continuously breaks inversion symmetry within the \emph{bulk} of the FM alloy along its thickness axis~\cite{Fallarino2021}. Specifically, we leverage a steep vertical compositional gradient in the FM, as illustrated in Fig.~\ref{fig:SHE}(b), allowing for the production of spin accumulation within the bulk of the FM. These asymmetric single-layer SOT devices are attractive because they may produce sizable SOTs in thicker FMs, e.g., $t_\mathrm{FM} \sim 10$ nm~\cite{Tang2020, Liu2020a, Tao2022FePt}. Such thick single-layer FMs may maintain low damping while attaining strong SOTs for power-efficient Type-Y SOT devices. 

Previous studies show SOT-induced perpendicular switching in ``Type-Z'' devices~\cite{Fukami2016b} consisting of $\sim$10 nm thick single-layer FePt with compositional gradients~\cite{Tang2020, Liu2020a, Tao2022FePt}. More broadly, SOTs have been reported in single-layer FM-HM alloys in which a 3$d$ transition-metal ferromagnet (e.g., Fe or Co) is mixed with a heavy element (e.g., Pt or a rare-earth metal)~\cite{Tang2020, Liu2020a, Tao2022FePt, Zhu2020c, Lee2020b, Zhang2020a, CespedesBerrocal2021, Zhu2021c, Liu2022BulkSOT}. Yet, these single-layer FM-HM systems possess high damping -- e.g., effective damping parameter of $\alpha_{\mathrm{eff}} \approx 3 \times 10^{-2}$ for FePt \cite{Chen2012FePt, Ma2015}, which is an order of magnitude greater than $\alpha_{\mathrm{eff}}$ for many other FMs~\cite{Schoen2017c}. Thus, the previous approach to single-layer FMs is unsuitable for Type-Y SOT devices that require low damping. 

To pursue low damping, we examined 10 nm thick single-layer FMs with asymmetric compositional gradients of ferromagnetic Fe and Ni [Fig.~\ref{fig:SHE}(b)]. We opted for alloys of Fe and Ni because (1) Fe has the lowest Gilbert damping among elemental FMs~\cite{Schoen2017c, Oogane2006} 
and (2) Ni has the strongest spin-orbit coupling among elemental FMs~\cite{Du2014b, Keller2019}. 
Considering these attributes, we hypothesized the vertically graded FeNi alloys to be viable single-layer FMs that exhibit low damping \emph{and} strong SOTs for power-efficient devices.

Our article is organized as follows. The nominal structures and the growth conditions of the FeNi films are described in Section~\ref{sec:growth}. We present our findings on damping in Section~\ref{sec:damping} and current-induced torques (with particular emphasis on the anti-damping SOT) in Section~\ref{sec:torques}. The observed effective damping remains remarkably low -- e.g., $\alpha_{\mathrm{eff}} \approx 4.5 \times 10^{-3}$ even with steep vertical Fe-Ni compositional gradients of $\sim$10 at.\%/nm. Moreover, the anti-damping SOT efficiencies in the FeNi alloys are comparable to those in HM/FM bilayers. 

Surprisingly, the SOT efficiency does not correlate with the nominal compositional gradient: we observe comparable magnitudes of anti-damping SOT in FeNi alloys with and \emph{without} the intentional compositional gradient. This observation indicates a nontrivial origin for the observed SOT. In Section~\ref{sec:depth}, we show depth-profile characterization of composition and lattice strain to gain further insights into the possible underlying mechanism of the SOT. In Section~\ref{sec:perspective}, we discuss the likely contribution of a strain gradient to the sizable SOT. Our work points to the crucial role of an atomic-scale \emph{structural gradient} in enabling a significant SOT. This revelation provides a fresh perspective for engineering low damping and strong SOTs for highly efficient nanomagnetic memories and oscillators.

\section{Film Growth}\label{sec:growth}
\begin{figure}[b]
 \centering
    \includegraphics[width = 3.3in, height = 1.2in]{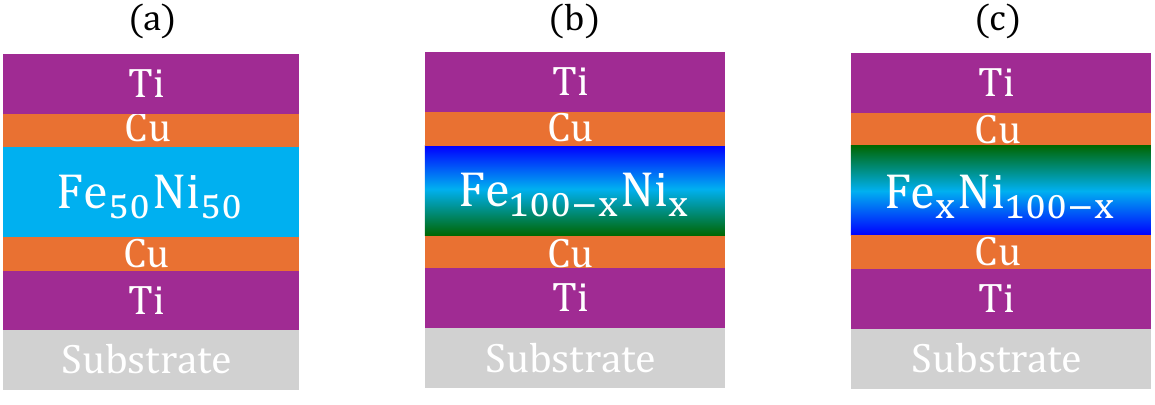}
    \caption{Nominal film structures for (a) compositionally symmetric \Hom, (b) compositionally asymmetric \FeBot, and (c) compositionally asymmetric \NiBot.}
    \label{fig:FSL}
\end{figure}

We focus on three types of 10 nm thick polycrystalline FM FeNi alloy films with an average Fe:Ni ratio\footnote{Strictly speaking, the Fe:Ni ratio of 50:50 here refers to the nominal \emph{volume} ratio. The corresponding \emph{atomic} ratio of Fe:Ni, using the tabulated densities and molar masses of Fe and Ni, would be 48:52.} of 50:50,  
as illustrated in Fig.~~\ref{fig:FSL}: 
\begin{itemize}
\item nominally symmetric, homogeneous alloy of Fe$_{50}$Ni$_{50}$ [Fig.~\ref{fig:FSL}(a)];
\item compositionally graded alloy with Fe on the bottom, denoted as \FeBot\ with $x=0$ at the bottom and $x = 100$ at the top of the FM   [Fig.~\ref{fig:FSL}(b)];
\item compositionally graded alloy with Ni on the bottom, denoted as \NiBot\ with $x=0$ at the bottom and $x = 100$ at the top of the FM   [Fig.~\ref{fig:FSL}(c)]. 
\end{itemize}
Each film was grown by dc magnetron sputtering on a Si substrate with a 50 nm thick thermally grown SiO$_2$ overlayer, unless otherwise noted. The base pressure prior to deposition was $\lesssim 3\times10^{-8}$ Torr, and the Ar sputtering gas pressure during deposition was 3 mTorr. The deposition rate of each sputtered material was calibrated by x-ray reflectometry. 

Each film was seeded by 3 nm thick Ti, followed by 1 nm thick Cu. The Ti layer promotes good adhesion of the film on the SiO$_2$ surface, whereas the Cu layer has been reported to reduce effective damping in 3$d$ FMs~\cite{Edwards2019}. The symmetric \Hom\ layer [Fig. \ref{fig:FSL}(a)] was grown by co-sputtering Fe and Ni targets at the same deposition rate. The compositionally graded \FeBot\ [Fig. \ref{fig:FSL}(b)] and \NiBot\ [Fig. \ref{fig:FSL}(c)] layers were grown by continuously ramping the sputtering power for the Fe and Ni targets. In particular, to grow \FeBot, the sputtering power of the Fe (Ni) target was ramped linearly from 97 W to 0 W (from 0 W to 80 W); to grow \NiBot, the sputtering power ramping directions were reversed. Experimentally, however, stable plasma was not achieved below 10 W of sputtering power. Thus, in both asymmetric films, there is likely a non-linear gradient at the beginning and end of the 10 nm deposition. As we will show in Sec.~\ref{sec:depth}, this deposition protocol resulted in an approximately linear compositional gradient in the bulk of \FeBot\ and \NiBot. Finally, each film was capped with 1 nm thick Cu and 3 nm thick Ti for protection from oxidation. As illustrated in Fig.~\ref{fig:FSL}, the Ti/Cu seed and Cu/Ti capping layers are nominally symmetric. With the stack structures in Fig.~\ref{fig:FSL}(b,c), the steep compositional gradient within FeNi is the only intentional source of symmetry breaking. 

\section{Damping} \label{sec:damping}

\begin{figure*}
 \centering
    \includegraphics[width = 7in, height = 2.6in]{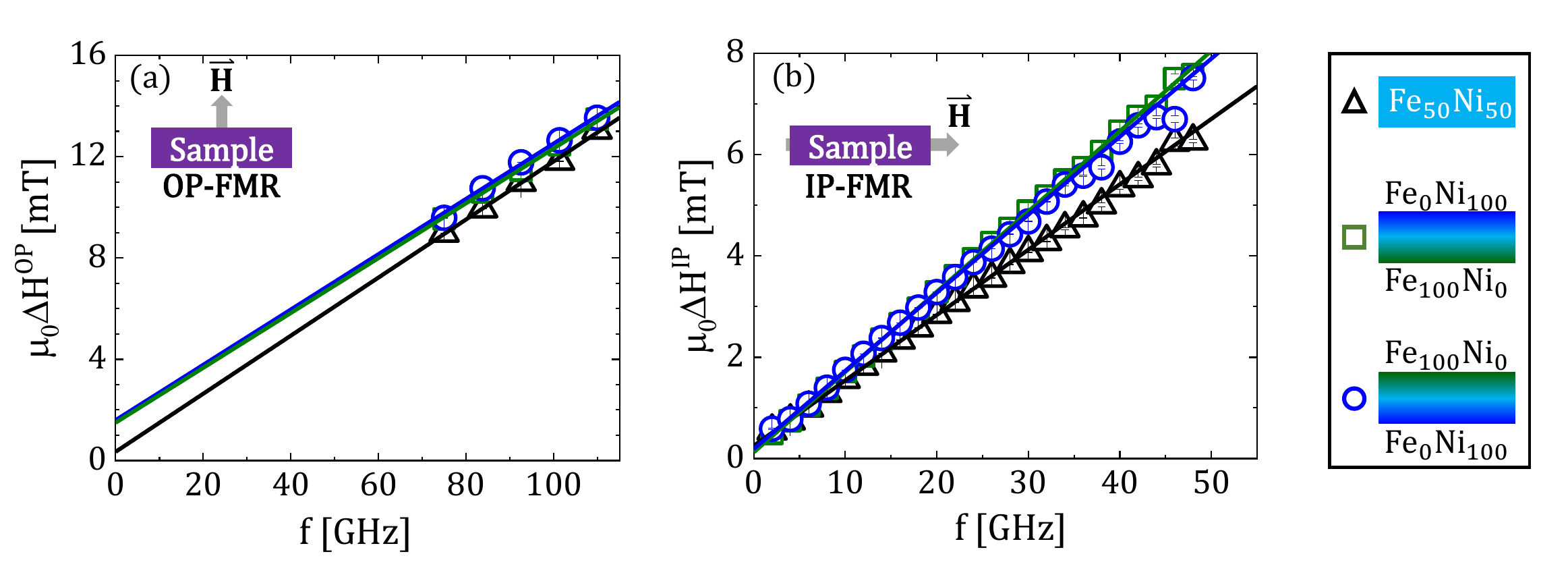}
    \caption{FMR linewidth vs frequency for $\mathrm{Fe_{50}Ni_{50}}$ (black triangle), $\mathrm{Fe_{100-x}Ni_{x}}$ (green square), and $\mathrm{Fe_{x}Ni_{100-x}}$ (blue circle) with FMR configuration (a) out-of-plane and (b) in-plane.}
    \label{fig:FMR}
\end{figure*}

We first present the impact of the steep compositional gradient on magnetic damping in the FeNi films with ferromagnetic resonance (FMR) spectroscopy. Since our study is restricted to room temperature, we might surmise that damping is dominated by the resistivity-like mechanism \cite{Gilmore2007, Khodadadi2020}, where damping increases with more electronic scattering. However, the sheet resistances of the three films are all within $\sim$10\% of $\approx$40 $\Omega /\square$ with no clear correlation with the compositional gradients -- so that the resistivity-like damping might be similar. In fact, it is not immediately clear how the steep compositional gradient should impact damping. While damping has been widely studied in bilayers, multilayers, and homogeneous alloy films~\cite{Schoen2017c, Heinrich2005, Shaw2011MLvsAlloys, Omelchenko2017}, there appear to be no published studies of damping in compositionally graded single-layer FM films. 

In our present study, we define the ``effective'' damping parameter $\alphaEff$ to be the sum of intrinsic and extrinsic contributions, 
\begin{equation}
\alphaEff = \overbrace{\alpha _{\text{int}}}^{\text{intrinsic}} + \overbrace{\alpha _{\text{TMS}}}^{\text{extrinsic}}.
\label{eq:EffDamp}
\end{equation}
The intrinsic component $\alphaInt$ in Eq.~\ref{eq:EffDamp} captures the viscous Gilbert damping, which causes the FMR linewidth to scale linearly with the excitation frequency~\cite{Heinrich2005, Mewes2015}. The damping contribution from spin pumping across interfaces is neglected because spin-orbit coupling (hence spin dissipation) is weak in Ti and Cu~\cite{Du2014b}. Since our FeNi films are significantly thinner than the skin depth for typical FeNi alloys, 
the eddy-current damping is also considered negligible. 

We attribute the extrinsic component $\alphaExt$ in Eq.~\ref{eq:EffDamp} to non-Gilbert magnetic relaxation -- namely, two-magnon scattering (TMS), a decay of the uniform FMR mode into magnon modes due to magnetic inhomogeneity in the film~\cite{McMichael2004, Zakeri2007}. Two-magnon scattering often manifests in a nonlinear frequency dependence of the FMR linewidth. However, when two-magnon scattering is small relative to the intrinsic damping, $\alphaExt$ may simply appear as a correction to $\alphaInt$, derived from the linear slope of the FMR linewidth over a limited frequency range~\cite{Wu2022, Zhu2019b}. 

We can disentangle the intrinsic and extrinsic damping parameters through different FMR measurement configurations~\cite{Wu2022, Edwards2019}. 
\begin{itemize}
    \item When the film is magnetized out-of-plane, there exist no magnon states degenerate with the FMR frequency, and two-magnon scattering is suppressed ($\alphaExt = 0$)~\cite{Heinrich2005, Mewes2015, McMichael2004}. Hence, $\alphaInt$ is measured in the out-of-plane FMR configuration [Sec.~\ref{subsec:OPFMR}]. 
    \item When the film is magnetized in-plane, there are magnon modes degenerate with the FMR mode, allowing for two-magnon scattering to occur~\cite{Heinrich2005, Mewes2015, McMichael2004}. Thus, the in-plane FMR configuration [Sec.~\ref{subsec:IPFMR}] measures $\alphaEff$ including both the intrinsic and extrinsic contributions [Eq.~\ref{eq:EffDamp}]. In the 10 nm thick FMs here, the source of two-magnon scattering is not necessarily restricted to the film interfaces~\cite{Zhu2019b} but may also emerge from inhomogeneities in the FM bulk~\cite{McMichael2004, Wu2022}. 
\end{itemize}
We assume that $\alphaInt$ is identical between the out-of-plane and in-plane configurations. This is likely reasonable because Gilbert damping is expected to isotropic in Fe and Ni at room temperature~\cite{Gilmore2010}.  While quantifying $\alphaInt$ is critical for examining the fundamental origin of damping, we note that $\alphaEff$ may be crucial for practical applications, especially Type-Y SOT devices, in which the magnetization lies in-plane~\cite{Fukami2016b, Hirohata2020, Shao2021}.

\subsection{Intrinsic damping}\label{subsec:OPFMR}
To conduct out-of-plane FMR measurements, each sample was placed on a W-band shorted waveguide in a superconducting magnet, allowing for a high applied field ($\gtrsim $ 4 T) to saturate the magnetization completely out-of-plane ~\cite{Wu2022, Smith2020c}. 
Figure~\ref{fig:FMR}(a) shows the frequency $f$ dependence of the half-width-at-half-maximum FMR linewidth $\Delta H^{\mathrm{OP}}$ for the three FeNi films. We quantify $\alphaInt$ from the linear fit of $\Delta H^{\mathrm{OP}}$ vs $f$,  
\begin{equation}\label{eq:OP}
\mu_0\Delta H^{\mathrm{OP}} = \mu_0\Delta H_0^{\mathrm{OP}} + \frac{2\pi}{\gamma}\alphaInt f,
\end{equation}
where $\mu_0$ is the permeability of free space, $\Delta H_0^{\mathrm{OP}}$ is the zero-frequency linewidth, and $\gamma/(2\pi)= 29.2$ GHz is the gyromagnetic ratio (derived from the $f$ dependence of the resonance field, fit with the Kittel equation~\cite{Wu2022, Smith2020c}). 

As summarized in Table~\ref{tab:damping}, we find that the intrinsic Gilbert damping parameters of the three films, $\alphaInt \approx (3.2-3.3)\times 10^{-3}$, are identical within the experimental uncertainty of $\approx 0.1\times 10^{-3}$. Remarkably, the steep compositional gradients do not impact intrinsic Gilbert damping in these films. 

From this invariance against the compositional gradient, we deduce that the average global composition (here, Fe:Ni ratio of 50:50), rather than the local inhomogeneities, predominantly governs $\alphaInt$ of the 10 nm thick FeNi films. The finding is also aligned with previous experiments reporting \emph{intrinsic} Gilbert damping to be invariant with compositional profiles~\cite{Shaw2011MLvsAlloys}.
Although a rigorous explanation of the underlying mechanism requires further work, we speculate that the ferromagnetic exchange length~\cite{Abo2013} may play a critical role. As long as the graded FM thickness is below or comparable to the exchange length (likely up to $\sim$10 nm in the alloys here~\cite{Abo2013, Niitsu2020}), the impact of local compositional variations on $\alphaInt$ may be averaged out. 

Regardless of the mechanism, we have demonstrated low intrinsic damping -- e.g., an order of magnitude lower than $\sim 10^{-2}$ in FePt~\cite{Chen2012FePt, Ma2015} used in graded SOT devices~\cite{Tang2020, Liu2020a, Tao2022FePt} -- even with a steep vertical compositional gradient of $\sim$10 at.\%/nm. This finding is highly promising for engineering symmetry-broken FMs for Type-Y SOT devices. 

While $\alphaInt$ is unaffected by the compositional gradient, we observe a clear difference in $\Delta H_0^{\mathrm{OP}}$ (vertical intercept in Fig.~\ref{fig:FMR}(a)) between the FeNi films with and without the gradient. In particular, $\Delta H_0^{\mathrm{OP}}$ is several times greater for the graded \FeBot\ and \NiBot\ samples compared to the homogeneous \Hom\ sample. This observation is in line with the notion that $\Delta H_0^{\mathrm{OP}}$ in the simple linear fit [Eq.~\ref{eq:OP}] is sensitive to local inhomogeneity~\cite{Heinrich2005, Mewes2015}. 

\begin{table}[b]
	\caption{Damping parameters and zero-frequency linewidths quantified from the FMR results in Fig.~\ref{fig:FMR}. The error bars are obtained from the square root of the diagonal of the covariance matrix associated with the weighted linear fit slopes and vertical intercepts [Fig.~\ref{fig:FMR}].} 
	\centering 
	\begin{tabular}{c c c c} 
		\hline 
           
		{}  & $\mathrm{Fe_{50}Ni_{50}}$ & $\mathrm{Fe_{100-x}Ni_{x}}$ & $\mathrm{Fe_{x}Ni_{100-x}}$\\ [3pt] 
		\hline 
         $\alpha_{\mathrm{int}}$ ($10^{-3}$) & $3.34 \pm 0.13$ & $3.15 \pm 0.09$ & $3.18 \pm 0.11$ \\ [6pt] 
         $\mu_0\Delta H_0^{\mathrm{OP}}$ (mT) & $0.4 \pm 0.5$ & $1.7 \pm 0.3$ & $1.8 \pm 0.4$\\ [6pt] 
		$\alpha_{\mathrm{eff}}$ ($10^{-3}$) & $3.77 \pm 0.02$ & $4.63 \pm 0.01$ & $4.50 \pm 0.03$ \\ [6pt]
         $\mu_0\Delta H_0^{\mathrm{IP}}$ (mT)  & $0.29 \pm 0.01$ & $0.14 \pm 0.01$ & $0.22 \pm 0.02$\\ [6pt] 
		\hline 
        \label{tab:damping}
	\end{tabular}
\end{table}

\subsection{Effective damping }\label{subsec:IPFMR}
We employed an FMR spectrometer based on a coplanar waveguide (details in Ref.~\cite{Smith2020c, Lim2021}). The frequency $f$ dependence of the in-plane half-width-at-half-maximum FMR linewidth $\Delta H^{\mathrm{IP}}$ is shown in Fig.~\ref{fig:FMR}(b). We quantify the effective damping parameter $\alphaEff$ through 
\begin{equation}\label{eq:IP}
\mu_0\Delta H^{\mathrm{IP}} = \mu_0\Delta H_0^{\mathrm{IP}} + \frac{2\pi}{\gamma}\alphaEff f. 
\end{equation}
As summarized in Table~\ref{tab:damping}, the values of $\alphaEff$ for \FeBot\ and \NiBot\ with steep vertical compositional gradients are about 20\% greater than the homogeneous \Hom\ film. The larger $\alphaEff$ may be accounted for by enhanced bulk two-magnon scattering induced by local magnetic inhomogeneities in the FM bulk~\cite{McMichael2004, Wu2022}, perhaps tied to the intentional compositional gradient within the FM. 
We note that $\alphaEff$ for \Hom\ is still greater than $\alphaInt$ by $\approx$10\%, which suggests that weak two-magnon scattering is also present in this nominally homogeneous sample.  

Despite the likely presence of two-magnon scattering, the effective damping remains low with $\alphaEff < 5\times10^{-3}$ for all three samples. Even with a steep, intentional compositional gradient, $\alphaEff$ here is an order of magnitude lower than the reported damping parameter of $\approx 3\times 10^{-2}$ in FePt~\cite{Chen2012FePt, Ma2015} used in graded SOT devices~\cite{Tang2020, Liu2020a, Tao2022FePt}. Furthermore, it is lower than $\alphaEff \approx 7\times 10^{-3}$ for the oft-studied prototypical soft FM of permalloy (Fe$_{20}$Ni$_{80}$)~\cite{Lim2021, Lim2022a, Lim2023}. The demonstrated low effective damping in the FeNi films here constitutes a crucial step toward power-efficient SOT devices.

\section{Current-Induced Torques}\label{sec:torques}

We demonstrated in Sec.~\ref{sec:damping} that the FeNi films meet the criterion of low damping for Type-Y SOT devices. We now show that a sizable anti-damping SOT -- also known as ``damping-like'' SOT -- emerges within the bulk of these FeNi films. Spin-torque FMR (ST-FMR) measurements were performed on 50 $\mathrm{\mu}$m wide lithographically patterned strips to examine torques induced by current, including the anti-damping SOT and the classical Oersted field torque~\cite{Liu2011}. An in-plane applied magnetic field at an angle $\phi$ from the current axis defines the precessional axis. Further details of the ST-FMR method are in Appendix~\ref{app:STFMR}. 

\begin{figure}[b]
 \centering
    \includegraphics[width = 3.1in]{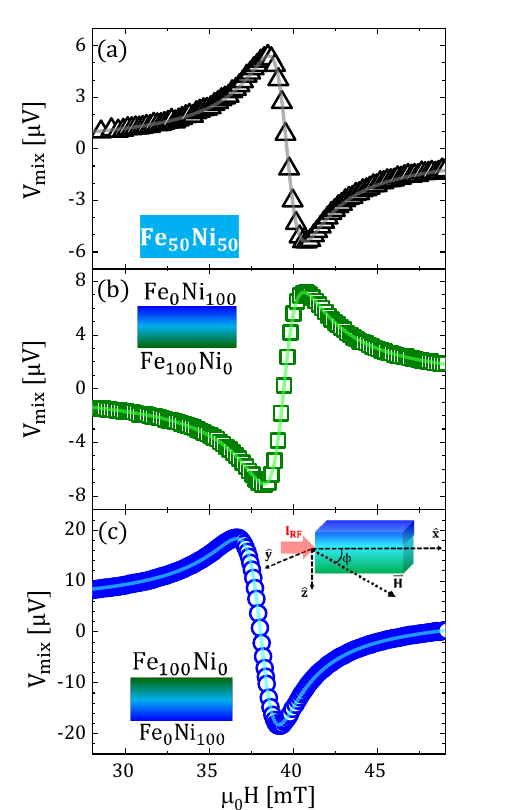}
    \caption{ST-FMR spectra at $\mathrm{\phi = 45^{o}}$ with respect to the current axis $\hat{x}$, and frequency $f = 7$ GHz for (a) \Hom, (b) \FeBot, and  (c) \NiBot.}
    \label{fig:stfmr}
\end{figure}

We highlight two key observations from the ST-FMR spectra in Fig.~\ref{fig:stfmr}. First, \FeBot\ and \NiBot\ exhibit opposite signal polarity [Fig. \ref{fig:stfmr}(b,c)]. This is unsurprising since the opposite compositional gradients are expected to yield opposite current-induced torques, captured in the polarity of the rectified voltage ${V_\mathrm{mix}}$. Second, \Hom\ yields a clear ST-FMR response [Fig. \ref{fig:stfmr}(a)], comparable in magnitude to \FeBot\ and \NiBot. This finding \emph{is} surprising because, in such a symmetric sample, the current-induced spin accumulations and Oersted field should average to zero. In other words, we would expect \Hom\ without any intentional symmetry breaking to exhibit little or no current-induced torques. 

In this section, we investigate how the current-induced torques depend on the compositional gradients -- and address how a sizable torque can emerge in the nominally symmetric \Hom\ sample. 
We evaluate the anti-damping SOT in Sec.~\ref{subsec:DL-SOT} and the Oersted field\footnote{There could also be a ``field-like'' SOT that acts similarly to the Oersted field torque, but we show that it is likely very small.}
in Sec.~\ref{subsec:FL-SOT}. Then, in Sec.~\ref{subsec:SOT-context}, we gain partial insights into the origin of the anti-damping SOT by comparing it with the Oersted field torque.

\begin{figure*}
 \centering
    \includegraphics[width = 7in, height = 3.7in]{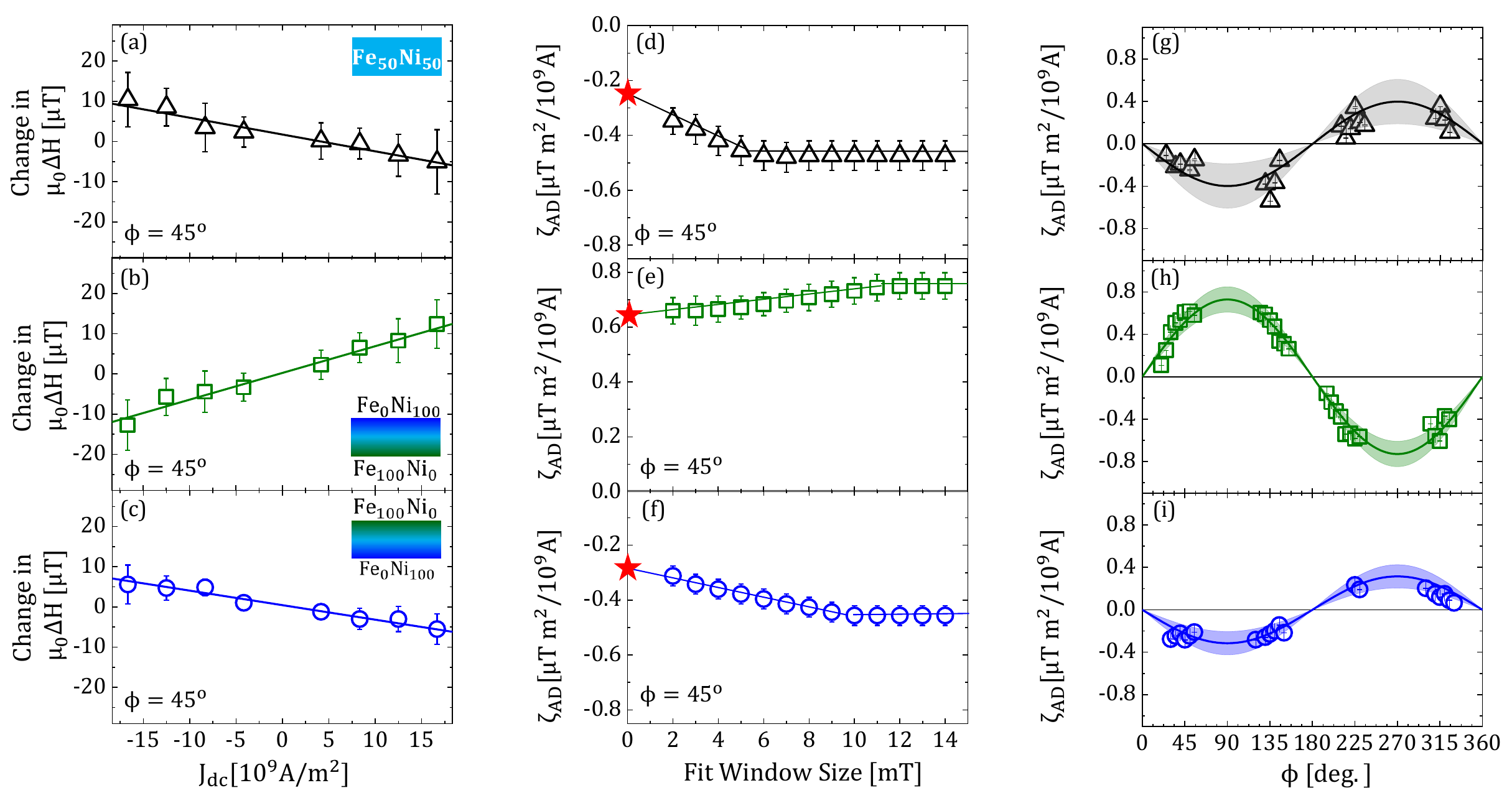}
    \caption{Left column (a-c): Change in linewidth $\Delta H$ due to dc bias current density $J_\mathrm{dc}$ at fixed in-plane field angle $\phi =45^\circ$ for (a) \Hom, (b) \FeBot, and (c)  \NiBot. The line indicates the linear fit to quantify the slope, $\zeta_\mathrm{AD}$. The error bars represent the standard deviation of 20 measurements. Center column (d-f): Change in $\zeta_\mathrm{AD}$ with fit window size for (d) \Hom, (e)\FeBot, and (f) \NiBot. The star at 0 $\mu$T indicates $\zeta_\mathrm{AD}$ as the fit window size goes to zero, following the protocol in Ref.~\cite{Karimeddiny2021}. Right column (g-i): $\zeta _\mathrm{AD}$ plotted against in-plane field angle $\phi$ for (g) \Hom, (h) \FeBot, and (i) \NiBot. The solid curve indicates the sinusoid $\propto \sin\phi$ whose amplitude is the mean of $\zeta_\mathrm{AD}/\sin\phi$ over all $\phi$, whereas the shaded region indicates $\pm$1 standard deviation.}
    \label{fig:DLT}
\end{figure*}

\subsection{Anti-damping SOT}
\label{subsec:DL-SOT}
Analyzing the shape of ST-FMR spectra (i.e., the ratio of the symmetric and antisymmetric Lorentzian components) is a common approach to quantify the anti-damping SOT~\cite{Liu2011, Nguyen2021}.  
However, ST-FMR spectra can contain spurious contributions from spin-pumping and thermoelectric voltage signals~\cite{Kondou2016, Okada2019, Schultheiss2012, Nguyen2021}. Moreover, the shape of a ST-FMR spectrum can be affected by a microwave current phase lag~\cite{TNan2024STFMR}. Therefore, we employed a more direct approach: injecting an additional dc current density $J_\mathrm{dc}$ and monitoring its effect on the linewidth $\Delta H$. A dc-current-induced anti-damping SOT manifests in a linear shift in $\Delta H$ with $J_\mathrm{dc}$~\cite{Liu2011, Kasai2014, Nan2015a}. 

Such a linear change is indeed observed for all three FeNi samples, as depicted in Fig.~\ref{fig:DLT}(a-c). 
The different signs of the slope for \FeBot\ and \NiBot\ appear consistent with opposite anti-damping SOTs due to opposite compositional gradients. 
Yet, we also observe a sizable slope for the compositionally homogeneous \Hom\ sample [Fig.~\ref{fig:DLT}(a)].

In the following, we closely inspect the linear slope of $\Delta H$ vs $J_\mathrm{dc}$, denoted as $\zeta_\mathrm{AD}$. 
The apparent value of $\Delta H$, hence $\zeta_\mathrm{AD}$, can be sensitive to the field range for fitting the ST-FMR spectra~\cite{Karimeddiny2021}, even if the spectral fits appear convincing.
Indeed, we find that the apparent $|\zeta_\mathrm{AD}|$ increases and then saturates with increasing fit window size [Fig.~\ref{fig:DLT}(d-f)], in qualitative agreement with Ref.~\cite{Karimeddiny2021}. Following the protocol in Ref.~\cite{Karimeddiny2021}, we report the lower-bound value of $|\zeta_\mathrm{AD}|$ extrapolated at zero fit window size, indicated as red stars in Fig.~\ref{fig:DLT}(d-f). 

Figure~\ref{fig:DLT}(g-i) summarizes $\zeta_\mathrm{AD}$ as a function of the in-plane applied field angle $\phi$. The data are adequately captured by the sinusoidal curve $\propto \sin \phi$. This observation can be attributed to a dc-induced spin accumulation polarized along the $\hat{y}$-axis (in-plane and transverse to the current axis, inset Fig.~\ref{fig:stfmr}(c)), which can emerge from a conventional spin Hall effect~\cite{Sinova2015}. In this case, the spin polarization is independent of the magnetization orientation, similar to recent reports of SOTs arising from FMs~\cite{Wang2019a, Soya2023}. 

Given the $\sin\phi$ angular dependence, we compute the dimensionless anti-damping SOT efficiency (sometimes called the ``effective spin Hall angle'') \cite{Liu2011}, 
\begin{equation}\label{DL_SE}
   \theta_\mathrm{AD} = \frac{2|e|\gamma}{h f} \left[H_\mathrm{res} + \frac{M_\mathrm{eff}}{2} \right] \mu _{0} M_\mathrm{s}t_\mathrm{FM} \frac{\zeta_\mathrm{AD}}{\sin\phi},
\end{equation}
where $\mu_0 M_\mathrm{s} = \mu_0 M_\mathrm{eff} \approx 1.5$ T. As summarized in Table~\ref{tab:DL}, the magnitudes of $\theta_\mathrm{AD}$ for the three FeNi samples approach $\sim$0.1. 
These values of $\theta_\mathrm{AD}$ are comparable to those for HM/FM bilayers~\cite{Shao2021, Manchon2019}. 

\begin{table}[b]
	\caption{Dimensionless anti-damping SOT  efficiency (effective spin Hall angle) $\theta_\mathrm{AD}$ quantified from Fig.~\ref{fig:DLT}(g-f). The error bars are obtained from the shaded regions in Fig.~\ref{fig:DLT}(g-f).} 
	\centering 
	\begin{tabular}{c c c c} 
		\hline 
            
		  & $\mathrm{Fe_{50}Ni_{50}}$ & $\mathrm{Fe_{100-x}Ni_{x}}$ & $\mathrm{Fe_{x}Ni_{100-x}}$\\ [4pt] 
		\hline 
          
            $\mathrm{\theta_\mathrm{AD}}$  & $-0.048 \pm 0.025$ & $0.088 \pm 0.016$ & $-0.038 \pm 0.013$\\
    
		\hline 

	\end{tabular}\label{tab:DL}
\end{table}

\begin{figure*}
 \centering
    \includegraphics[width = 7in, height = 3.7in]{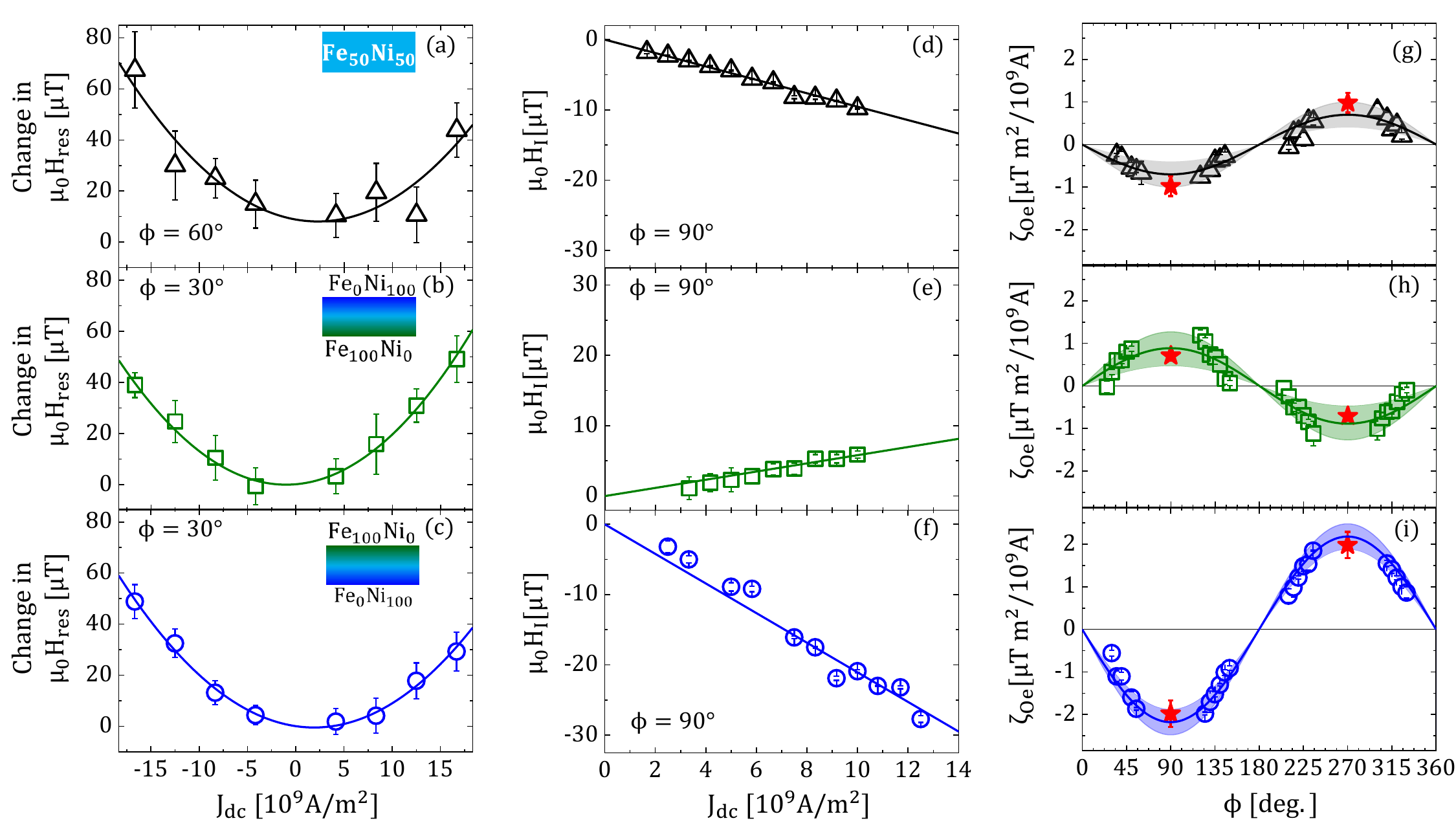}
    \caption{Left column (a-c): Change in resonance field $H_\mathrm{res}$ with $J_\mathrm{dc}$ for (a) \Hom, (b) \FeBot, and (c) \NiBot. The curve indicates the quadratic fit of the form  $\beta _\mathrm{heat} {J_\mathrm{dc}}^{2} + \zeta_\mathrm{Oe} J_\mathrm{dc}$, with $\zeta_\mathrm{Oe}$ representing the linear current-induced shift of $H_\mathrm{res}$. The error bars represent the standard deviation of 20 measurements. 
    Center column (d-f): Linear change in $\hat{y}$-oriented current-induced field $H_\mathrm{I}$ with $J_\mathrm{dc}$ from second-order PHE measurements for $\phi =$ $90^{\circ}$ for (d) \Hom, (e) \FeBot, and (f) \NiBot. Right column (g-i): $\zeta _\mathrm{Oe}$ plotted against in-plane field angle $\phi$ for (g) \Hom, (h) \FeBot, and (i) \NiBot. The solid curve indicates the sinusoid  $\propto \sin\phi$ whose amplitude is the mean of $\zeta _\mathrm{Oe}/\sin\phi$ over all $\phi$, whereas the shaded region indicates $\pm$1 standard deviation.
    The open symbols indicate $\zeta_\mathrm{Oe}$ 
    from ST-FMR measurements (left column), whereas the red star symbols indicate $\zeta^\mathrm{max}_\mathrm{Oe}$ from the second-order PHE measurements (center column) at 
    $90^\mathrm{o}$ and $270^\mathrm{o}$.}
    \label{fig:FLT}
\end{figure*}

We find opposite signs of $\theta_\mathrm{AD}$ with opposite compositional gradients\footnote{Here, $\theta_\mathrm{AD}<0$ $(>0)$ indicates that the polarity of the SOT is consistent with FM/Pt with Pt on top (bottom).}. 
This finding alone might suggest that the anti-damping SOT originates from the steep Fe-Ni compositional gradient. However, this simple scenario is not supported by the quantified values of $\theta_\mathrm{AD}$ [Table~\ref{tab:DL}]. First, $|\theta_\mathrm{AD}|$ is a factor of 2 greater for \FeBot\ than \NiBot; reversing the gradient not only reverses the sign of the SOT but also greatly influences its magnitude. More importantly, \Hom\ without any intentional compositional gradient exhibits a sizable $|\theta_\mathrm{AD}|$, similar in magnitude to the compositionally graded samples.  

These observations point to significant anti-damping SOTs with no clear correlation with the intentional compositional gradients. The key question is then: What is the mechanism giving rise to the observed anti-damping SOT? We address this question in the remainder of this article.

\subsection{Oersted field}
\label{subsec:FL-SOT}

The dc bias current (generating the dc anti-damping SOT examined in Sec.~\ref{subsec:DL-SOT}) can also generate an in-plane dc Oersted field oriented transverse to the current axis. This Oersted field causes the resonance field $H_\mathrm{res}$ to shift linearly with $J_\mathrm{dc}$~\cite{Nan2015a, Kim2018b}. In reality, as seen in Fig.~\ref{fig:FLT}(a-c), we also observe a significant quadratic shift in $H_\mathrm{res}$ with $J_\mathrm{dc}$, attributed to Joule heating. The overall shift in $H_\mathrm{res}$ with $J_\mathrm{dc}$ is fit with $\beta _\mathrm{heat} {J_\mathrm{dc}}^{2} + \zeta_\mathrm{Oe} J_\mathrm{dc}$. The linear coefficient $\zeta_\mathrm{Oe}$ captures the shift due to the Oersted field. 
Figure~\ref{fig:FLT}(g-i) summarizes $\zeta_\mathrm{Oe}$ obtained from dc-bias ST-FMR measurements at various values of $\phi$. 

Around $\phi \approx 90^\circ$ and $270^\circ$, we could not attain sufficient signal-to-noise ratios from dc-bias ST-FMR measurements. To fill in these gaps, the current-induced field $H_\mathrm{I}$, oriented in-plane and transverse to the current, was measured using the second-order planar Hall effect (PHE) method~\cite{Fan2014, Emori2016, Greening2020}. These PHE measurements were carried out on 500 $\micro$m wide strips, lithographically patterned at the same time as the rectangular ST-FMR strips. This PHE method measures the dc applied field needed to null the second-order planar Hall signal from $H_\mathrm{I}$; further details are found in Appendix~\ref{app:PHE} and Refs.~\cite{Fan2014, Emori2016, Greening2020}. Examples of $H_\mathrm{I}$ vs dc current density $J_\mathrm{dc}$ are displayed in Fig.~\ref{fig:FLT}(d-f). The observed slope is linear, which implies that the measured response is dominated by the current-induced field, rather than Joule heating. The values of the linear slopes from the PHE measurements are indicated as red stars at $\phi = 90^\circ$ and $270^\circ$ in Fig.~\ref{fig:FLT}(g-i). They are in good agreement with the $\phi$ dependence of $\zeta_\mathrm{Oe}$ from the dc bias ST-FMR method. 

The combined ST-FMR and PHE results in Fig.~\ref{fig:FLT} are adequately fit with the sinusoid $\propto \sin\phi$. This angular dependence is consistent with the symmetry of the dc Oersted field. To represent this $\hat{y}$-oriented Oersted field, we normalize $\zeta_\mathrm{Oe}$ to
\begin{equation}\label{eq:ZFL}
Z_\mathrm{Oe} = \frac{\zeta_\mathrm{Oe}}{\sin\phi}.
\end{equation}
The resulting values of $Z_\mathrm{Oe}$ from the sinusoidal fits of $\zeta_{\mathrm{Oe}}$ are summarized in Table~\ref{tab:fieldlike}. \FeBot\ and \NiBot\ with opposite compositional gradients exhibit opposite signs, as intuitively expected. Yet, $|Z_\mathrm{Oe}|$ is a factor of 2 smaller for \FeBot\ compared to \NiBot. 
Furthermore, \Hom\ with nominally no compositional gradient also exhibits non-negligible $Z_\mathrm{Oe}$ with the same sign as \NiBot. 

\begin{table}[b]
	\caption{Current-induced field (assumed to be the Oersted field) per unit current density $Z_\mathrm{Oe}$. The error bars are obtained from the shaded regions in Fig.~\ref{fig:FLT}(g-f). The average Joule heating coefficient $\beta _{\mathrm{heat}}$ over the full angular range is also shown, with the error bars indicating $\pm$1 standard deviation.}  
	\centering 
	\begin{tabular}{c c c c} 
		\hline 
           
		{}  & \Hom & \FeBot & \NiBot\\ [3pt] 
		\hline 
         $Z_\mathrm{Oe}$ $\mathrm{\left( \frac{\mu T m^{2}}{10^{9} A} \right)}$ & $-0.70 \pm 0.06$ & $0.87 \pm 0.06$ & $-2.18 \pm 0.04$ \\ [6pt] 
        $\beta _\mathrm{heat}$ $\mathrm{\left( \frac{\mu T m^{4}}{10^{18} A^{2}}\right)}$ & $12\pm 6$ & $12\pm 6$ & $2 \pm 7$\\ [6pt] 
		\hline 
        \label{tab:fieldlike}
	\end{tabular}
\end{table}

\begin{figure*}
 \centering
    \includegraphics[width = 7in]{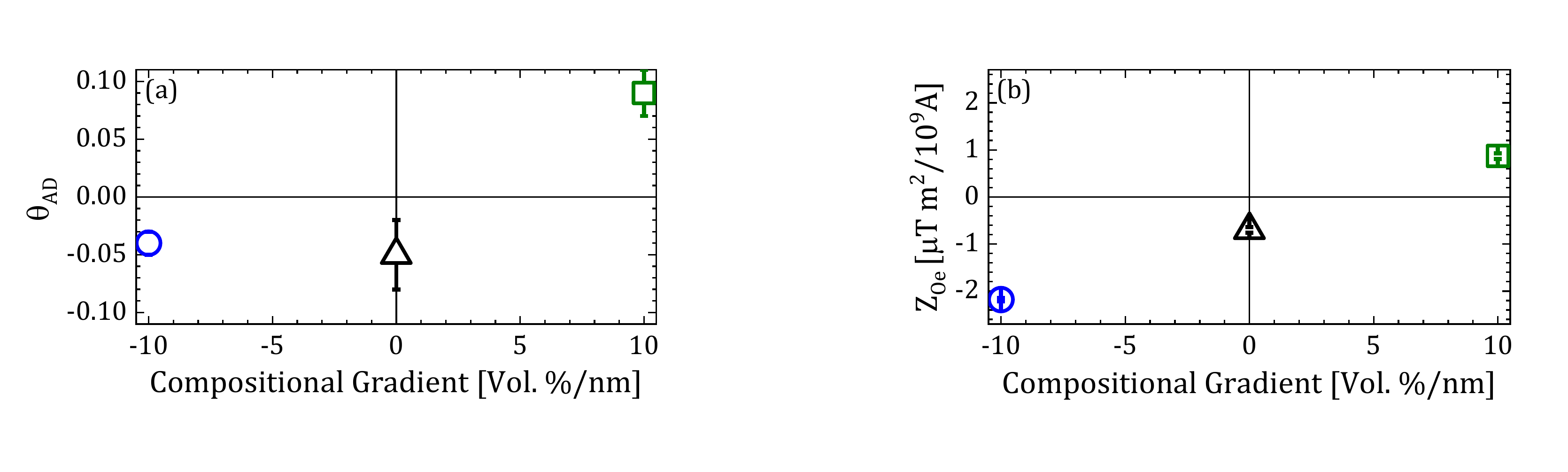}
    \caption{Compositional gradient dependence of (a)  anti-damping SOT efficiency $\theta_\mathrm{AD}$ and (b) current-induced field per unit current density $Z_\mathrm{Oe}$. The error bars indicate $\pm$1 standard deviation computed over the full angular dependence (shaded region in Figs.~\ref{fig:DLT}(g-i) and \ref{fig:FLT}(g-i)).}
    \label{fig:ThetaandGrad}
\end{figure*}

The maximum possible dc Oersted field -- in the most extreme case with all of the bias current flowing immediately above (or below) the FM layer -- is $|H^\mathrm{max}_\mathrm{Oe}| = |J_\mathrm{dc}| t_\mathrm{FM}/2$. Then, the maximum magnitude of $Z_\mathrm{Oe}$ here is  
\begin{equation}
    |Z^\mathrm{max}_\mathrm{Oe}| = \frac{\mu_0 |H^\mathrm{max}_\mathrm{Oe}|} {|J_\mathrm{dc}|} 
    = \frac{\mu_0 t_\mathrm{FM}}{2} \approx 6.3~\frac{\mathrm{\mu T m^2}}{10^9 \mathrm{A}}.
\end{equation}
The magnitudes of $Z_\mathrm{Oe}$ [Table~\ref{tab:fieldlike}] are all well below $|Z^\mathrm{max}_\mathrm{Oe}|$. Our results imply that the Oersted field can account for the entire observed current-induced field. In other words, only a small asymmetry in current distribution is needed, with a slight difference in conductivity between the top and bottom portions of the film. For example, the Ni-rich bottom of \NiBot\ needs to be just $\approx$30\% more conductive than the Fe-rich top\footnote{We find that the sheet resistance is $\approx 20 \Omega/\Box$ for a Ti/Cu-seeded and capped 10 nm thick Ni film and $\approx 24 \Omega/\Box$ for a similarly seeded and capped 10 nm thick Fe film, corroborating that Ni is slightly more conductive than Fe.}. 
Moreover, the observed current-induced field in \Hom\ requires the bottom to be only $\approx$10\% more conductive than the top. For instance, even though \Hom\ has no intentional symmetry breaking, the bottom Cu/\Hom\ interface may be smoother – hence exhibiting slightly less electronic scattering and higher conductivity – than the top \Hom/Cu interface.

Some portion of the current-induced field possibly originates from spin-orbit effects (i.e., ``field-like'' SOT)~\cite{Manchon2019, Fan2014, Greening2020}. However, we emphasize that an uncompensated Oersted field adequately explains our observation, without invoking any additional spin-orbit field. It is natural to deduce that the field-like SOT is likely much weaker than the classical Oersted field torque in the FeNi samples examined here.

\subsection{Current-induced torques vs nominal compositional gradient}
\label{subsec:SOT-context}
To gain a broader perspective on the current-induced torques, we compare how the anti-damping SOT and the Oersted field depend on the nominal compositional gradient in Fig.~\ref{fig:ThetaandGrad}. 
The anti-damping SOT efficiency $\theta_\mathrm{AD}$ exhibits a nonlinear trend [Fig.~\ref{fig:ThetaandGrad}(a)]. In contrast, the Oersted field parameter $Z_\mathrm{Oe}$ appears to scale linearly with the compositional gradient [Fig.~\ref{fig:ThetaandGrad}(b)], albeit with an offset yielding nonzero $Z_\mathrm{Oe}$ for \Hom. 
As discussed in Sec.~\ref{subsec:FL-SOT}, the linear scaling for $Z_\mathrm{Oe}$ is readily explained by an asymmetric current distribution that approximately tracks the compositional gradient. Yet, the origin of the anti-damping SOT is far less clear from the results in Fig.~\ref{fig:ThetaandGrad} alone. 

Nevertheless, it is instructive to re-emphasize key points from the results in Fig.~\ref{fig:ThetaandGrad}. First and foremost, the intentional compositional gradient cannot fully account for the observed anti-damping SOT. Similarly, an uncompensated spin accumulation from any current imbalance (i.e., conductivity asymmetry from compositional asymmetry) is not the primary contribution to the anti-damping SOT; if it were, $\theta_\mathrm{AD}$ would have to show a similar scaling as  $Z_\mathrm{Oe}$. 
The above points suggest that we must consider an alternative source of symmetry breaking for the anti-damping SOT.

\section{Depth-Resolved Properties}\label{sec:depth}
We now examine the origin of the anti-damping SOT without any clear scaling with the intentional Fe-Ni compositional gradient. We posit two possibilities of \emph{unintentional} symmetry breaking that can generate the unexpected SOT:

First, the actual compositional profile along the film thickness might be significantly different from the intended one illustrated in Fig.~\ref{fig:FSL}. For instance, the nominally homogeneous \Hom\ film might exhibit a significant compositional gradient~\cite{Inyang2023} due to phase segregation. Alternatively, atomic intermixing during deposition can lead to non-negligible gradients at film interfaces. In Sec.~\ref{subsec:PNR}, we present spin-polarized neutron reflectometry (PNR) measurements that verify the compositional and magnetic depth profiles of the three FeNi films. 

Second, a change in the lattice parameter -- i.e., a strain gradient -- along the FM thickness might provide the required symmetry breaking for the SOT. Such a strain gradient might arise during film growth (e.g., governed by mismatch in lattice parameters among the different film layers). 
Indeed, previous experiments have suggested strain gradients as a possible mechanism for SOTs in nominally homogeneous FMs~\cite{Zhu2021c}, although no direct evidence has been presented for such strain gradients. In Sec.~\ref{subsec:GIXRD}, we directly quantify the change of the lattice parameter along the film thickness through grazing-incidence x-ray diffraction (GI-XRD). 

PNR and GI-XRD are \emph{non-destructive} methods, enabling depth-resolved characterization without any irreversible sample damage. This is in contrast to cross-sectional transmission electron microscopy, in which the milling for sample preparation can irreversibly relax the built-in strain in the film. Our PNR and GI-XRD measurements were performed on unpatterned films with lateral dimensions $> 2\times2~\mathrm{cm}^2$. Some films were subjected to a process that emulated the heating during the microfabrication of the patterned samples for the SOT experiments; the films were coated with photoresist, baked, and ion-milled for the same duration as the patterned samples in Sec.~\ref{sec:torques}. The films with and without heating exhibit no systematic difference in the PNR and GI-XRD results. 
In the following, we present PNR and GI-XRD results from the ``heated'' samples.

\subsection{Compositional gradient}\label{subsec:PNR}
\begin{figure*}
 \centering
    \includegraphics[width = 7.0in]{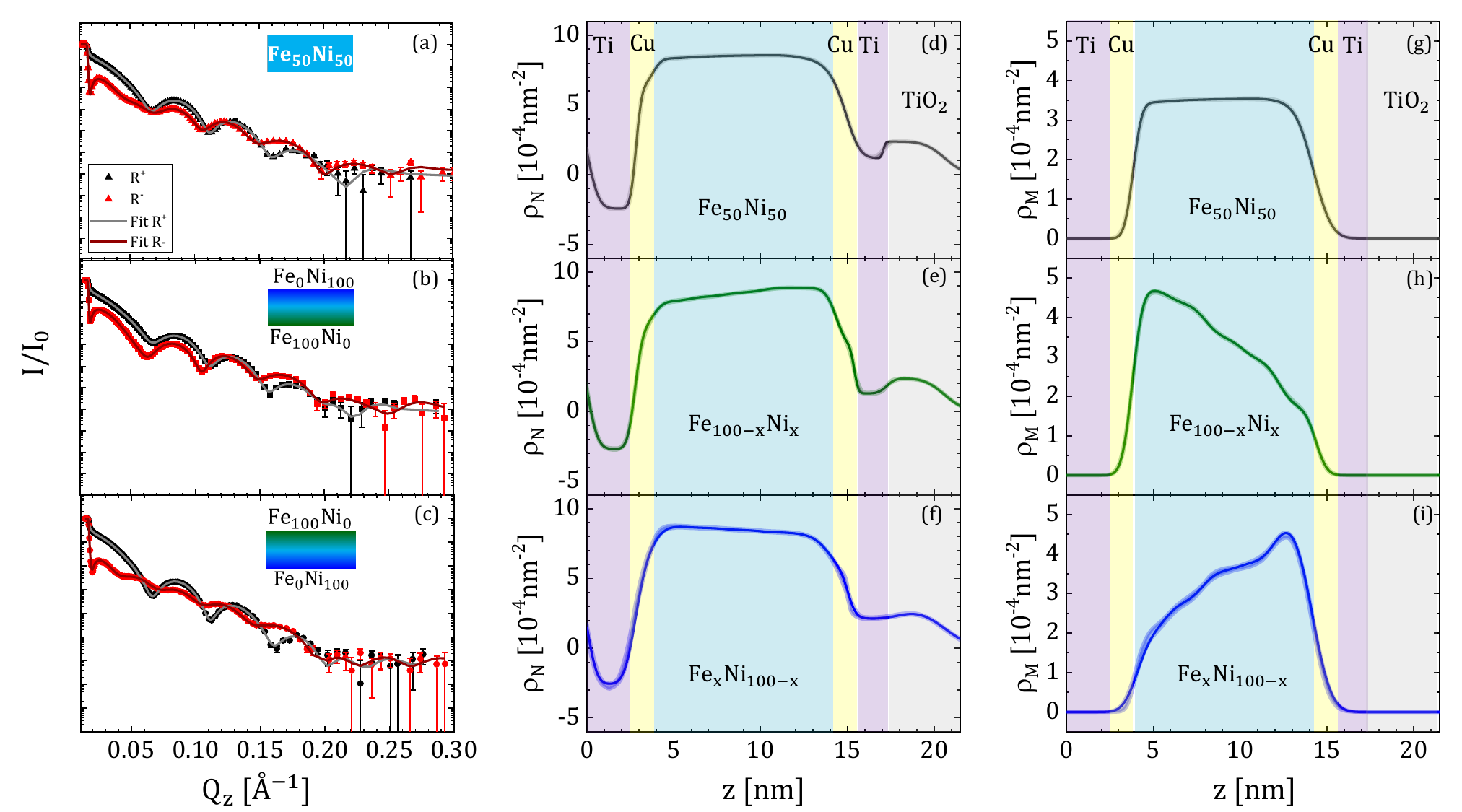}
    \caption{Left column (a-c): Normalized reflectivity in reciprocal space for polarized neutrons spin up, $R^{+}$, or spin down, $R^{-}$, (closed symbols) for (a) $\mathrm{Fe_{50}Ni_{50}}$, (b) $\mathrm{Fe_{100-x}Ni_{x}}$, (c) $\mathrm{Fe_{x}Ni_{100-x}}$. Theoretical fits solid line grey ($R^{+}$), maroon ($R^{-}$).  Center column (d-f): Nuclear scattering length density $\rho_\mathrm{N}$ with film thickness $z$ for (d) \Hom, (e) \FeBot, and (f) \NiBot. Right column (g-i): magnetic scattering length density $\rho_\mathrm{M}$ and corresponding magnetization $M$ (1 kA/m = $2.91 \times 10^{-7}$ nm$^{-2}$) with film thickness $z$, for (g) $\mathrm{Fe_{50}Ni_{50}}$ (h) $\mathrm{Fe_{100-x}Ni_{x}}$, and (i) $\mathrm{Fe_{x}Ni_{100-x}}$.  Error bars represent $\pm$ 1 standard deviation. Shaded bands indicate the 95\% confidence bands of the best-fit depth profiles, determined by Markov chain Monte Carlo calculations.}
    \label{fig:PNR}
\end{figure*}
To verify the compositional and magnetic gradients, PNR was performed using the Polref instrument at the ISIS Neutron and Muon Source. The probed films were grown on thermally-oxidized Si and (0001)-oriented sapphire substrates. The results were essentially identical irrespective of the substrate; here, we show PNR results for the films on sapphire substrates, which provide better nuclear scattering length density contrast with the film. The measurements were conducted under an in-plane applied magnetic field of 0.7 T, sufficient to saturate the samples. The neutron beam was spin-polarized parallel or antiparallel to the field, and the corresponding reflectivity cross sections ($R^+$ and $R^-$) were measured as a function of the perpendicular scattering wavevector $Q_\mathrm{z}$. The obtained PNR data are shown in Fig.~\ref{fig:PNR}(a-c). 

PNR depends on the depth profiles of the nuclear scattering length density $\rho_\mathrm{N}$ and magnetic scattering length density $\rho_\mathrm{M}$. We use the Refl1D package~\cite{Ref1Dweb} to fit the PNR data [Fig.~\ref{fig:PNR}(a-c)] and extract the profiles of the composition ($\propto \rho_\mathrm{N}$ in Fig.~\ref{fig:PNR}(d-f)) and net in-plane magnetization ($\propto \rho_\mathrm{M}$ in Fig.~\ref{fig:PNR}(g-i)) along the vertical coordinate $z$. In modeling each sample, the FM layer is represented by five equal-thickness sections with equally rough interfaces to approximate a smooth gradient of alloy composition and magnetization. The fitting reproduces the PNR results well, as seen in Fig.~\ref{fig:PNR}(a-c). The derived depth profiles, as summarized in Fig.~\ref{fig:PNR}(d-i), are also in good agreement with the nominal film stack structures [Fig.~\ref{fig:FSL}]. The only major exception is that the top Ti layer is partially oxidized, which is reasonable as the films are exposed to ambient air. The parameters in our modeling are summarized in Table~\ref{tab:SLD}. 

The nominally asymmetric \FeBot\ and \NiBot\ samples show clear linear slopes in $\rho_\mathrm{N}$ [Fig.~\ref{fig:PNR}(e,f)] and $\rho_\mathrm{M}$ [Fig.~\ref{fig:PNR}(h,i)] with opposite direction within the FM layer. The direction of the slope in $\rho_\mathrm{N}$ agrees with the intended compositional profile. 
Similarly, the slope in $\rho_\mathrm{M}$ is also consistent with the compositional gradient, which naturally leads to a magnetic gradient. Given the greater magnetism of Fe compared to Ni, the Fe-rich bottom of \FeBot\ leads to larger $\rho_\mathrm{M}$ at lower $z$. Indeed, the linearly extrapolated values of $\rho_\mathrm{M}$ at the top ($z = 10$ nm) and bottom ($z = 0$ nm) interfaces agree well with the tabulated values of $\approx 5\times10^{-4}~\mathrm{nm}^{-2}$ for Fe and $\approx 1.5\times10^{-4}~\mathrm{nm}^{-2}$ for Ni~\cite{Fitzsimmons2005} -- further giving credence to the quantitative accuracy of our PNR fit results. 

\begin{table}[b]
	\caption{Material parameters and scattering length densities for various compositions and densities from Figs.~\ref{fig:PNR} and \ref{fig:gixrdfcc}.} 
	\centering 
	\begin{tabular}{l l |c |c |c}
		\hline 
           
		{}  & & \Hom & \FeBot & \NiBot \\ 
		\hline 
         & atomic composition & Fe$_{48}$Ni$_{52}$ & Ni & Fe \\ 
		& $a_\textrm{fcc}$ (nm) & 0.3557 & 0.3536 & 0.3560 \\ 
	Top	& $\rho$ (g/cm$^3$) & 8.460 & 8.818 & 8.22 \\ 
		& $\rho_\mathrm{N}$ (10$^{-4}$ nm$^{-2}$) & 8.791 & 9.319 & 8.377 \\ 
            & $\mu_0M_\mathrm{s}$ (T)~\cite{Schoen2017b} & 1.51 & 0.45 & 2.05  \\
		& $\rho_\mathrm{M}$ (10$^{-4}$ nm$^{-2}$) & 3.723 & 1.15 &  4.90\\ 
		\hline 
           & atomic composition & Fe$_{48}$Ni$_{52}$ & Fe & Ni \\ 
		& $a_\textrm{fcc}$ (nm) & 0.3573 & 0.3546 & 0.3560 \\ 
	Bot.	& $\rho$ (g/cm$^3$) & 8.347 & 8.319 & 8.640 \\ 
		& $\rho_\mathrm{N}$ (10$^{-4}$ nm$^{-2}$) & 8.673 & 8.478 & 9.131 \\ 
              & $\mu_0M_\mathrm{s}$ (T)~\cite{Schoen2017b} & 1.51 & 2.05 & 0.45 \\
		& $\rho_\mathrm{M}$ (10$^{-4}$ nm$^{-2}$) & 3.673 & 4.96 & 1.13 \\ 
 \hline 
          &  $\chi^2$ of PNR fit & 1.165(55) & 1.090(18) & 1.217(69) \\  
		\hline 
        \label{tab:SLD}
	\end{tabular}
\end{table}

The PNR results for \Hom\ are also largely consistent with the intended film stack structure, showing a nearly uniform depth profile within bulk of the FM layer. Although the gradients of $\rho_\mathrm{N}$ and $\rho_\mathrm{M}$ are in fact nonzero in the bulk of \Hom 
 [Table~\ref{tab:SLD}], they are tiny -- more than an order of magnitude smaller than the gradients in \FeBot\ and \NiBot. Thus, the unintentional bulk compositional or magnetic gradient is highly unlikely to be the primary source of asymmetry for the SOT in \Hom. 

We do find non-negligible asymmetry between the bottom Cu/\Hom\ and top \Hom/Cu interfaces. In particular, the top interface exhibits shallower gradients in both $\rho_\mathrm{N}$ and $\rho_\mathrm{M}$, which can arise due to greater atomic intermixing~\cite{Bandiera2011}. Such asymmetry between the interfaces has been reported to give rise to an uncompensated SOT and Dzyaloshinkii-Moriya interaction in sub-nm thin FM sandwiched between Pt~\cite{Haazen2013, Je2013}.  
However, we find it difficult to justify that the asymmetric interfaces sufficiently account for the sizable SOT in \Hom. The large FM thickness of 10 nm would likely make it difficult for the uncompensated interfacial spin accumulations to generate a sizable torque. More crucially, as shown in Fig.~\ref{fig:PNR}, the overall compositional and magnetic asymmetries are clearly much greater in the graded \FeBot\ and \NiBot\ than in \Hom. It is hence doubtful that the asymmetries apparent in the PNR results [Fig.~\ref{fig:PNR}] alone are responsible for the SOT in \Hom. 
Rather than a mechanism originating from the \emph{compositional} profile (and, by association, magnetic profile) in the bulk or at the interfaces, we suspect \emph{structural} asymmetry to play a critical role, particularly in \Hom.

\subsection{Strain gradient} 
\label{subsec:GIXRD}
\begin{figure*}
 \centering
    \includegraphics[width = 6.0in]{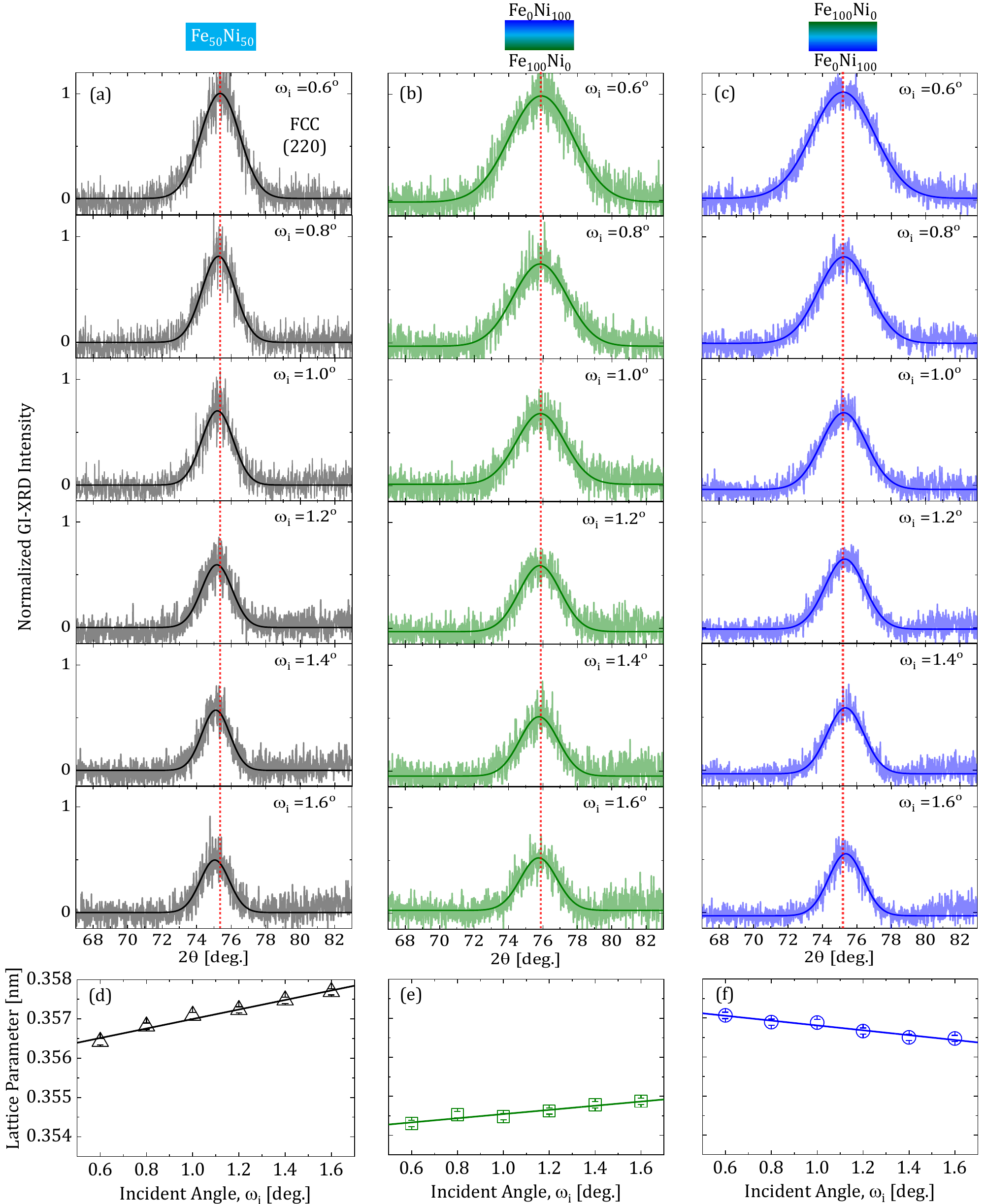}
    \caption{(Bottom Panel) (a-c) Grazing-incidence XRD spectra, with the shift from refraction corrected (see Appendix~\ref{app:refraction} for details), from incident angle $\omega _{i} = 0.60^{\circ} $ (top side of the FM) to $\omega _{i} = 1.60^{\circ} $ (bottom side of the FM) for (a) \Hom, (b) \FeBot, and (c) \NiBot. The intensity is normalized by the amplitude of the Gaussian fit peak at $\omega_{i}= 0.60^{\circ}$. (d-f) Change in lattice parameter with $\omega_{i}$ for (d) \Hom, (e) \FeBot, and (f) \NiBot. The error bars, propagated from the uncertainty of $\approx 0.02^\circ$ in the peak center in (a-c), are of the order of the symbol size.}
    \label{fig:gixrdfcc}
\end{figure*}
We employed GI-XRD to evaluate the lattice parameter as a function of depth. 
Unlike conventional XRD where the Bragg diffraction angle $2\theta$ is scanned and the incident angle $\omega_\mathrm{i}$ is set to $\theta$ (i.e., the incident and diffracted beams are symmetric),  GI-XRD is performed by scanning $2\theta$ while setting $\omega_\mathrm{i}$ to a much smaller value, typically around $\sim 1^\circ$ (i.e., the incident and diffracted beams are asymmetric). While conventional XRD is primarily sensitive to crystal planes parallel to the sample surface, GI-XRD can pick up diffraction from crystal planes tilted from the sample surface. 

When $\omega_\mathrm{i}$ is below the critical angle in x-ray reflectometry (($2\theta_\mathrm{c})/2 \approx 0.4^\circ$ here), most of the x-ray beam does not penetrate into the sample. 
Increasing $\omega_\mathrm{i}$ beyond the critical angle permits the x-ray beam to penetrate deeper into the film, thereby yielding a diffraction signal for different depths along the film thickness. 
At $\omega_\mathrm{i} \gtrsim 1.8^\circ$, the diffraction signal is dominated by the substrate plane, Si(311), indicating that the beam penetrates through the film. Therefore, we focus on GI-XRD measurements at $\omega_\mathrm{i}$ between $0.6^\circ$ and $1.6^\circ$, with an $\omega_\mathrm{i}$ step size of $0.2^\circ$, to acquire the diffraction response primarily from the 10 nm thick FM. 

We begin by discussing results from \Hom, for which GI-XRD provides the most useful insight into the symmetry breaking presumably underlying the anti-damping SOT. As shown in Fig.~\ref{fig:gixrdfcc}(a), we observe a pronounced diffraction peak around $2\theta \approx 75^\circ$, attributed to the (220) plane of face-centered cubic (fcc) \Hom~\cite{Schoen2017b}. We do not observe a diffraction peak around $2\theta \approx 44^\circ$ that would arise from the presence of an Fe-rich body-centered cubic (bcc) phase~\cite{Wu2022}. Thus, our GI-XRD results corroborate a homogeneous crystal phase across the thickness of \Hom. 

Despite the homogeneous crystal phase in \Hom, a systematic linear shift in the position of the diffraction peak [Fig.~\ref{fig:gixrdfcc}(a)] is observed. 
We remark that the refraction of the x-ray beam can cause the diffraction peak to shift~\cite{Noma1999, Colombi2006}, but this refraction-induced peak shift is readily computed, as explained in Appendix~\ref{app:refraction}. In Fig.~\ref{fig:gixrdfcc}, the GI-XRD results are shown with the refraction-induced peak shift subtracted.  
By applying Bragg's law, we obtain a lattice parameter of $\approx 0.3565$ nm near the top of the \Hom\ film ($\omega_\mathrm{i} = 0.6^\circ$), which is close to the bulk \Hom\ lattice parameter of $\approx$ 0.356 nm. We find a larger lattice parameter of $\approx 0.3576$ nm near the bottom of the film ($\omega_\mathrm{i} = 1.6^\circ$), which is closer to the bulk lattice parameter $\approx$ 0.361 nm of fcc Cu\footnote{The ultrathin Cu layer itself is likely strained, so its lattice parameter is not necessarily equal to the bulk value}.
Our observation is consistent with the \Hom\ lattice strained on the bottom to match the underlying Cu seed layer and progressively relaxing toward the top. Assuming that $\omega_\mathrm{i} = 0.6-1.6^\circ$ primarily probes the film as explained previously, we estimate a linear strain gradient of $\sim 0.3\%$ over the \Hom\ thickness of 10 nm. Moreover, this strain gradient is qualitatively consistent with the slight density gradient in \Hom\ seen using PNR [Table~\ref{tab:SLD}].

We observe similar gradients for an as-grown \Hom\ sample and the \Hom\ sample subjected to heating and milling that emulate the microfabrication steps -- corroborating the notion that the strain gradient arises during film growth, rather than any post processes. 
This growth-induced residual strain gradient appears to be the most plausible symmetry-breaking mechanism that accounts for the sizable anti-damping SOT in \Hom. 
Our GI-XRD work demonstrates that a strain gradient provides the needed symmetry breaking for the SOT, even in a single-layer FM without any significant compositional asymmetry.  

The compositionally graded \FeBot\ and \NiBot\ samples exhibit both bcc (110) and fcc (220) diffraction peaks. The intentional compositional gradients lead to mixed crystalline phases, with Fe-rich regions being bcc and Ni-rich regions being fcc~\cite{Schoen2017b}. However, the signal-to-noise ratio of the bcc (110) peak is insufficient for precisely quantifying the lattice parameter. Therefore, we focus on analyzing the fcc (220) diffraction peak [Fig.~\ref{fig:gixrdfcc}(b,c)].  
We observe strain gradients of $\sim 0.15$\% of opposite polarity in \FeBot\ and \NiBot\ [Fig.~\ref{fig:gixrdfcc}(e,f)]. This trend is qualitatively reasonable: the lattice parameter towards the top of \FeBot\ approaches that of Ni ($\approx 0.353$ nm), whereas the lattice parameter towards the top of \NiBot\ approaches that of Fe-rich fcc FeNi alloy ($\approx 0.360$ nm)~\cite{Schoen2017b}.


\section{Discussion}
\label{sec:perspective}
It is instructive to consider how the strain gradient permits uncompensated current-induced spin accumulations that exert torques on the magnetization. 
A local strain may sufficiently modify the electronic band structure, and hence the intrinsic spin Hall effect, as corroborated by studies of SOTs in films on piezoelectric and bendable substrates~\cite{Filianina2020, Wong2021strainSOT}. 
In our case with FM films on rigid substrates, a built-in residual strain gradient can establish a gradient of the intrinsic spin Hall effect. 
While a uniform spin Hall effect in the FM would yield a net zero non-equilibrium spin accumulation~\cite{Wang2019a}, the strain-induced \emph{graded} spin Hall effect can generate a net nonzero spin accumulation throughout the FM thickness. 
Alternatively, the strain gradient may produce a graded orbital Hall effect~\cite{Go2018}, yielding a net orbital accumulation that is converted to a spin accumulation via spin-orbit coupling~\cite{Go2020a}. This is an intriguing possibility, given that the lattice strain should couple more directly to the orbital (rather than spin) degree of freedom in the FM. 

In \FeBot\ and \NiBot, the steep compositional gradients may still be the primary source of asymmetry for the anti-damping SOT. Yet, as summarized in Fig.~\ref{fig:ThetaandGrad}(a) and Table~\ref{tab:DL}, \FeBot\ with a ``positive'' strain gradient [Fig.~\ref{fig:gixrdfcc}(e)] exhibits a factor of 2 greater anti-damping SOT than \NiBot\ with a ``negative'' strain gradient [Fig.~\ref{fig:gixrdfcc}(f)]. This observation suggests that a vertical strain gradient of a certain polarity may be more effective in enhancing the anti-damping SOT, although the reason for this is yet unclear. 
It is possible that the magnitude of the graded spin Hall effect depends on the strain state, governed by whether Fe or Ni is on the bottom.
For further enhancement of the SOT, the strain gradient may be deliberately tuned by varying the film thickness, the sputtering gas pressure, the substrate, or the seed layer composition. It would be also useful to investigate the tunability and robustness of the strain gradient effect against annealing.   
More broadly, our present study indicates that asymmetry in chemical composition is not the only route to SOTs. Rather, asymmetry in microstructure may be another effective approach to enhance SOTs.  

One remaining mystery is the polarity of the anti-damping SOT in \Hom. In particular, \Hom\ and \FeBot\ share the same strain gradient direction [Fig.~\ref{fig:gixrdfcc}(d,e)], but they exhibit opposite anti-damping SOT polarities [Fig.~\ref{fig:ThetaandGrad}]. This discrepancy suggests the need for a more nuanced explanation, perhaps involving the qualitative structural difference between the homogeneous \Hom\ alloy (purely fcc) and the graded \FeBot\ alloy (mixed bcc + fcc phases). We may not be able to make a straightforward comparison of anti-damping SOT mechanisms between these two distinct alloys. 
Further studies are warranted to uncover the full interplay of compositional and strain gradients in FMs. 

\section{Summary}
Our work demonstrates that low damping and a sizable SOT can coexist in symmetry-broken single-layer polycrystalline FMs. In 10 nm thick FeNi with and without intentional compositional gradients, we have quantified effective damping parameters of $\alphaEff < 5\times 10^{-3}$ (lower than $\approx 7 \times 10^{-3}$ in permalloy) and anti-damping SOT efficiencies approaching $|\theta_\mathrm{AD}|$ of order 0.1 (on par with oft-studied HM/FM bilayers). These findings illuminate a new path toward developing low-loss Type-Y spintronic memories and oscillators. Moreover, the sizable SOT in these single-layer all-3$d$ FMs supports the recent notion that the HM in HM/FM bilayers is not the sole source of current-induced spin accumulation -- but rather the FM can host its own current-induced spin accumulation that generates a ``self torque'' within itself~\cite{Davidson2020a, Kim2020, Wang2019a}. 

Another notable implication is that we may not necessarily need compositional asymmetry to attain a sizable SOT. 
In particular, a built-in \emph{strain gradient} in \Hom\ here may contribute to an anti-damping SOT comparable to that from compositional gradients. 
Such a strain gradient within the FM bulk could possibly explain SOTs reported in compositionally symmetric single-layer FMs~\cite{Seki2021, Fu2022}. Future studies may deliberately engineer the strain gradient to enhance the SOT efficiency. 
Our work motivates further endeavors to uncover the fundamental impacts of compositional and strain gradients on spintronic (and potentially orbitronic) phenomena for energy-efficient nanomagnetic devices.

\begin{acknowledgments}
R.E.M., J.L.J., and S.E. were supported by the National Science Foundation under Grant No. ECCS-2144333. D.A.S. and W.C.T. were supported by the National Science Foundation under Grant No. DMR-2003914. Y.L. was supported by the Air Force Office of
Scientific Research (AFOSR) under Grant No. FA9550-21-1-0365. P.P.B. was supported by an NRC RAP award. We thank Vivek P. Amin for helpful discussions. We thank the ISIS Neutron and Muon Source for the provision of beamtime.  Raw PNR data from the Polref instrument can be accessed at \href{https://doi.org/10.5286/ISIS.E.RB2210102}{https://doi.org/10.5286/ISIS.E.RB2210102}.

The authors contributed to the following components of this work. 
Conceptualization: S.E. 
Project management: R.E.M and S.E. 
Film growth: R.E.M. and S.W.
In-plane FMR: R.E.M. and S.W. 
Out-of-plane FMR: P.N., B.N., and T.M.
Microfabrication: R.E.M., S.W., D.A.S, and Y.L. 
ST-FMR: R.E.M. and S.E.
Second-order PHE: R.E.M., J.L.J., W.C.T., and S.E. 
PNR: P.P.B., A.J.G., C.J.K., and A.J.C.
GI-XRD: R.E.M., J.Z., F.M.M., and S.E.
Initial draft: R.E.M., P.P.B., A.J.G., and S.E.
Review of the initial draft: all.
Final draft: R.E.M., S.W., P.P.B., A.G.J., T.M., and S.E.

\end{acknowledgments}

\appendix
\section{Spin-Torque FMR Methods}
\label{app:STFMR}
For ST-FMR measurements, rectangular strips were patterned by photolithography and Ar ion milling. 
An additional layer of photolithography and liftoff was performed to contact these strips with electrodes, consisting of Ti(3 nm)/Cu(100 nm)/Pt(2 nm). 
A ground-signal-ground probe was used to apply a 7 GHz microwave current at a power of +8 dBm. An in-plane applied field is swept at in-plane angle $\phi$ with respect to the current axis. The microwave current generates oscillatory torques that drive the magnetization about the field, leading to a change in the anisotropic magnetoresistance at the frequency of 7 GHz. The mixing of the microwave current and oscillating resistance produces a rectified voltage signal $V_{\mathrm{mix}}$~\cite{Liu2011}. We detected $V_{\mathrm{mix}}$ using a lock-in amplifier with a reference frequency of 1777.77 Hz for amplitude modulation of the microwave current. Figure~\ref{fig:stfmr} shows examples of ST-FMR spectra, obtained by acquiring $V_\mathrm{mix}$ while sweeping the in-plane applied field $H$. These spectra are fit with the generalized Lorentzian of the form 
\begin{equation}\label{eq:Vmix}
    V_\mathrm{mix} = \frac{\Delta H}{\Delta H^{2} +\left( H - H_\mathrm{res} \right)^{2}} \left[S\Delta H + A( H - H_\mathrm{res}) \right],
\end{equation}
where $S$ and $A$ are the coefficients for the symmetric and antisymmetric parts of the Lorentzian. From Eq.~\ref{eq:Vmix}, we quantify the resonance field $H_\mathrm{res}$ and the half-width-at-half-maximum FMR linewidth $\Delta H$. 


Our experimental setup limits the maximum dc current amplitude to 10 mA. By taking the width 50 $\micro$m and conductive thickness 12 nm (i.e., accounting for the FeNi layer and the Cu seed and capping layers, while ignoring the highly resistive Ti layers), we have an average current density of up to $|J_\mathrm{dc}| = 17\times 10^{9}$ A/m$^2$. At each value of $J_\mathrm{dc}$, we averaged 20 measurements to quantify $\Delta H$ and $H_\mathrm{res}$.

\begin{figure}[t]
 \centering
    \includegraphics[width = 3.3in]{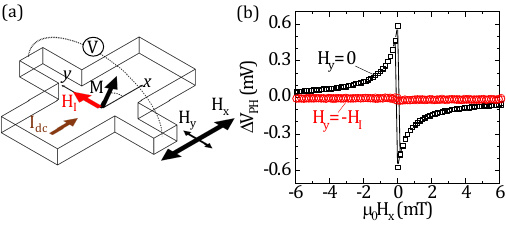}
    \caption{(a) Schematic of the second-order PHE measurement. (b) Examples of second-order planar Hall voltage $\Delta H_\mathrm{PH}$ vs swept field $H_\mathrm{x}$, where the current-induced field $H_\mathrm{I}$ is uncompensated ($H_\mathrm{y}=0$) and almost entirely compensated ($H_\mathrm{y}=-H_\mathrm{I}$).}
    \label{fig:PHE}
\end{figure}

\section{Second-Order Planar Hall Effect Methods}
\label{app:PHE}
The second-order PHE technique, illustrated in Fig.~\ref{fig:PHE}, sensitively quantifies the total current-induced field $H_\mathrm{I}$ oriented in-plane and transverse to the current axis. Hall crosses were patterned by photolithography on the same wafers as the ST-FMR devices [Appendix~\ref{app:STFMR}]. The measured Hall cross was placed in the middle of a two-axis Helmholtz-coil setup. As depicted in Fig.~\ref{fig:PHE}(a), we swept a field $H_\mathrm{x}$ of up to $\sim$10 mT along the $x$-axis; we also applied a small constant field $H_\mathrm{y}$ of +0.1 or -0.1 mT in a subset of measurements. A DC current $I_\mathrm{dc}$ along the $x$-axis generated $H_\mathrm{I}$, attributed to the Oersted field as explained in Sec.~\ref{subsec:FL-SOT}. $H_\mathrm{I}$ pulls the magnetization away from the long axis of the Hall cross ($x$-axis), thereby producing a nonzero $y$-component of the magnetization. We measured the transverse Hall voltage $V_\mathrm{H}$ dominated by the planar Hall effect (i.e., the transverse voltage response originating from the same physics as the anisotropic magnetoresistance in the FM). 
We then obtained the second-order planar Hall voltage, 
\begin{equation}
\Delta V_\mathrm{PH}(|I_\mathrm{dc}|) = V_\mathrm{H}(I_\mathrm{dc}>0)+V_\mathrm{H}(I_\mathrm{dc}<0).  
\end{equation}
When the $y$-component of the magnetization is small, $\Delta V_\mathrm{PH}$ is proportional to $H_\mathrm{I}/H_\mathrm{x}$~\cite{Fan2013}. Applying a specific magnitude and direction of $H_\mathrm{y}$ can null $\Delta V_\mathrm{PH}(H_\mathrm{x})$, as illustrated in Fig.~\ref{fig:PHE}(b). In other words, we can quantify $H_\mathrm{I}$ from the condition $H_\mathrm{I} = -H_\mathrm{y}$. Rather than finely stepping through various values of $H_\mathrm{y}$ as done in Ref.~\cite{Fan2013}, we applied a constant transverse field $\mu_0|H_\mathrm{y}| = 0.1$ mT and extrapolated the critical value of $H_\mathrm{y}$ that nulls $\Delta V_\mathrm{PH}(H_\mathrm{x})$ as done in Refs.~\cite{Fan2014, Emori2016, Greening2020}. The results in Fig.~\ref{fig:FLT}(d-f) summarize the values of $H_\mathrm{I}$ at several different dc current densities $J_\mathrm{dc}$ for each FeNi sample.

\section{GI-XRD Peak Shift due to Refraction}
\label{app:refraction}
\begin{figure}
 \centering
    \includegraphics[width = 3.0in]{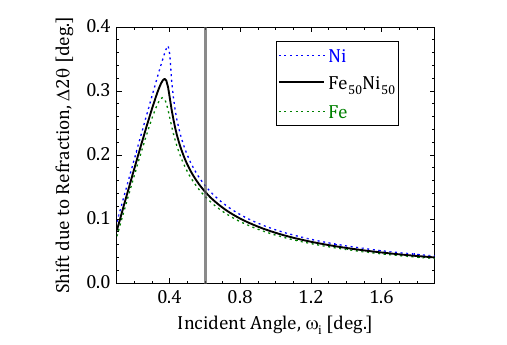}
    \caption{Diffraction peak shift $\Delta 2\theta$ due to refraction as a function of incident angle $\omega_\mathrm{i}$, computed with Eq.~\ref{eq:refraction}}. 
    \label{fig:refraction}
\end{figure}

In GI-XRD, a diffraction peak shifts with the incident angle $\omega_\mathrm{i}$ due to the refraction of the x-ray beam~\cite{Noma1999, Colombi2006}. The GI-XRD spectra displayed in Fig.~\ref{fig:gixrdfcc} account for this effect -- i.e., the refraction-induced shift $\Delta 2\theta$ is subtracted from $2\theta$. 
The shift $\Delta 2\theta$ as a function of $\omega_\mathrm{i}$ is computed with~\cite{Noma1999}
\begin{equation}
\Delta 2\theta = \omega_\mathrm{i} - \frac{1}{\sqrt{2}}
\sqrt{[(\omega_\mathrm{i}^2 - 2\delta)^2 + 4\beta^2]^{1/2} 
+(\omega_\mathrm{i}^2 - 2\delta)},
\label{eq:refraction}
\end{equation}
where $\delta$ and $\beta$ are the dispersive and absorptive components, respectively, of the refractive index of the film material, $r = 1-\delta-i\beta$. 
The values of $\delta$ and $\beta$, which depend on the material composition and density, can be readily found from databases and calculators available online~\cite{TUWienXrayCalculator}. For example, for \Hom\ with density 8.40 g/cm$^3$ [Table~\ref{tab:SLD}], we have $\delta = 2.3\times10^{-5}$ and $\beta = 2\times10^{-6}$ at Cu K$\alpha$1 x-ray wavelength 0.15406 nm (or photon energy 8048 eV). The resulting refraction-induced GI-XRD peak shift $\Delta 2\theta$ is shown in Fig.~\ref{fig:refraction}. 

The precise calculation of $\Delta 2\theta$ would be much more difficult for compositionally graded films because, strictly speaking, $\delta$ and $\beta$ would depend on the position along the film thickness. Fortunately, as shown in Fig.~\ref{fig:refraction}, \Hom, Fe, and Ni exhibit similar values of $\Delta 2\theta$ in the incident angle range of interest to us ($\omega_{i} > 0.6^\circ$). For example, at $\omega_{i} = 0.6^\circ$, we have $\Delta 2\theta = 0.143^\circ$ for \Hom; $\Delta 2\theta$ deviates from that value by only $<0.010^\circ$ for Fe and Ni. As a good approximation, $\Delta 2\theta$ computed for \Hom\ is applied to all three materials examined in this study. 


\begin{thebibliography}{79}%
\makeatletter
\providecommand \@ifxundefined [1]{%
 \@ifx{#1\undefined}
}%
\providecommand \@ifnum [1]{%
 \ifnum #1\expandafter \@firstoftwo
 \else \expandafter \@secondoftwo
 \fi
}%
\providecommand \@ifx [1]{%
 \ifx #1\expandafter \@firstoftwo
 \else \expandafter \@secondoftwo
 \fi
}%
\providecommand \natexlab [1]{#1}%
\providecommand \enquote  [1]{``#1''}%
\providecommand \bibnamefont  [1]{#1}%
\providecommand \bibfnamefont [1]{#1}%
\providecommand \citenamefont [1]{#1}%
\providecommand \href@noop [0]{\@secondoftwo}%
\providecommand \href [0]{\begingroup \@sanitize@url \@href}%
\providecommand \@href[1]{\@@startlink{#1}\@@href}%
\providecommand \@@href[1]{\endgroup#1\@@endlink}%
\providecommand \@sanitize@url [0]{\catcode `\\12\catcode `\$12\catcode `\&12\catcode `\#12\catcode `\^12\catcode `\_12\catcode `\%12\relax}%
\providecommand \@@startlink[1]{}%
\providecommand \@@endlink[0]{}%
\providecommand \url  [0]{\begingroup\@sanitize@url \@url }%
\providecommand \@url [1]{\endgroup\@href {#1}{\urlprefix }}%
\providecommand \urlprefix  [0]{URL }%
\providecommand \Eprint [0]{\href }%
\providecommand \doibase [0]{http://dx.doi.org/}%
\providecommand \selectlanguage [0]{\@gobble}%
\providecommand \bibinfo  [0]{\@secondoftwo}%
\providecommand \bibfield  [0]{\@secondoftwo}%
\providecommand \translation [1]{[#1]}%
\providecommand \BibitemOpen [0]{}%
\providecommand \bibitemStop [0]{}%
\providecommand \bibitemNoStop [0]{.\EOS\space}%
\providecommand \EOS [0]{\spacefactor3000\relax}%
\providecommand \BibitemShut  [1]{\csname bibitem#1\endcsname}%
\let\auto@bib@innerbib\@empty
\bibitem [{\citenamefont {Hirohata}\ \emph {et~al.}(2020)\citenamefont {Hirohata}, \citenamefont {Yamada}, \citenamefont {Nakatani}, \citenamefont {Prejbeanu}, \citenamefont {Diény}, \citenamefont {Pirro},\ and\ \citenamefont {Hillebrands}}]{Hirohata2020}%
  \BibitemOpen
  \bibfield  {author} {\bibinfo {author} {\bibfnamefont {Atsufumi}\ \bibnamefont {Hirohata}}, \bibinfo {author} {\bibfnamefont {Keisuke}\ \bibnamefont {Yamada}}, \bibinfo {author} {\bibfnamefont {Yoshinobu}\ \bibnamefont {Nakatani}}, \bibinfo {author} {\bibfnamefont {Lucian}\ \bibnamefont {Prejbeanu}}, \bibinfo {author} {\bibfnamefont {Bernard}\ \bibnamefont {Diény}}, \bibinfo {author} {\bibfnamefont {Philipp}\ \bibnamefont {Pirro}}, \ and\ \bibinfo {author} {\bibfnamefont {Burkard}\ \bibnamefont {Hillebrands}},\ }\bibfield  {title} {\enquote {\bibinfo {title} {Review on spintronics: Principles and device applications},}\ }\href {\doibase 10.1016/J.JMMM.2020.166711} {\bibfield  {journal} {\bibinfo  {journal} {Journal of Magnetism and Magnetic Materials}\ }\textbf {\bibinfo {volume} {509}},\ \bibinfo {pages} {166711} (\bibinfo {year} {2020})}\BibitemShut {NoStop}%
\bibitem [{\citenamefont {Shao}\ \emph {et~al.}(2021)\citenamefont {Shao}, \citenamefont {Li}, \citenamefont {Liu}, \citenamefont {Yang}, \citenamefont {Fukami}, \citenamefont {Razavi}, \citenamefont {Wu}, \citenamefont {Wang}, \citenamefont {Freimuth}, \citenamefont {Mokrousov}, \citenamefont {Stiles}, \citenamefont {Emori}, \citenamefont {Hoffmann}, \citenamefont {Akerman}, \citenamefont {Roy}, \citenamefont {Wang}, \citenamefont {Yang}, \citenamefont {Garellob},\ and\ \citenamefont {Zhang}}]{Shao2021}%
  \BibitemOpen
  \bibfield  {author} {\bibinfo {author} {\bibfnamefont {Qiming}\ \bibnamefont {Shao}}, \bibinfo {author} {\bibfnamefont {Peng}\ \bibnamefont {Li}}, \bibinfo {author} {\bibfnamefont {Luqiao}\ \bibnamefont {Liu}}, \bibinfo {author} {\bibfnamefont {Hyunsoo}\ \bibnamefont {Yang}}, \bibinfo {author} {\bibfnamefont {Shunsuke}\ \bibnamefont {Fukami}}, \bibinfo {author} {\bibfnamefont {Armin}\ \bibnamefont {Razavi}}, \bibinfo {author} {\bibfnamefont {Hao}\ \bibnamefont {Wu}}, \bibinfo {author} {\bibfnamefont {Kang}\ \bibnamefont {Wang}}, \bibinfo {author} {\bibfnamefont {Frank}\ \bibnamefont {Freimuth}}, \bibinfo {author} {\bibfnamefont {Yuriy}\ \bibnamefont {Mokrousov}}, \bibinfo {author} {\bibfnamefont {Mark~D.}\ \bibnamefont {Stiles}}, \bibinfo {author} {\bibfnamefont {Satoru}\ \bibnamefont {Emori}}, \bibinfo {author} {\bibfnamefont {Axel}\ \bibnamefont {Hoffmann}}, \bibinfo {author} {\bibfnamefont {Johan}\ \bibnamefont {Akerman}}, \bibinfo {author} {\bibfnamefont {Kaushik}\ \bibnamefont {Roy}}, \bibinfo {author}
  {\bibfnamefont {Jian-Ping}\ \bibnamefont {Wang}}, \bibinfo {author} {\bibfnamefont {See-Hun}\ \bibnamefont {Yang}}, \bibinfo {author} {\bibfnamefont {Kevin}\ \bibnamefont {Garellob}}, \ and\ \bibinfo {author} {\bibfnamefont {Wei}\ \bibnamefont {Zhang}},\ }\bibfield  {title} {\enquote {\bibinfo {title} {Roadmap of spin-orbit torques},}\ }\href {\doibase 10.1109/tmag.2021.3078583} {\bibfield  {journal} {\bibinfo  {journal} {IEEE Transactions on Magnetics}\ \textbf {\bibinfo {volume} {57}},\ \bibinfo {pages} {800439}} (\bibinfo {year} {2021})}\BibitemShut {NoStop}%
\bibitem [{\citenamefont {Manchon}\ \emph {et~al.}(2019)\citenamefont {Manchon}, \citenamefont {Železný}, \citenamefont {Miron}, \citenamefont {Jungwirth}, \citenamefont {Sinova}, \citenamefont {Thiaville}, \citenamefont {Garello},\ and\ \citenamefont {Gambardella}}]{Manchon2019}%
  \BibitemOpen
  \bibfield  {author} {\bibinfo {author} {\bibfnamefont {A.}~\bibnamefont {Manchon}}, \bibinfo {author} {\bibfnamefont {J.}~\bibnamefont {Železný}}, \bibinfo {author} {\bibfnamefont {I.~M.}\ \bibnamefont {Miron}}, \bibinfo {author} {\bibfnamefont {T.}~\bibnamefont {Jungwirth}}, \bibinfo {author} {\bibfnamefont {J.}~\bibnamefont {Sinova}}, \bibinfo {author} {\bibfnamefont {A.}~\bibnamefont {Thiaville}}, \bibinfo {author} {\bibfnamefont {K.}~\bibnamefont {Garello}}, \ and\ \bibinfo {author} {\bibfnamefont {P.}~\bibnamefont {Gambardella}},\ }\bibfield  {title} {\enquote {\bibinfo {title} {Current-induced spin-orbit torques in ferromagnetic and antiferromagnetic systems},}\ }\href {\doibase 10.1103/RevModPhys.91.035004} {\bibfield  {journal} {\bibinfo  {journal} {Reviews of Modern Physics}\ }\textbf {\bibinfo {volume} {91}},\ \bibinfo {pages} {035004} (\bibinfo {year} {2019})}\BibitemShut {NoStop}%
\bibitem [{\citenamefont {Fukami}\ \emph {et~al.}(2016)\citenamefont {Fukami}, \citenamefont {Anekawa}, \citenamefont {Zhang},\ and\ \citenamefont {Ohno}}]{Fukami2016b}%
  \BibitemOpen
  \bibfield  {author} {\bibinfo {author} {\bibfnamefont {S.}~\bibnamefont {Fukami}}, \bibinfo {author} {\bibfnamefont {T.}~\bibnamefont {Anekawa}}, \bibinfo {author} {\bibfnamefont {C.}~\bibnamefont {Zhang}}, \ and\ \bibinfo {author} {\bibfnamefont {H.}~\bibnamefont {Ohno}},\ }\bibfield  {title} {\enquote {\bibinfo {title} {A spin–orbit torque switching scheme with collinear magnetic easy axis and current configuration},}\ }\href {\doibase 10.1038/nnano.2016.29} {\bibfield  {journal} {\bibinfo  {journal} {Nature Nanotechnology}\ }\textbf {\bibinfo {volume} {11}},\ \bibinfo {pages} {621} (\bibinfo {year} {2016})}\BibitemShut {NoStop}%
\bibitem [{\citenamefont {Davidson}\ \emph {et~al.}(2020)\citenamefont {Davidson}, \citenamefont {Amin}, \citenamefont {Aljuaid}, \citenamefont {Haney},\ and\ \citenamefont {Fan}}]{Davidson2020a}%
  \BibitemOpen
  \bibfield  {author} {\bibinfo {author} {\bibfnamefont {Angie}\ \bibnamefont {Davidson}}, \bibinfo {author} {\bibfnamefont {Vivek~P.}\ \bibnamefont {Amin}}, \bibinfo {author} {\bibfnamefont {Wafa~S.}\ \bibnamefont {Aljuaid}}, \bibinfo {author} {\bibfnamefont {Paul~M.}\ \bibnamefont {Haney}}, \ and\ \bibinfo {author} {\bibfnamefont {Xin}\ \bibnamefont {Fan}},\ }\bibfield  {title} {\enquote {\bibinfo {title} {Perspectives of electrically generated spin currents in ferromagnetic materials},}\ }\href {\doibase 10.1016/j.physleta.2019.126228} {\bibfield  {journal} {\bibinfo  {journal} {Physics Letters A}\ }\textbf {\bibinfo {volume} {384}},\ \bibinfo {pages} {126228} (\bibinfo {year} {2020})}\BibitemShut {NoStop}%
\bibitem [{\citenamefont {Sinova}\ \emph {et~al.}(2015)\citenamefont {Sinova}, \citenamefont {Valenzuela}, \citenamefont {Wunderlich}, \citenamefont {Back},\ and\ \citenamefont {Jungwirth}}]{Sinova2015}%
  \BibitemOpen
  \bibfield  {author} {\bibinfo {author} {\bibfnamefont {Jairo}\ \bibnamefont {Sinova}}, \bibinfo {author} {\bibfnamefont {Sergio~O.}\ \bibnamefont {Valenzuela}}, \bibinfo {author} {\bibfnamefont {J.}~\bibnamefont {Wunderlich}}, \bibinfo {author} {\bibfnamefont {C.~H.}\ \bibnamefont {Back}}, \ and\ \bibinfo {author} {\bibfnamefont {T.}~\bibnamefont {Jungwirth}},\ }\bibfield  {title} {\enquote {\bibinfo {title} {Spin Hall effects},}\ }\href {\doibase 10.1103/RevModPhys.87.1213} {\bibfield  {journal} {\bibinfo  {journal} {Reviews of Modern Physics}\ }\textbf {\bibinfo {volume} {87}},\ \bibinfo {pages} {1213--1260} (\bibinfo {year} {2015})}\BibitemShut {NoStop}%
\bibitem [{\citenamefont {Kim}\ and\ \citenamefont {Lee}(2020)}]{Kim2020}%
  \BibitemOpen
  \bibfield  {author} {\bibinfo {author} {\bibfnamefont {Kyoung~Whan}\ \bibnamefont {Kim}}\ and\ \bibinfo {author} {\bibfnamefont {Kyung~Jin}\ \bibnamefont {Lee}},\ }\bibfield  {title} {\enquote {\bibinfo {title} {Generalized spin drift-diffusion formalism in the presence of spin-orbit interaction of ferromagnets},}\ }\href {\doibase 10.1103/PhysRevLett.125.207205} {\bibfield  {journal} {\bibinfo  {journal} {Physical Review Letters}\ }\textbf {\bibinfo {volume} {125}},\ \bibinfo {pages} {207205} (\bibinfo {year} {2020})}\BibitemShut {NoStop}%
\bibitem [{\citenamefont {Taniguchi}\ \emph {et~al.}(2015)\citenamefont {Taniguchi}, \citenamefont {Grollier},\ and\ \citenamefont {Stiles}}]{Taniguchi2015}%
  \BibitemOpen
  \bibfield  {author} {\bibinfo {author} {\bibfnamefont {Tomohiro}\ \bibnamefont {Taniguchi}}, \bibinfo {author} {\bibfnamefont {J.}~\bibnamefont {Grollier}}, \ and\ \bibinfo {author} {\bibfnamefont {M.~D.}\ \bibnamefont {Stiles}},\ }\bibfield  {title} {\enquote {\bibinfo {title} {Spin-transfer torques generated by the anomalous hall effect and anisotropic magnetoresistance},}\ }\href {\doibase 10.1103/PhysRevApplied.3.044001} {\bibfield  {journal} {\bibinfo  {journal} {Physical Review Applied}\ }\textbf {\bibinfo {volume} {3}},\ \bibinfo {pages} {044001} (\bibinfo {year} {2015})}\BibitemShut {NoStop}%
\bibitem [{\citenamefont {heon C.~Baek}\ \emph {et~al.}(2018)\citenamefont {heon C.~Baek}, \citenamefont {Amin}, \citenamefont {Oh}, \citenamefont {Go}, \citenamefont {Lee}, \citenamefont {Lee}, \citenamefont {Kim}, \citenamefont {Stiles}, \citenamefont {Park},\ and\ \citenamefont {Lee}}]{Baek2018}%
  \BibitemOpen
  \bibfield  {author} {\bibinfo {author} {\bibfnamefont {Seung}\ \bibnamefont {heon C.~Baek}}, \bibinfo {author} {\bibfnamefont {Vivek~P.}\ \bibnamefont {Amin}}, \bibinfo {author} {\bibfnamefont {Young-Wan}\ \bibnamefont {Oh}}, \bibinfo {author} {\bibfnamefont {Gyungchoon}\ \bibnamefont {Go}}, \bibinfo {author} {\bibfnamefont {Seung-Jae}\ \bibnamefont {Lee}}, \bibinfo {author} {\bibfnamefont {Geun-Hee}\ \bibnamefont {Lee}}, \bibinfo {author} {\bibfnamefont {Kab-Jin}\ \bibnamefont {Kim}}, \bibinfo {author} {\bibfnamefont {M.~D.}\ \bibnamefont {Stiles}}, \bibinfo {author} {\bibfnamefont {Byong-Guk}\ \bibnamefont {Park}}, \ and\ \bibinfo {author} {\bibfnamefont {Kyung-Jin}\ \bibnamefont {Lee}},\ }\bibfield  {title} {\enquote {\bibinfo {title} {Spin currents and spin–orbit torques in ferromagnetic trilayers},}\ }\href {\doibase 10.1038/s41563-018-0041-5} {\bibfield  {journal} {\bibinfo  {journal} {Nature Materials}\ \textbf {\bibinfo {volume} {17}},\ \bibinfo {pages} {509}} (\bibinfo {year} {2018})}\BibitemShut {NoStop}%
\bibitem [{\citenamefont {Chuang}\ \emph {et~al.}(2019)\citenamefont {Chuang}, \citenamefont {Pai},\ and\ \citenamefont {Huang}}]{Chuang2019}%
  \BibitemOpen
  \bibfield  {author} {\bibinfo {author} {\bibfnamefont {T.~C.}\ \bibnamefont {Chuang}}, \bibinfo {author} {\bibfnamefont {C.~F.}\ \bibnamefont {Pai}}, \ and\ \bibinfo {author} {\bibfnamefont {S.~Y.}\ \bibnamefont {Huang}},\ }\bibfield  {title} {\enquote {\bibinfo {title} {Cr-induced perpendicular magnetic anisotropy and field-free spin-orbit-torque switching},}\ }\href {\doibase 10.1103/PhysRevApplied.11.061005} {\bibfield  {journal} {\bibinfo  {journal} {Physical Review Applied}\ }\textbf {\bibinfo {volume} {11}},\ \bibinfo {pages} {061005} (\bibinfo {year} {2019})}\BibitemShut {NoStop}%
\bibitem [{\citenamefont {Montoya}\ \emph {et~al.}(2024)\citenamefont {Montoya}, \citenamefont {Pei},\ and\ \citenamefont {Krivorotov}}]{Montoya2024}%
  \BibitemOpen
  \bibfield  {author} {\bibinfo {author} {\bibfnamefont {Eric~Arturo}\ \bibnamefont {Montoya}}, \bibinfo {author} {\bibfnamefont {Xinyao}\ \bibnamefont {Pei}}, \ and\ \bibinfo {author} {\bibfnamefont {Ilya~N.}\ \bibnamefont {Krivorotov}},\ }\bibfield  {title} {\enquote {\bibinfo {title} {Self-generated spin-orbit torque driven by anomalous Hall current},}\ }\href {https://arxiv.org/abs/2401.05006v1} {\  (\bibinfo {year} {2024})}\BibitemShut {NoStop}%
\bibitem [{\citenamefont {Haney}\ \emph {et~al.}(2013)\citenamefont {Haney}, \citenamefont {Lee}, \citenamefont {Lee}, \citenamefont {Manchon},\ and\ \citenamefont {Stiles}}]{Haney2013a}%
  \BibitemOpen
  \bibfield  {author} {\bibinfo {author} {\bibfnamefont {Paul~M.}\ \bibnamefont {Haney}}, \bibinfo {author} {\bibfnamefont {Hyun-Woo}\ \bibnamefont {Lee}}, \bibinfo {author} {\bibfnamefont {Kyung-Jin}\ \bibnamefont {Lee}}, \bibinfo {author} {\bibfnamefont {Aurélien}\ \bibnamefont {Manchon}}, \ and\ \bibinfo {author} {\bibfnamefont {M.~D.}\ \bibnamefont {Stiles}},\ }\bibfield  {title} {\enquote {\bibinfo {title} {Current induced torques and interfacial spin-orbit coupling: Semiclassical modeling},}\ }\href {\doibase 10.1103/PhysRevB.87.174411} {\bibfield  {journal} {\bibinfo  {journal} {Physical Review B}\ }\textbf {\bibinfo {volume} {87}},\ \bibinfo {pages} {174411} (\bibinfo {year} {2013})}\BibitemShut {NoStop}%
\bibitem [{\citenamefont {Fan}\ \emph {et~al.}(2014)\citenamefont {Fan}, \citenamefont {Celik}, \citenamefont {Wu}, \citenamefont {Ni}, \citenamefont {Lee}, \citenamefont {Lorenz},\ and\ \citenamefont {Xiao}}]{Fan2014}%
  \BibitemOpen
  \bibfield  {author} {\bibinfo {author} {\bibfnamefont {Xin}\ \bibnamefont {Fan}}, \bibinfo {author} {\bibfnamefont {Halise}\ \bibnamefont {Celik}}, \bibinfo {author} {\bibfnamefont {Jun}\ \bibnamefont {Wu}}, \bibinfo {author} {\bibfnamefont {Chaoying}\ \bibnamefont {Ni}}, \bibinfo {author} {\bibfnamefont {Kyung-Jin}\ \bibnamefont {Lee}}, \bibinfo {author} {\bibfnamefont {Virginia~O}\ \bibnamefont {Lorenz}}, \ and\ \bibinfo {author} {\bibfnamefont {John~Q}\ \bibnamefont {Xiao}},\ }\bibfield  {title} {\enquote {\bibinfo {title} {Quantifying interface and bulk contributions to spin-orbit torque in magnetic bilayers.}}\ }\href {\doibase 10.1038/ncomms4042} {\bibfield  {journal} {\bibinfo  {journal} {Nature Communications}\ }\textbf {\bibinfo {volume} {5}},\ \bibinfo {pages} {3042} (\bibinfo {year} {2014})}\BibitemShut {NoStop}%
\bibitem [{\citenamefont {Tserkovnyak}\ \emph {et~al.}(2005)\citenamefont {Tserkovnyak}, \citenamefont {Brataas}, \citenamefont {Bauer},\ and\ \citenamefont {Halperin}}]{Tserkovnyak2005}%
  \BibitemOpen
  \bibfield  {author} {\bibinfo {author} {\bibfnamefont {Yaroslav}\ \bibnamefont {Tserkovnyak}}, \bibinfo {author} {\bibfnamefont {Arne}\ \bibnamefont {Brataas}}, \bibinfo {author} {\bibfnamefont {Gerrit E.~W.}\ \bibnamefont {Bauer}}, \ and\ \bibinfo {author} {\bibfnamefont {Bertrand~I.}\ \bibnamefont {Halperin}},\ }\bibfield  {title} {\enquote {\bibinfo {title} {Nonlocal magnetization dynamics in ferromagnetic heterostructures},}\ }\href {\doibase 10.1103/RevModPhys.77.1375} {\bibfield  {journal} {\bibinfo  {journal} {Reviews of Modern Physics}\ }\textbf {\bibinfo {volume} {77}},\ \bibinfo {pages} {1375--1421} (\bibinfo {year} {2005})}\BibitemShut {NoStop}%
\bibitem [{\citenamefont {McMichael}\ and\ \citenamefont {Krivosik}(2004)}]{McMichael2004}%
  \BibitemOpen
  \bibfield  {author} {\bibinfo {author} {\bibfnamefont {R.D.}\ \bibnamefont {McMichael}}\ and\ \bibinfo {author} {\bibfnamefont {P.}~\bibnamefont {Krivosik}},\ }\bibfield  {title} {\enquote {\bibinfo {title} {Classical model of extrinsic ferromagnetic resonance linewidth in ultrathin films},}\ }\href {\doibase 10.1109/TMAG.2003.821564} {\bibfield  {journal} {\bibinfo  {journal} {IEEE Transactions on Magnetics}\ }\textbf {\bibinfo {volume} {40}},\ \bibinfo {pages} {2--11} (\bibinfo {year} {2004})}\BibitemShut {NoStop}%
\bibitem [{\citenamefont {Zhu}\ \emph {et~al.}(2019)\citenamefont {Zhu}, \citenamefont {Ralph},\ and\ \citenamefont {Buhrman}}]{Zhu2019b}%
  \BibitemOpen
  \bibfield  {author} {\bibinfo {author} {\bibfnamefont {Lijun}\ \bibnamefont {Zhu}}, \bibinfo {author} {\bibfnamefont {Daniel~C.}\ \bibnamefont {Ralph}}, \ and\ \bibinfo {author} {\bibfnamefont {Robert~A.}\ \bibnamefont {Buhrman}},\ }\bibfield  {title} {\enquote {\bibinfo {title} {Effective spin-mixing conductance of heavy-metal–ferromagnet interfaces},}\ }\href {\doibase 10.1103/PhysRevLett.123.057203} {\bibfield  {journal} {\bibinfo  {journal} {Physical Review Letters}\ }\textbf {\bibinfo {volume} {123}},\ \bibinfo {pages} {057203} (\bibinfo {year} {2019})}\BibitemShut {NoStop}%
\bibitem [{\citenamefont {Fallarino}\ \emph {et~al.}(2021)\citenamefont {Fallarino}, \citenamefont {Kirby},\ and\ \citenamefont {Fullerton}}]{Fallarino2021}%
  \BibitemOpen
  \bibfield  {author} {\bibinfo {author} {\bibfnamefont {Lorenzo}\ \bibnamefont {Fallarino}}, \bibinfo {author} {\bibfnamefont {Brian}\ \bibnamefont {Kirby}}, \ and\ \bibinfo {author} {\bibfnamefont {Eric}\ \bibnamefont {Fullerton}},\ }\bibfield  {title} {\enquote {\bibinfo {title} {Graded magnetic materials},}\ }\href {\doibase 10.1088/1361-6463/abfad3} {\bibfield  {journal} {\bibinfo  {journal} {Journal of Physics D: Applied Physics}\ }\textbf {\bibinfo {volume} {54}},\ \bibinfo {pages} {303002} (\bibinfo {year} {2021})}\BibitemShut {NoStop}%
\bibitem [{\citenamefont {Tang}\ \emph {et~al.}(2020)\citenamefont {Tang}, \citenamefont {Shen}, \citenamefont {Xu}, \citenamefont {Yang}, \citenamefont {Hu}, \citenamefont {Lü}, \citenamefont {Li}, \citenamefont {Li}, \citenamefont {Yuan}, \citenamefont {Pennycook}, \citenamefont {Xia}, \citenamefont {Manchon}, \citenamefont {Zhou},\ and\ \citenamefont {Qiu}}]{Tang2020}%
  \BibitemOpen
  \bibfield  {author} {\bibinfo {author} {\bibfnamefont {Meng}\ \bibnamefont {Tang}}, \bibinfo {author} {\bibfnamefont {Ka}~\bibnamefont {Shen}}, \bibinfo {author} {\bibfnamefont {Shijie}\ \bibnamefont {Xu}}, \bibinfo {author} {\bibfnamefont {Huanglin}\ \bibnamefont {Yang}}, \bibinfo {author} {\bibfnamefont {Shuai}\ \bibnamefont {Hu}}, \bibinfo {author} {\bibfnamefont {Weiming}\ \bibnamefont {Lü}}, \bibinfo {author} {\bibfnamefont {Changjian}\ \bibnamefont {Li}}, \bibinfo {author} {\bibfnamefont {Mengsha}\ \bibnamefont {Li}}, \bibinfo {author} {\bibfnamefont {Zhe}\ \bibnamefont {Yuan}}, \bibinfo {author} {\bibfnamefont {Stephen~J.}\ \bibnamefont {Pennycook}}, \bibinfo {author} {\bibfnamefont {Ke}~\bibnamefont {Xia}}, \bibinfo {author} {\bibfnamefont {Aurelien}\ \bibnamefont {Manchon}}, \bibinfo {author} {\bibfnamefont {Shiming}\ \bibnamefont {Zhou}}, \ and\ \bibinfo {author} {\bibfnamefont {Xuepeng}\ \bibnamefont {Qiu}},\ }\bibfield  {title} {\enquote {\bibinfo {title} {Bulk spin torque‐driven
  perpendicular magnetization switching in L1$_0$ FePt single layer},}\ }\href {\doibase 10.1002/adma.202002607} {\bibfield  {journal} {\bibinfo  {journal} {Advanced Materials}\ }\textbf {\bibinfo {volume} {32}},\ \bibinfo {pages} {2002607} (\bibinfo {year} {2020})}\BibitemShut {NoStop}%
\bibitem [{\citenamefont {Liu}\ \emph {et~al.}(2020)\citenamefont {Liu}, \citenamefont {Yu}, \citenamefont {González-Hernández}, \citenamefont {Li}, \citenamefont {Deng}, \citenamefont {Lin}, \citenamefont {Zhou}, \citenamefont {Zhou}, \citenamefont {Zhou}, \citenamefont {Wang}, \citenamefont {Guo}, \citenamefont {Yoong}, \citenamefont {Chow}, \citenamefont {Han}, \citenamefont {Dupé}, \citenamefont {Železný}, \citenamefont {Sinova},\ and\ \citenamefont {Chen}}]{Liu2020a}%
  \BibitemOpen
  \bibfield  {author} {\bibinfo {author} {\bibfnamefont {Liang}\ \bibnamefont {Liu}}, \bibinfo {author} {\bibfnamefont {Jihang}\ \bibnamefont {Yu}}, \bibinfo {author} {\bibfnamefont {Rafael}\ \bibnamefont {González-Hernández}}, \bibinfo {author} {\bibfnamefont {Changjian}\ \bibnamefont {Li}}, \bibinfo {author} {\bibfnamefont {Jinyu}\ \bibnamefont {Deng}}, \bibinfo {author} {\bibfnamefont {Weinan}\ \bibnamefont {Lin}}, \bibinfo {author} {\bibfnamefont {Chenghang}\ \bibnamefont {Zhou}}, \bibinfo {author} {\bibfnamefont {Tiejun}\ \bibnamefont {Zhou}}, \bibinfo {author} {\bibfnamefont {Jing}\ \bibnamefont {Zhou}}, \bibinfo {author} {\bibfnamefont {Han}\ \bibnamefont {Wang}}, \bibinfo {author} {\bibfnamefont {Rui}\ \bibnamefont {Guo}}, \bibinfo {author} {\bibfnamefont {Herng~Yau}\ \bibnamefont {Yoong}}, \bibinfo {author} {\bibfnamefont {Gan~Moog}\ \bibnamefont {Chow}}, \bibinfo {author} {\bibfnamefont {Xiufeng}\ \bibnamefont {Han}}, \bibinfo {author} {\bibfnamefont {Bertrand}\ \bibnamefont {Dupé}}, \bibinfo
  {author} {\bibfnamefont {Jakub}\ \bibnamefont {Železný}}, \bibinfo {author} {\bibfnamefont {Jairo}\ \bibnamefont {Sinova}}, \ and\ \bibinfo {author} {\bibfnamefont {Jingsheng}\ \bibnamefont {Chen}},\ }\bibfield  {title} {\enquote {\bibinfo {title} {Electrical switching of perpendicular magnetization in a single ferromagnetic layer},}\ }\href {\doibase 10.1103/PhysRevB.101.220402} {\bibfield  {journal} {\bibinfo  {journal} {Physical Review B}\ }\textbf {\bibinfo {volume} {101}},\ \bibinfo {pages} {220402} (\bibinfo {year} {2020})}\BibitemShut {NoStop}%
\bibitem [{\citenamefont {Tao}\ \emph {et~al.}(2022)\citenamefont {Tao}, \citenamefont {Sun}, \citenamefont {Li}, \citenamefont {Yang}, \citenamefont {Jin}, \citenamefont {Hui}, \citenamefont {Li}, \citenamefont {Wang},\ and\ \citenamefont {Dong}}]{Tao2022FePt}%
  \BibitemOpen
  \bibfield  {author} {\bibinfo {author} {\bibfnamefont {Ying}\ \bibnamefont {Tao}}, \bibinfo {author} {\bibfnamefont {Chao}\ \bibnamefont {Sun}}, \bibinfo {author} {\bibfnamefont {Wendi}\ \bibnamefont {Li}}, \bibinfo {author} {\bibfnamefont {Liu}\ \bibnamefont {Yang}}, \bibinfo {author} {\bibfnamefont {Fang}\ \bibnamefont {Jin}}, \bibinfo {author} {\bibfnamefont {Yajuan}\ \bibnamefont {Hui}}, \bibinfo {author} {\bibfnamefont {Huihui}\ \bibnamefont {Li}}, \bibinfo {author} {\bibfnamefont {Xiaoguang}\ \bibnamefont {Wang}}, \ and\ \bibinfo {author} {\bibfnamefont {Kaifeng}\ \bibnamefont {Dong}},\ }\bibfield  {title} {\enquote {\bibinfo {title} {Field-free spin-orbit torque switching in L1$_{0}$-FePt single layer with tilted anisotropy},}\ }\href {\doibase 10.1063/5.0077465/2833059} {\bibfield  {journal} {\bibinfo  {journal} {Applied Physics Letters}\ }\textbf {\bibinfo {volume} {120}},\ \bibinfo {pages} {102405} (\bibinfo {year} {2022})}\BibitemShut {NoStop}%
\bibitem [{\citenamefont {Zhu}\ \emph {et~al.}(2020)\citenamefont {Zhu}, \citenamefont {Zhang}, \citenamefont {Muller}, \citenamefont {Ralph},\ and\ \citenamefont {Buhrman}}]{Zhu2020c}%
  \BibitemOpen
  \bibfield  {author} {\bibinfo {author} {\bibfnamefont {Lijun}\ \bibnamefont {Zhu}}, \bibinfo {author} {\bibfnamefont {Xiyue~S.}\ \bibnamefont {Zhang}}, \bibinfo {author} {\bibfnamefont {David~A.}\ \bibnamefont {Muller}}, \bibinfo {author} {\bibfnamefont {Daniel~C.}\ \bibnamefont {Ralph}}, \ and\ \bibinfo {author} {\bibfnamefont {Robert~A.}\ \bibnamefont {Buhrman}},\ }\bibfield  {title} {\enquote {\bibinfo {title} {Observation of strong bulk damping‐like spin‐orbit torque in chemically disordered ferromagnetic single layers},}\ }\href {\doibase 10.1002/adfm.202005201} {\bibfield  {journal} {\bibinfo  {journal} {Advanced Functional Materials}\ }\textbf {\bibinfo {volume} {30}},\ \bibinfo {pages} {2005201} (\bibinfo {year} {2020})}\BibitemShut {NoStop}%
\bibitem [{\citenamefont {Lee}\ \emph {et~al.}(2020)\citenamefont {Lee}, \citenamefont {Park}, \citenamefont {Yuk},\ and\ \citenamefont {Park}}]{Lee2020b}%
  \BibitemOpen
  \bibfield  {author} {\bibinfo {author} {\bibfnamefont {Jae~Wook}\ \bibnamefont {Lee}}, \bibinfo {author} {\bibfnamefont {Jae~Yeol}\ \bibnamefont {Park}}, \bibinfo {author} {\bibfnamefont {Jong~Min}\ \bibnamefont {Yuk}}, \ and\ \bibinfo {author} {\bibfnamefont {Byong~Guk}\ \bibnamefont {Park}},\ }\bibfield  {title} {\enquote {\bibinfo {title} {Spin-orbit torque in a perpendicularly magnetized ferrimagnetic Tb-Co single layer},}\ }\href {\doibase 10.1103/PhysRevApplied.13.044030} {\bibfield  {journal} {\bibinfo  {journal} {Physical Review Applied}\ }\textbf {\bibinfo {volume} {13}},\ \bibinfo {pages} {044030} (\bibinfo {year} {2020})}\BibitemShut {NoStop}%
\bibitem [{\citenamefont {Zhang}\ \emph {et~al.}(2020)\citenamefont {Zhang}, \citenamefont {Liao}, \citenamefont {Chen}, \citenamefont {Xu}, \citenamefont {Cai}, \citenamefont {Guo}, \citenamefont {Bai}, \citenamefont {Sun}, \citenamefont {Xue}, \citenamefont {Su}, \citenamefont {Wang}, \citenamefont {Wan}, \citenamefont {Bai}, \citenamefont {Song}, \citenamefont {Chen}, \citenamefont {Chen}, \citenamefont {Jiang}, \citenamefont {Kou}, \citenamefont {Cai}, \citenamefont {Wu}, \citenamefont {Pan},\ and\ \citenamefont {Song}}]{Zhang2020a}%
  \BibitemOpen
  \bibfield  {author} {\bibinfo {author} {\bibfnamefont {R.~Q.}\ \bibnamefont {Zhang}}, \bibinfo {author} {\bibfnamefont {L.~Y.}\ \bibnamefont {Liao}}, \bibinfo {author} {\bibfnamefont {X.~Z.}\ \bibnamefont {Chen}}, \bibinfo {author} {\bibfnamefont {T.}~\bibnamefont {Xu}}, \bibinfo {author} {\bibfnamefont {L.}~\bibnamefont {Cai}}, \bibinfo {author} {\bibfnamefont {M.~H.}\ \bibnamefont {Guo}}, \bibinfo {author} {\bibfnamefont {Hao}\ \bibnamefont {Bai}}, \bibinfo {author} {\bibfnamefont {L.}~\bibnamefont {Sun}}, \bibinfo {author} {\bibfnamefont {F.~H.}\ \bibnamefont {Xue}}, \bibinfo {author} {\bibfnamefont {J.}~\bibnamefont {Su}}, \bibinfo {author} {\bibfnamefont {X.}~\bibnamefont {Wang}}, \bibinfo {author} {\bibfnamefont {C.~H.}\ \bibnamefont {Wan}}, \bibinfo {author} {\bibfnamefont {Hua}\ \bibnamefont {Bai}}, \bibinfo {author} {\bibfnamefont {Y.~X.}\ \bibnamefont {Song}}, \bibinfo {author} {\bibfnamefont {R.~Y.}\ \bibnamefont {Chen}}, \bibinfo {author} {\bibfnamefont {N.}~\bibnamefont {Chen}}, \bibinfo
  {author} {\bibfnamefont {W.~J.}\ \bibnamefont {Jiang}}, \bibinfo {author} {\bibfnamefont {X.~F.}\ \bibnamefont {Kou}}, \bibinfo {author} {\bibfnamefont {J.~W.}\ \bibnamefont {Cai}}, \bibinfo {author} {\bibfnamefont {H.~Q.}\ \bibnamefont {Wu}}, \bibinfo {author} {\bibfnamefont {F.}~\bibnamefont {Pan}}, \ and\ \bibinfo {author} {\bibfnamefont {C.}~\bibnamefont {Song}},\ }\bibfield  {title} {\enquote {\bibinfo {title} {Current-induced magnetization switching in a cotb amorphous single layer},}\ }\href {\doibase 10.1103/PhysRevB.101.214418} {\bibfield  {journal} {\bibinfo  {journal} {Physical Review B}\ }\textbf {\bibinfo {volume} {101}},\ \bibinfo {pages} {214418} (\bibinfo {year} {2020})}\BibitemShut {NoStop}%
\bibitem [{\citenamefont {Céspedes‐Berrocal}\ \emph {et~al.}(2021)\citenamefont {Céspedes‐Berrocal}, \citenamefont {Damas}, \citenamefont {Petit‐Watelot}, \citenamefont {Maccariello}, \citenamefont {Tang}, \citenamefont {Arriola‐Córdova}, \citenamefont {Vallobra}, \citenamefont {Xu}, \citenamefont {Bello}, \citenamefont {Martin}, \citenamefont {Migot}, \citenamefont {Ghanbaja}, \citenamefont {Zhang}, \citenamefont {Hehn}, \citenamefont {Mangin}, \citenamefont {Panagopoulos}, \citenamefont {Cros}, \citenamefont {Fert},\ and\ \citenamefont {Rojas‐Sánchez}}]{CespedesBerrocal2021}%
  \BibitemOpen
  \bibfield  {author} {\bibinfo {author} {\bibfnamefont {David}\ \bibnamefont {Céspedes‐Berrocal}}, \bibinfo {author} {\bibfnamefont {Heloïse}\ \bibnamefont {Damas}}, \bibinfo {author} {\bibfnamefont {Sébastien}\ \bibnamefont {Petit‐Watelot}}, \bibinfo {author} {\bibfnamefont {Davide}\ \bibnamefont {Maccariello}}, \bibinfo {author} {\bibfnamefont {Ping}\ \bibnamefont {Tang}}, \bibinfo {author} {\bibfnamefont {Aldo}\ \bibnamefont {Arriola‐Córdova}}, \bibinfo {author} {\bibfnamefont {Pierre}\ \bibnamefont {Vallobra}}, \bibinfo {author} {\bibfnamefont {Yong}\ \bibnamefont {Xu}}, \bibinfo {author} {\bibfnamefont {Jean‐Loïs}\ \bibnamefont {Bello}}, \bibinfo {author} {\bibfnamefont {Elodie}\ \bibnamefont {Martin}}, \bibinfo {author} {\bibfnamefont {Sylvie}\ \bibnamefont {Migot}}, \bibinfo {author} {\bibfnamefont {Jaafar}\ \bibnamefont {Ghanbaja}}, \bibinfo {author} {\bibfnamefont {Shufeng}\ \bibnamefont {Zhang}}, \bibinfo {author} {\bibfnamefont {Michel}\ \bibnamefont {Hehn}}, \bibinfo {author}
  {\bibfnamefont {Stéphane}\ \bibnamefont {Mangin}}, \bibinfo {author} {\bibfnamefont {Christos}\ \bibnamefont {Panagopoulos}}, \bibinfo {author} {\bibfnamefont {Vincent}\ \bibnamefont {Cros}}, \bibinfo {author} {\bibfnamefont {Albert}\ \bibnamefont {Fert}}, \ and\ \bibinfo {author} {\bibfnamefont {Juan‐Carlos}\ \bibnamefont {Rojas‐Sánchez}},\ }\bibfield  {title} {\enquote {\bibinfo {title} {Current‐induced spin torques on single gdfeco magnetic layers},}\ }\href {\doibase 10.1002/adma.202007047} {\bibfield  {journal} {\bibinfo  {journal} {Advanced Materials}\ }\textbf {\bibinfo {volume} {33}},\ \bibinfo {pages} {2007047} (\bibinfo {year} {2021})}\BibitemShut {NoStop}%
\bibitem [{\citenamefont {Zhu}\ \emph {et~al.}(2021)\citenamefont {Zhu}, \citenamefont {Ralph},\ and\ \citenamefont {Buhrman}}]{Zhu2021c}%
  \BibitemOpen
  \bibfield  {author} {\bibinfo {author} {\bibfnamefont {Lijun}\ \bibnamefont {Zhu}}, \bibinfo {author} {\bibfnamefont {Daniel~C.}\ \bibnamefont {Ralph}}, \ and\ \bibinfo {author} {\bibfnamefont {Robert~A.}\ \bibnamefont {Buhrman}},\ }\bibfield  {title} {\enquote {\bibinfo {title} {Unveiling the mechanism of bulk spin-orbit torques within chemically disordered Fe$_x$Pt$_{1-x}$ single layers},}\ }\href {\doibase 10.1002/ADFM.202103898} {\bibfield  {journal} {\bibinfo  {journal} {Advanced Functional Materials}\ ,\ \bibinfo {pages} {2103898}} (\bibinfo {year} {2021})}\BibitemShut {NoStop}%
\bibitem [{\citenamefont {Liu}\ \emph {et~al.}(2022)\citenamefont {Liu}, \citenamefont {Zhu}, \citenamefont {Zhang}, \citenamefont {Muller},\ and\ \citenamefont {Ralph}}]{Liu2022BulkSOT}%
  \BibitemOpen
  \bibfield  {author} {\bibinfo {author} {\bibfnamefont {Qianbiao}\ \bibnamefont {Liu}}, \bibinfo {author} {\bibfnamefont {Lijun}\ \bibnamefont {Zhu}}, \bibinfo {author} {\bibfnamefont {Xiyue~S.}\ \bibnamefont {Zhang}}, \bibinfo {author} {\bibfnamefont {David~A.}\ \bibnamefont {Muller}}, \ and\ \bibinfo {author} {\bibfnamefont {Daniel~C.}\ \bibnamefont {Ralph}},\ }\bibfield  {title} {\enquote {\bibinfo {title} {Giant bulk spin-orbit torque and efficient electrical switching in single ferrimagnetic fetb layers with strong perpendicular magnetic anisotropy},}\ }\href {\doibase 10.1063/5.0087260/2835395} {\bibfield  {journal} {\bibinfo  {journal} {Applied Physics Reviews}\ }\textbf {\bibinfo {volume} {9}} (\bibinfo {year} {2022})}\BibitemShut {NoStop}%
\bibitem [{\citenamefont {Chen}\ \emph {et~al.}(2012)\citenamefont {Chen}, \citenamefont {Yi}, \citenamefont {Chen}, \citenamefont {Li}, \citenamefont {Zhou},\ and\ \citenamefont {Lai}}]{Chen2012FePt}%
  \BibitemOpen
  \bibfield  {author} {\bibinfo {author} {\bibfnamefont {Zhifeng}\ \bibnamefont {Chen}}, \bibinfo {author} {\bibfnamefont {Ming}\ \bibnamefont {Yi}}, \bibinfo {author} {\bibfnamefont {Ming}\ \bibnamefont {Chen}}, \bibinfo {author} {\bibfnamefont {Shufa}\ \bibnamefont {Li}}, \bibinfo {author} {\bibfnamefont {Shiming}\ \bibnamefont {Zhou}}, \ and\ \bibinfo {author} {\bibfnamefont {Tianshu}\ \bibnamefont {Lai}},\ }\bibfield  {title} {\enquote {\bibinfo {title} {Spin waves and small intrinsic damping in an in-plane magnetized FePt film},}\ }\href {\doibase 10.1063/1.4768787/128310} {\bibfield  {journal} {\bibinfo  {journal} {Applied Physics Letters}\ }\textbf {\bibinfo {volume} {101}} (\bibinfo {year} {2012})}\BibitemShut {NoStop}%
\bibitem [{\citenamefont {Ma}\ \emph {et~al.}(2015)\citenamefont {Ma}, \citenamefont {Ma}, \citenamefont {He}, \citenamefont {Zhao}, \citenamefont {Zhou},\ and\ \citenamefont {Lüpke}}]{Ma2015}%
  \BibitemOpen
  \bibfield  {author} {\bibinfo {author} {\bibfnamefont {X.}~\bibnamefont {Ma}}, \bibinfo {author} {\bibfnamefont {L.}~\bibnamefont {Ma}}, \bibinfo {author} {\bibfnamefont {P.}~\bibnamefont {He}}, \bibinfo {author} {\bibfnamefont {H.~B.}\ \bibnamefont {Zhao}}, \bibinfo {author} {\bibfnamefont {S.~M.}\ \bibnamefont {Zhou}}, \ and\ \bibinfo {author} {\bibfnamefont {G.}~\bibnamefont {Lüpke}},\ }\bibfield  {title} {\enquote {\bibinfo {title} {Role of antisite disorder on intrinsic gilbert damping in L1$_0$ FePt films},}\ }\href {\doibase 10.1103/PhysRevB.91.014438} {\bibfield  {journal} {\bibinfo  {journal} {Physical Review B}\ }\textbf {\bibinfo {volume} {91}},\ \bibinfo {pages} {014438} (\bibinfo {year} {2015})}\BibitemShut {NoStop}%
\bibitem [{\citenamefont {Schoen}\ \emph {et~al.}(2017{\natexlab{a}})\citenamefont {Schoen}, \citenamefont {Lucassen}, \citenamefont {Nembach}, \citenamefont {Silva}, \citenamefont {Koopmans}, \citenamefont {Back},\ and\ \citenamefont {Shaw}}]{Schoen2017c}%
  \BibitemOpen
  \bibfield  {author} {\bibinfo {author} {\bibfnamefont {Martin A.~W.}\ \bibnamefont {Schoen}}, \bibinfo {author} {\bibfnamefont {Juriaan}\ \bibnamefont {Lucassen}}, \bibinfo {author} {\bibfnamefont {Hans~T.}\ \bibnamefont {Nembach}}, \bibinfo {author} {\bibfnamefont {T.~J.}\ \bibnamefont {Silva}}, \bibinfo {author} {\bibfnamefont {Bert}\ \bibnamefont {Koopmans}}, \bibinfo {author} {\bibfnamefont {Christian~H.}\ \bibnamefont {Back}}, \ and\ \bibinfo {author} {\bibfnamefont {Justin~M.}\ \bibnamefont {Shaw}},\ }\bibfield  {title} {\enquote {\bibinfo {title} {Magnetic properties in ultrathin 3d transition-metal binary alloys. ii. experimental verification of quantitative theories of damping and spin pumping},}\ }\href {\doibase 10.1103/PhysRevB.95.134411} {\bibfield  {journal} {\bibinfo  {journal} {Physical Review B}\ }\textbf {\bibinfo {volume} {95}},\ \bibinfo {pages} {134411} (\bibinfo {year} {2017}{\natexlab{a}})}\BibitemShut {NoStop}%
\bibitem [{\citenamefont {Oogane}\ \emph {et~al.}(2006)\citenamefont {Oogane}, \citenamefont {Wakitani}, \citenamefont {Yakata}, \citenamefont {Yilgin}, \citenamefont {Ando}, \citenamefont {Sakuma},\ and\ \citenamefont {Miyazaki}}]{Oogane2006}%
  \BibitemOpen
  \bibfield  {author} {\bibinfo {author} {\bibfnamefont {Mikihiko}\ \bibnamefont {Oogane}}, \bibinfo {author} {\bibfnamefont {Takeshi}\ \bibnamefont {Wakitani}}, \bibinfo {author} {\bibfnamefont {Satoshi}\ \bibnamefont {Yakata}}, \bibinfo {author} {\bibfnamefont {Resul}\ \bibnamefont {Yilgin}}, \bibinfo {author} {\bibfnamefont {Yasuo}\ \bibnamefont {Ando}}, \bibinfo {author} {\bibfnamefont {Akimasa}\ \bibnamefont {Sakuma}}, \ and\ \bibinfo {author} {\bibfnamefont {Terunobu}\ \bibnamefont {Miyazaki}},\ }\bibfield  {title} {\enquote {\bibinfo {title} {Magnetic damping in ferromagnetic thin films},}\ }\href {\doibase 10.1143/JJAP.45.3889} {\bibfield  {journal} {\bibinfo  {journal} {Japanese Journal of Applied Physics}\ }\textbf {\bibinfo {volume} {45}},\ \bibinfo {pages} {3889--3891} (\bibinfo {year} {2006})}\BibitemShut {NoStop}%
\bibitem [{\citenamefont {Du}\ \emph {et~al.}(2014)\citenamefont {Du}, \citenamefont {Wang}, \citenamefont {Yang},\ and\ \citenamefont {Hammel}}]{Du2014b}%
  \BibitemOpen
  \bibfield  {author} {\bibinfo {author} {\bibfnamefont {Chunhui}\ \bibnamefont {Du}}, \bibinfo {author} {\bibfnamefont {Hailong}\ \bibnamefont {Wang}}, \bibinfo {author} {\bibfnamefont {Fengyuan}\ \bibnamefont {Yang}}, \ and\ \bibinfo {author} {\bibfnamefont {P.~Chris}\ \bibnamefont {Hammel}},\ }\bibfield  {title} {\enquote {\bibinfo {title} {Systematic variation of spin-orbit coupling with d-orbital filling: Large inverse spin Hall effect in 3d transition metals},}\ }\href {\doibase 10.1103/PhysRevB.90.140407} {\bibfield  {journal} {\bibinfo  {journal} {Physical Review B}\ }\textbf {\bibinfo {volume} {90}},\ \bibinfo {pages} {140407} (\bibinfo {year} {2014})}\BibitemShut {NoStop}%
\bibitem [{\citenamefont {Keller}\ \emph {et~al.}(2019)\citenamefont {Keller}, \citenamefont {Gerace}, \citenamefont {Arora}, \citenamefont {Delczeg-Czirjak}, \citenamefont {Shaw},\ and\ \citenamefont {Silva}}]{Keller2019}%
  \BibitemOpen
  \bibfield  {author} {\bibinfo {author} {\bibfnamefont {Mark~W.}\ \bibnamefont {Keller}}, \bibinfo {author} {\bibfnamefont {Katy~S.}\ \bibnamefont {Gerace}}, \bibinfo {author} {\bibfnamefont {Monika}\ \bibnamefont {Arora}}, \bibinfo {author} {\bibfnamefont {Erna~Krisztina}\ \bibnamefont {Delczeg-Czirjak}}, \bibinfo {author} {\bibfnamefont {Justin~M.}\ \bibnamefont {Shaw}}, \ and\ \bibinfo {author} {\bibfnamefont {T.~J.}\ \bibnamefont {Silva}},\ }\bibfield  {title} {\enquote {\bibinfo {title} {Near-unity spin Hall ratio in Ni$_x$Cu$_{1-x}$ alloys},}\ }\href {\doibase 10.1103/PhysRevB.99.214411} {\bibfield  {journal} {\bibinfo  {journal} {Physical Review B}\ }\textbf {\bibinfo {volume} {99}},\ \bibinfo {pages} {214411} (\bibinfo {year} {2019})}\BibitemShut {NoStop}%
\bibitem [{\citenamefont {Edwards}\ \emph {et~al.}(2019)\citenamefont {Edwards}, \citenamefont {Nembach},\ and\ \citenamefont {Shaw}}]{Edwards2019}%
  \BibitemOpen
  \bibfield  {author} {\bibinfo {author} {\bibfnamefont {Eric~R.J.}\ \bibnamefont {Edwards}}, \bibinfo {author} {\bibfnamefont {Hans~T.}\ \bibnamefont {Nembach}}, \ and\ \bibinfo {author} {\bibfnamefont {Justin~M.}\ \bibnamefont {Shaw}},\ }\bibfield  {title} {\enquote {\bibinfo {title} {Co$_{25}$Fe$_{75}$ thin films with ultralow total damping of ferromagnetic resonance},}\ }\href {\doibase 10.1103/PhysRevApplied.11.054036} {\bibfield  {journal} {\bibinfo  {journal} {Physical Review Applied}\ }\textbf {\bibinfo {volume} {11}},\ \bibinfo {pages} {054036} (\bibinfo {year} {2019})}\BibitemShut {NoStop}%
\bibitem [{\citenamefont {Gilmore}\ \emph {et~al.}(2007)\citenamefont {Gilmore}, \citenamefont {Idzerda},\ and\ \citenamefont {Stiles}}]{Gilmore2007}%
  \BibitemOpen
  \bibfield  {author} {\bibinfo {author} {\bibfnamefont {K.}~\bibnamefont {Gilmore}}, \bibinfo {author} {\bibfnamefont {Y.~U.}\ \bibnamefont {Idzerda}}, \ and\ \bibinfo {author} {\bibfnamefont {M.~D.}\ \bibnamefont {Stiles}},\ }\bibfield  {title} {\enquote {\bibinfo {title} {Identification of the dominant precession-damping mechanism in Fe, Co, and Ni by first-principles calculations},}\ }\href {\doibase 10.1103/PhysRevLett.99.027204} {\bibfield  {journal} {\bibinfo  {journal} {Physical Review Letters}\ }\textbf {\bibinfo {volume} {99}},\ \bibinfo {pages} {027204} (\bibinfo {year} {2007})}\BibitemShut {NoStop}%
\bibitem [{\citenamefont {Khodadadi}\ \emph {et~al.}(2020)\citenamefont {Khodadadi}, \citenamefont {Rai}, \citenamefont {Sapkota}, \citenamefont {Srivastava}, \citenamefont {Nepal}, \citenamefont {Lim}, \citenamefont {Smith}, \citenamefont {Mewes}, \citenamefont {Budhathoki}, \citenamefont {Hauser}, \citenamefont {Gao}, \citenamefont {Li}, \citenamefont {Viehland}, \citenamefont {Jiang}, \citenamefont {Heremans}, \citenamefont {Balachandran}, \citenamefont {Mewes},\ and\ \citenamefont {Emori}}]{Khodadadi2020}%
  \BibitemOpen
  \bibfield  {author} {\bibinfo {author} {\bibfnamefont {Behrouz}\ \bibnamefont {Khodadadi}}, \bibinfo {author} {\bibfnamefont {Anish}\ \bibnamefont {Rai}}, \bibinfo {author} {\bibfnamefont {Arjun}\ \bibnamefont {Sapkota}}, \bibinfo {author} {\bibfnamefont {Abhishek}\ \bibnamefont {Srivastava}}, \bibinfo {author} {\bibfnamefont {Bhuwan}\ \bibnamefont {Nepal}}, \bibinfo {author} {\bibfnamefont {Youngmin}\ \bibnamefont {Lim}}, \bibinfo {author} {\bibfnamefont {David~A.}\ \bibnamefont {Smith}}, \bibinfo {author} {\bibfnamefont {Claudia}\ \bibnamefont {Mewes}}, \bibinfo {author} {\bibfnamefont {Sujan}\ \bibnamefont {Budhathoki}}, \bibinfo {author} {\bibfnamefont {Adam~J.}\ \bibnamefont {Hauser}}, \bibinfo {author} {\bibfnamefont {Min}\ \bibnamefont {Gao}}, \bibinfo {author} {\bibfnamefont {Jie-Fang}\ \bibnamefont {Li}}, \bibinfo {author} {\bibfnamefont {Dwight~D.}\ \bibnamefont {Viehland}}, \bibinfo {author} {\bibfnamefont {Zijian}\ \bibnamefont {Jiang}}, \bibinfo {author} {\bibfnamefont {Jean~J.}\ \bibnamefont
  {Heremans}}, \bibinfo {author} {\bibfnamefont {Prasanna~V.}\ \bibnamefont {Balachandran}}, \bibinfo {author} {\bibfnamefont {Tim}\ \bibnamefont {Mewes}}, \ and\ \bibinfo {author} {\bibfnamefont {Satoru}\ \bibnamefont {Emori}},\ }\bibfield  {title} {\enquote {\bibinfo {title} {Conductivitylike gilbert damping due to intraband scattering in epitaxial iron},}\ }\href {\doibase 10.1103/PhysRevLett.124.157201} {\bibfield  {journal} {\bibinfo  {journal} {Physical Review Letters}\ }\textbf {\bibinfo {volume} {124}},\ \bibinfo {pages} {157201} (\bibinfo {year} {2020})}\BibitemShut {NoStop}%
\bibitem [{\citenamefont {Heinrich}(2005)}]{Heinrich2005}%
  \BibitemOpen
  \bibfield  {author} {\bibinfo {author} {\bibfnamefont {B.}~\bibnamefont {Heinrich}},\ }\href {\doibase 10.1007/b138703} {\enquote {\bibinfo {title} {Spin relaxation in magnetic metallic layers and multilayers},}\ } (\bibinfo {year} {2005})\BibitemShut {NoStop}%
\bibitem [{\citenamefont {Shaw}\ \emph {et~al.}(2011)\citenamefont {Shaw}, \citenamefont {Nembach},\ and\ \citenamefont {Silva}}]{Shaw2011MLvsAlloys}%
  \BibitemOpen
  \bibfield  {author} {\bibinfo {author} {\bibfnamefont {Justin~M.}\ \bibnamefont {Shaw}}, \bibinfo {author} {\bibfnamefont {Hans~T.}\ \bibnamefont {Nembach}}, \ and\ \bibinfo {author} {\bibfnamefont {T.~J.}\ \bibnamefont {Silva}},\ }\bibfield  {title} {\enquote {\bibinfo {title} {Damping phenomena in Co$_{90}$Fe$_{10}$/Ni multilayers and alloys},}\ }\href {\doibase 10.1063/1.3607278/237722} {\bibfield  {journal} {\bibinfo  {journal} {Applied Physics Letters}\ }\textbf {\bibinfo {volume} {99}} ,\ \bibinfo {pages} {012503}  (\bibinfo {year} {2011})}\BibitemShut {NoStop}%
\bibitem [{\citenamefont {Omelchenko}\ \emph {et~al.}(2017)\citenamefont {Omelchenko}, \citenamefont {Montoya}, \citenamefont {Coutts}, \citenamefont {Heinrich},\ and\ \citenamefont {Girt}}]{Omelchenko2017}%
  \BibitemOpen
  \bibfield  {author} {\bibinfo {author} {\bibfnamefont {Pavlo}\ \bibnamefont {Omelchenko}}, \bibinfo {author} {\bibfnamefont {Eric~Arturo}\ \bibnamefont {Montoya}}, \bibinfo {author} {\bibfnamefont {Chris}\ \bibnamefont {Coutts}}, \bibinfo {author} {\bibfnamefont {Bret}\ \bibnamefont {Heinrich}}, \ and\ \bibinfo {author} {\bibfnamefont {Erol}\ \bibnamefont {Girt}},\ }\bibfield  {title} {\enquote {\bibinfo {title} {Tunable magnetization and damping of sputter-deposited, exchange coupled Py|Fe bilayers},}\ }\href {\doibase 10.1038/s41598-017-05030-8} 
  {\bibfield  {journal} {\bibinfo  {journal} {Scientific Reports}\ }\textbf {\bibinfo {volume} {7}},\ \bibinfo {pages} {4861} (\bibinfo {year} {2017})} \BibitemShut {NoStop}%
\bibitem [{\citenamefont {Mewes}\ and\ \citenamefont {Mewes}(2015)}]{Mewes2015}%
  \BibitemOpen
  \bibfield  {author} {\bibinfo {author} {\bibfnamefont {C.~K.~A.}\ \bibnamefont {Mewes}}\ and\ \bibinfo {author} {\bibfnamefont {Tim}\ \bibnamefont {Mewes}},\ }\href {https://www.taylorfrancis.com/books/e/9780429069215/chapters/10.1201/b18942-6} {\enquote {\bibinfo {title} {Relaxation in magnetic materials for spintronics},}\ } (\bibinfo {year} {2015})\BibitemShut {NoStop}%
\bibitem [{\citenamefont {Zakeri}\ \emph {et~al.}(2007)\citenamefont {Zakeri}, \citenamefont {Lindner}, \citenamefont {Barsukov}, \citenamefont {Meckenstock}, \citenamefont {Farle}, \citenamefont {von Hörsten}, \citenamefont {Wende}, \citenamefont {Keune}, \citenamefont {Rocker}, \citenamefont {Kalarickal}, \citenamefont {Lenz}, \citenamefont {Kuch}, \citenamefont {Baberschke},\ and\ \citenamefont {Frait}}]{Zakeri2007}%
  \BibitemOpen
  \bibfield  {author} {\bibinfo {author} {\bibfnamefont {Kh.}\ \bibnamefont {Zakeri}}, \bibinfo {author} {\bibfnamefont {J.}~\bibnamefont {Lindner}}, \bibinfo {author} {\bibfnamefont {I.}~\bibnamefont {Barsukov}}, \bibinfo {author} {\bibfnamefont {R.}~\bibnamefont {Meckenstock}}, \bibinfo {author} {\bibfnamefont {M.}~\bibnamefont {Farle}}, \bibinfo {author} {\bibfnamefont {U.}~\bibnamefont {von Hörsten}}, \bibinfo {author} {\bibfnamefont {H.}~\bibnamefont {Wende}}, \bibinfo {author} {\bibfnamefont {W.}~\bibnamefont {Keune}}, \bibinfo {author} {\bibfnamefont {J.}~\bibnamefont {Rocker}}, \bibinfo {author} {\bibfnamefont {S.~S.}\ \bibnamefont {Kalarickal}}, \bibinfo {author} {\bibfnamefont {K.}~\bibnamefont {Lenz}}, \bibinfo {author} {\bibfnamefont {W.}~\bibnamefont {Kuch}}, \bibinfo {author} {\bibfnamefont {K.}~\bibnamefont {Baberschke}}, \ and\ \bibinfo {author} {\bibfnamefont {Z.}~\bibnamefont {Frait}},\ }\bibfield  {title} {\enquote {\bibinfo {title} {Spin dynamics in ferromagnets: Gilbert damping and
  two-magnon scattering},}\ }\href {\doibase 10.1103/PhysRevB.76.104416} {\bibfield  {journal} {\bibinfo  {journal} {Physical Review B}\ }\textbf {\bibinfo {volume} {76}},\ \bibinfo {pages} {104416} (\bibinfo {year} {2007})}\BibitemShut {NoStop}%
\bibitem [{\citenamefont {Wu}\ \emph {et~al.}(2022)\citenamefont {Wu}, \citenamefont {Smith}, \citenamefont {Nakarmi}, \citenamefont {Rai}, \citenamefont {Clavel}, \citenamefont {Hudait}, \citenamefont {Zhao}, \citenamefont {Michel}, \citenamefont {Mewes}, \citenamefont {Mewes},\ and\ \citenamefont {Emori}}]{Wu2022}%
  \BibitemOpen
  \bibfield  {author} {\bibinfo {author} {\bibfnamefont {Shuang}\ \bibnamefont {Wu}}, \bibinfo {author} {\bibfnamefont {David~A.}\ \bibnamefont {Smith}}, \bibinfo {author} {\bibfnamefont {Prabandha}\ \bibnamefont {Nakarmi}}, \bibinfo {author} {\bibfnamefont {Anish}\ \bibnamefont {Rai}}, \bibinfo {author} {\bibfnamefont {Michael}\ \bibnamefont {Clavel}}, \bibinfo {author} {\bibfnamefont {Mantu~K.}\ \bibnamefont {Hudait}}, \bibinfo {author} {\bibfnamefont {Jing}\ \bibnamefont {Zhao}}, \bibinfo {author} {\bibfnamefont {F.~Marc}\ \bibnamefont {Michel}}, \bibinfo {author} {\bibfnamefont {Claudia}\ \bibnamefont {Mewes}}, \bibinfo {author} {\bibfnamefont {Tim}\ \bibnamefont {Mewes}}, \ and\ \bibinfo {author} {\bibfnamefont {Satoru}\ \bibnamefont {Emori}},\ }\bibfield  {title} {\enquote {\bibinfo {title} {Room-temperature intrinsic and extrinsic damping in polycrystalline Fe thin films},}\ }\href {\doibase 10.1103/PHYSREVB.105.174408/FIGURES/7/MEDIUM} {\bibfield  {journal} {\bibinfo  {journal} {Physical Review B}\
  }\textbf {\bibinfo {volume} {105}},\ \bibinfo {pages} {174408} (\bibinfo {year} {2022})}\BibitemShut {NoStop}%
\bibitem [{\citenamefont {Gilmore}\ \emph {et~al.}(2010)\citenamefont {Gilmore}, \citenamefont {Stiles}, \citenamefont {Seib}, \citenamefont {Steiauf},\ and\ \citenamefont {Fähnle}}]{Gilmore2010}%
  \BibitemOpen
  \bibfield  {author} {\bibinfo {author} {\bibfnamefont {Keith}\ \bibnamefont {Gilmore}}, \bibinfo {author} {\bibfnamefont {M.~D.}\ \bibnamefont {Stiles}}, \bibinfo {author} {\bibfnamefont {Jonas}\ \bibnamefont {Seib}}, \bibinfo {author} {\bibfnamefont {Daniel}\ \bibnamefont {Steiauf}}, \ and\ \bibinfo {author} {\bibfnamefont {Manfred}\ \bibnamefont {Fähnle}},\ }\bibfield  {title} {\enquote {\bibinfo {title} {Anisotropic damping of the magnetization dynamics in Ni, Co, and Fe},}\ }\href {\doibase 10.1103/PhysRevB.81.174414} {\bibfield  {journal} {\bibinfo  {journal} {Physical Review B}\ }\textbf {\bibinfo {volume} {81}},\ \bibinfo {pages} {174414} (\bibinfo {year} {2010})}\BibitemShut {NoStop}%
\bibitem [{\citenamefont {Smith}\ \emph {et~al.}(2020)\citenamefont {Smith}, \citenamefont {Rai}, \citenamefont {Lim}, \citenamefont {Hartnett}, \citenamefont {Sapkota}, \citenamefont {Srivastava}, \citenamefont {Mewes}, \citenamefont {Jiang}, \citenamefont {Clavel}, \citenamefont {Hudait}, \citenamefont {Viehland}, \citenamefont {Heremans}, \citenamefont {Balachandran}, \citenamefont {Mewes},\ and\ \citenamefont {Emori}}]{Smith2020c}%
  \BibitemOpen
  \bibfield  {author} {\bibinfo {author} {\bibfnamefont {David~A.}\ \bibnamefont {Smith}}, \bibinfo {author} {\bibfnamefont {Anish}\ \bibnamefont {Rai}}, \bibinfo {author} {\bibfnamefont {Youngmin}\ \bibnamefont {Lim}}, \bibinfo {author} {\bibfnamefont {Timothy~Q.}\ \bibnamefont {Hartnett}}, \bibinfo {author} {\bibfnamefont {Arjun}\ \bibnamefont {Sapkota}}, \bibinfo {author} {\bibfnamefont {Abhishek}\ \bibnamefont {Srivastava}}, \bibinfo {author} {\bibfnamefont {Claudia}\ \bibnamefont {Mewes}}, \bibinfo {author} {\bibfnamefont {Zijian}\ \bibnamefont {Jiang}}, \bibinfo {author} {\bibfnamefont {Michael}\ \bibnamefont {Clavel}}, \bibinfo {author} {\bibfnamefont {Mantu~K.}\ \bibnamefont {Hudait}}, \bibinfo {author} {\bibfnamefont {Dwight~D.}\ \bibnamefont {Viehland}}, \bibinfo {author} {\bibfnamefont {Jean~J.}\ \bibnamefont {Heremans}}, \bibinfo {author} {\bibfnamefont {Prasanna~V.}\ \bibnamefont {Balachandran}}, \bibinfo {author} {\bibfnamefont {Tim}\ \bibnamefont {Mewes}}, \ and\ \bibinfo {author}
  {\bibfnamefont {Satoru}\ \bibnamefont {Emori}},\ }\bibfield  {title} {\enquote {\bibinfo {title} {Magnetic damping in epitaxial iron alloyed with vanadium and aluminum},}\ }\href {\doibase 10.1103/PhysRevApplied.14.034042} {\bibfield  {journal} {\bibinfo  {journal} {Physical Review Applied}\ }\textbf {\bibinfo {volume} {14}},\ \bibinfo {pages} {034042} (\bibinfo {year} {2020})}\BibitemShut {NoStop}%
\bibitem [{\citenamefont {Abo}\ \emph {et~al.}(2013)\citenamefont {Abo}, \citenamefont {Hong}, \citenamefont {Park}, \citenamefont {Lee}, \citenamefont {Lee},\ and\ \citenamefont {Choi}}]{Abo2013}%
  \BibitemOpen
  \bibfield  {author} {\bibinfo {author} {\bibfnamefont {Gavin~S.}\ \bibnamefont {Abo}}, \bibinfo {author} {\bibfnamefont {Yang-Ki}\ \bibnamefont {Hong}}, \bibinfo {author} {\bibfnamefont {Jihoon}\ \bibnamefont {Park}}, \bibinfo {author} {\bibfnamefont {Jaejin}\ \bibnamefont {Lee}}, \bibinfo {author} {\bibfnamefont {Woncheol}\ \bibnamefont {Lee}}, \ and\ \bibinfo {author} {\bibfnamefont {Byoung-Chul}\ \bibnamefont {Choi}},\ }\bibfield  {title} {\enquote {\bibinfo {title} {Definition of magnetic exchange length},}\ }\href {https://ieeexplore.ieee.org/document/6497624/} {\bibfield  {journal} {\bibinfo  {journal} {IEEE Transactions on Magnetics}\ }\textbf {\bibinfo {volume} {49}},\ \bibinfo {pages} {4937 -- 4939} (\bibinfo {year} {2013})}\BibitemShut {NoStop}%
\bibitem [{\citenamefont {Niitsu}(2020)}]{Niitsu2020}%
  \BibitemOpen
  \bibfield  {author} {\bibinfo {author} {\bibfnamefont {Kodai}\ \bibnamefont {Niitsu}},\ }\bibfield  {title} {\enquote {\bibinfo {title} {Temperature dependence of magnetic exchange stiffness in iron and nickel},}\ }\href {\doibase 10.1088/1361-6463/AB9672} {\bibfield  {journal} {\bibinfo  {journal} {Journal of Physics D: Applied Physics}\ }\textbf {\bibinfo {volume} {53}},\ \bibinfo {pages} {39LT01} (\bibinfo {year} {2020})}\BibitemShut {NoStop}%
\bibitem [{\citenamefont {Lim}\ \emph {et~al.}(2021)\citenamefont {Lim}, \citenamefont {Khodadadi}, \citenamefont {Li}, \citenamefont {Viehland}, \citenamefont {Manchon},\ and\ \citenamefont {Emori}}]{Lim2021}%
  \BibitemOpen
  \bibfield  {author} {\bibinfo {author} {\bibfnamefont {Youngmin}\ \bibnamefont {Lim}}, \bibinfo {author} {\bibfnamefont {Behrouz}\ \bibnamefont {Khodadadi}}, \bibinfo {author} {\bibfnamefont {Jie-Fang}\ \bibnamefont {Li}}, \bibinfo {author} {\bibfnamefont {Dwight}\ \bibnamefont {Viehland}}, \bibinfo {author} {\bibfnamefont {Aurelien}\ \bibnamefont {Manchon}}, \ and\ \bibinfo {author} {\bibfnamefont {Satoru}\ \bibnamefont {Emori}},\ }\bibfield  {title} {\enquote {\bibinfo {title} {Dephasing of transverse spin current in ferrimagnetic alloys},}\ }\href {\doibase 10.1103/PhysRevB.103.024443} {\bibfield  {journal} {\bibinfo  {journal} {Physical Review B}\ }\textbf {\bibinfo {volume} {103}},\ \bibinfo {pages} {024443} (\bibinfo {year} {2021})}\BibitemShut {NoStop}%
\bibitem [{\citenamefont {Lim}\ \emph {et~al.}(2022)\citenamefont {Lim}, \citenamefont {Wu}, \citenamefont {Smith}, \citenamefont {Klewe}, \citenamefont {Shafer},\ and\ \citenamefont {Emori}}]{Lim2022a}%
  \BibitemOpen
  \bibfield  {author} {\bibinfo {author} {\bibfnamefont {Youngmin}\ \bibnamefont {Lim}}, \bibinfo {author} {\bibfnamefont {Shuang}\ \bibnamefont {Wu}}, \bibinfo {author} {\bibfnamefont {David~A.}\ \bibnamefont {Smith}}, \bibinfo {author} {\bibfnamefont {Christoph}\ \bibnamefont {Klewe}}, \bibinfo {author} {\bibfnamefont {Padraic}\ \bibnamefont {Shafer}}, \ and\ \bibinfo {author} {\bibfnamefont {Satoru}\ \bibnamefont {Emori}},\ }\bibfield  {title} {\enquote {\bibinfo {title} {Absorption of transverse spin current in ferromagnetic nicu: Dominance of bulk dephasing over spin-flip scattering},}\ }\href {\doibase 10.1063/5.0120865} {\bibfield  {journal} {\bibinfo  {journal} {Applied Physics Letters}\ }\textbf {\bibinfo {volume} {121}},\ \bibinfo {pages} {222403} (\bibinfo {year} {2022})}\BibitemShut {NoStop}%
\bibitem [{\citenamefont {Lim}\ \emph {et~al.}(2023)\citenamefont {Lim}, \citenamefont {Nepal}, \citenamefont {Smith}, \citenamefont {Wu}, \citenamefont {Srivastava}, \citenamefont {Nakarmi}, \citenamefont {Mewes}, \citenamefont {Jiang}, \citenamefont {Gupta}, \citenamefont {Viehland}, \citenamefont {Klewe}, \citenamefont {Shafer}, \citenamefont {Park}, \citenamefont {Mabe}, \citenamefont {Amin}, \citenamefont {Heremans}, \citenamefont {Mewes}, \citenamefont {Emori},\ and\ \citenamefont {Gupta}}]{Lim2023}%
  \BibitemOpen
  \bibfield  {author} {\bibinfo {author} {\bibfnamefont {Youngmin}\ \bibnamefont {Lim}}, \bibinfo {author} {\bibfnamefont {Bhuwan}\ \bibnamefont {Nepal}}, \bibinfo {author} {\bibfnamefont {David~A}\ \bibnamefont {Smith}}, \bibinfo {author} {\bibfnamefont {Shuang}\ \bibnamefont {Wu}}, \bibinfo {author} {\bibfnamefont {Abhishek}\ \bibnamefont {Srivastava}}, \bibinfo {author} {\bibfnamefont {Prabandha}\ \bibnamefont {Nakarmi}}, \bibinfo {author} {\bibfnamefont {Claudia}\ \bibnamefont {Mewes}}, \bibinfo {author} {\bibfnamefont {Zijian}\ \bibnamefont {Jiang}}, \bibinfo {author} {\bibfnamefont {;~Adbhut}\ \bibnamefont {Gupta}}, \bibinfo {author} {\bibfnamefont {Dwight~D}\ \bibnamefont {Viehland}}, \bibinfo {author} {\bibfnamefont {Christoph}\ \bibnamefont {Klewe}}, \bibinfo {author} {\bibfnamefont {Padraic}\ \bibnamefont {Shafer}}, \bibinfo {author} {\bibfnamefont {In~Jun}\ \bibnamefont {Park}}, \bibinfo {author} {\bibfnamefont {Timothy}\ \bibnamefont {Mabe}}, \bibinfo {author} {\bibfnamefont {Vivek~P}\
  \bibnamefont {Amin}}, \bibinfo {author} {\bibfnamefont {Jean~J}\ \bibnamefont {Heremans}}, \bibinfo {author} {\bibfnamefont {Tim}\ \bibnamefont {Mewes}}, \bibinfo {author} {\bibfnamefont {Satoru}\ \bibnamefont {Emori}}, \ and\ \bibinfo {author} {\bibfnamefont {Adbhut}\ \bibnamefont {Gupta}},\ }\bibfield  {title} {\enquote {\bibinfo {title} {Suppression of spin pumping at metal interfaces},}\ }\href {\doibase 10.1063/5.0156429} {\bibfield  {journal} {\bibinfo  {journal} {APL Materials}\ }\textbf {\bibinfo {volume} {11}},\ \bibinfo {pages} {101121} (\bibinfo {year} {2023})}\BibitemShut {NoStop}%
\bibitem [{\citenamefont {Liu}\ \emph {et~al.}(2011)\citenamefont {Liu}, \citenamefont {Moriyama}, \citenamefont {Ralph},\ and\ \citenamefont {Buhrman}}]{Liu2011}%
  \BibitemOpen
  \bibfield  {author} {\bibinfo {author} {\bibfnamefont {Luqiao}\ \bibnamefont {Liu}}, \bibinfo {author} {\bibfnamefont {Takahiro}\ \bibnamefont {Moriyama}}, \bibinfo {author} {\bibfnamefont {D.~C.}\ \bibnamefont {Ralph}}, \ and\ \bibinfo {author} {\bibfnamefont {R.~A.}\ \bibnamefont {Buhrman}},\ }\bibfield  {title} {\enquote {\bibinfo {title} {Spin-torque ferromagnetic resonance induced by the spin Hall effect},}\ }\href {\doibase 10.1103/PhysRevLett.106.036601} {\bibfield  {journal} {\bibinfo  {journal} {Physical Review Letters}\ }\textbf {\bibinfo {volume} {106}},\ \bibinfo {pages} {036601} (\bibinfo {year} {2011})}\BibitemShut {NoStop}%
\bibitem [{\citenamefont {Karimeddiny}\ and\ \citenamefont {Ralph}(2021)}]{Karimeddiny2021}%
  \BibitemOpen
  \bibfield  {author} {\bibinfo {author} {\bibfnamefont {Saba}\ \bibnamefont {Karimeddiny}}\ and\ \bibinfo {author} {\bibfnamefont {Daniel~C.}\ \bibnamefont {Ralph}},\ }\bibfield  {title} {\enquote {\bibinfo {title} {Resolving discrepancies in spin-torque ferromagnetic resonance measurements: Lineshape versus linewidth analyses},}\ }\href {\doibase 10.1103/PHYSREVAPPLIED.15.064017/FIGURES/7/MEDIUM} {\bibfield  {journal} {\bibinfo  {journal} {Physical Review Applied}\ }\textbf {\bibinfo {volume} {15}},\ \bibinfo {pages} {064017} (\bibinfo {year} {2021})}\BibitemShut {NoStop}%
\bibitem [{\citenamefont {Nguyen}\ and\ \citenamefont {Pai}(2021)}]{Nguyen2021}%
  \BibitemOpen
  \bibfield  {author} {\bibinfo {author} {\bibfnamefont {Minh~Hai}\ \bibnamefont {Nguyen}}\ and\ \bibinfo {author} {\bibfnamefont {Chi~Feng}\ \bibnamefont {Pai}},\ }\bibfield  {title} {\enquote {\bibinfo {title} {Spin-orbit torque characterization in a nutshell},}\ }\href {\doibase 10.1063/5.0041123/892395} {\bibfield  {journal} {\bibinfo  {journal} {APL Materials}\ }\textbf {\bibinfo {volume} {9}} (\bibinfo {year} {2021})}\BibitemShut {NoStop}%
\bibitem [{\citenamefont {Kondou}\ \emph {et~al.}(2016)\citenamefont {Kondou}, \citenamefont {Sukegawa}, \citenamefont {Kasai}, \citenamefont {Mitani}, \citenamefont {Niimi},\ and\ \citenamefont {Otani}}]{Kondou2016}%
  \BibitemOpen
  \bibfield  {author} {\bibinfo {author} {\bibfnamefont {Kouta}\ \bibnamefont {Kondou}}, \bibinfo {author} {\bibfnamefont {Hiroaki}\ \bibnamefont {Sukegawa}}, \bibinfo {author} {\bibfnamefont {Shinya}\ \bibnamefont {Kasai}}, \bibinfo {author} {\bibfnamefont {Seiji}\ \bibnamefont {Mitani}}, \bibinfo {author} {\bibfnamefont {Yasuhiro}\ \bibnamefont {Niimi}}, \ and\ \bibinfo {author} {\bibfnamefont {YoshiChika}\ \bibnamefont {Otani}},\ }\bibfield  {title} {\enquote {\bibinfo {title} {Influence of inverse spin Hall effect in spin-torque ferromagnetic resonance measurements},}\ }\href {\doibase 10.7567/APEX.9.023002} {\bibfield  {journal} {\bibinfo  {journal} {Applied Physics Express}\ }\textbf {\bibinfo {volume} {9}},\ \bibinfo {pages} {023002} (\bibinfo {year} {2016})}\BibitemShut {NoStop}%
\bibitem [{\citenamefont {Okada}\ \emph {et~al.}(2019)\citenamefont {Okada}, \citenamefont {Takeuchi}, \citenamefont {Furuya}, \citenamefont {Zhang}, \citenamefont {Sato}, \citenamefont {Fukami},\ and\ \citenamefont {Ohno}}]{Okada2019}%
  \BibitemOpen
  \bibfield  {author} {\bibinfo {author} {\bibfnamefont {Atsushi}\ \bibnamefont {Okada}}, \bibinfo {author} {\bibfnamefont {Yutaro}\ \bibnamefont {Takeuchi}}, \bibinfo {author} {\bibfnamefont {Kaito}\ \bibnamefont {Furuya}}, \bibinfo {author} {\bibfnamefont {Chaoliang}\ \bibnamefont {Zhang}}, \bibinfo {author} {\bibfnamefont {Hideo}\ \bibnamefont {Sato}}, \bibinfo {author} {\bibfnamefont {Shunsuke}\ \bibnamefont {Fukami}}, \ and\ \bibinfo {author} {\bibfnamefont {Hideo}\ \bibnamefont {Ohno}},\ }\bibfield  {title} {\enquote {\bibinfo {title} {Spin-pumping-free determination of spin-orbit torque efficiency from spin-torque ferromagnetic resonance},}\ }\href {\doibase 10.1103/PhysRevApplied.12.014040} {\bibfield  {journal} {\bibinfo  {journal} {Physical Review Applied}\ }\textbf {\bibinfo {volume} {12}},\ \bibinfo {pages} {014040} (\bibinfo {year} {2019})}\BibitemShut {NoStop}%
\bibitem [{\citenamefont {Schultheiss}\ \emph {et~al.}(2012)\citenamefont {Schultheiss}, \citenamefont {Pearson}, \citenamefont {Bader},\ and\ \citenamefont {Hoffmann}}]{Schultheiss2012}%
  \BibitemOpen
  \bibfield  {author} {\bibinfo {author} {\bibfnamefont {H.}~\bibnamefont {Schultheiss}}, \bibinfo {author} {\bibfnamefont {J.~E.}\ \bibnamefont {Pearson}}, \bibinfo {author} {\bibfnamefont {S.~D.}\ \bibnamefont {Bader}}, \ and\ \bibinfo {author} {\bibfnamefont {A.}~\bibnamefont {Hoffmann}},\ }\bibfield  {title} {\enquote {\bibinfo {title} {Thermoelectric detection of spin waves},}\ }\href {\doibase 10.1103/PhysRevLett.109.237204} {\bibfield  {journal} {\bibinfo  {journal} {Physical Review Letters}\ }\textbf {\bibinfo {volume} {109}},\ \bibinfo {pages} {237204} (\bibinfo {year} {2012})}\BibitemShut {NoStop}%
\bibitem [{\citenamefont {Jiang}\ \emph {et~al.}(2024)\citenamefont {Jiang}, \citenamefont {Chen}, \citenamefont {Ji}, \citenamefont {Chai}, \citenamefont {Zhang}, \citenamefont {Liang}, \citenamefont {Liu}, \citenamefont {Skowroński}, \citenamefont {Yu}, \citenamefont {Yi},\ and\ \citenamefont {Nan}}]{TNan2024STFMR}%
  \BibitemOpen
  \bibfield  {author} {\bibinfo {author} {\bibfnamefont {Dingsong}\ \bibnamefont {Jiang}}, \bibinfo {author} {\bibfnamefont {Hetian}\ \bibnamefont {Chen}}, \bibinfo {author} {\bibfnamefont {Guiping}\ \bibnamefont {Ji}}, \bibinfo {author} {\bibfnamefont {Yahong}\ \bibnamefont {Chai}}, \bibinfo {author} {\bibfnamefont {Chenye}\ \bibnamefont {Zhang}}, \bibinfo {author} {\bibfnamefont {Yuhan}\ \bibnamefont {Liang}}, \bibinfo {author} {\bibfnamefont {Jingchun}\ \bibnamefont {Liu}}, \bibinfo {author} {\bibfnamefont {Witold}\ \bibnamefont {Skowroński}}, \bibinfo {author} {\bibfnamefont {Pu}~\bibnamefont {Yu}}, \bibinfo {author} {\bibfnamefont {Di}~\bibnamefont {Yi}}, \ and\ \bibinfo {author} {\bibfnamefont {Tianxiang}\ \bibnamefont {Nan}},\ }\bibfield  {title} {\enquote {\bibinfo {title} {Substrate-induced spin-torque-like signal in spin-torque ferromagnetic resonance measurement},}\ }\href {\doibase 10.1103/PHYSREVAPPLIED.21.024021/FIGURES/4/MEDIUM} {\bibfield  {journal} {\bibinfo  {journal} {Physical Review
  Applied}\ }\textbf {\bibinfo {volume} {21}},\ \bibinfo {pages} {024021} (\bibinfo {year} {2024})}\BibitemShut {NoStop}%
\bibitem [{\citenamefont {Kasai}\ \emph {et~al.}(2014)\citenamefont {Kasai}, \citenamefont {Kondou}, \citenamefont {Sukegawa}, \citenamefont {Mitani}, \citenamefont {Tsukagoshi},\ and\ \citenamefont {Otani}}]{Kasai2014}%
  \BibitemOpen
  \bibfield  {author} {\bibinfo {author} {\bibfnamefont {Shinya}\ \bibnamefont {Kasai}}, \bibinfo {author} {\bibfnamefont {Kouta}\ \bibnamefont {Kondou}}, \bibinfo {author} {\bibfnamefont {Hiroaki}\ \bibnamefont {Sukegawa}}, \bibinfo {author} {\bibfnamefont {Seiji}\ \bibnamefont {Mitani}}, \bibinfo {author} {\bibfnamefont {Kazuhito}\ \bibnamefont {Tsukagoshi}}, \ and\ \bibinfo {author} {\bibfnamefont {Yoshichika}\ \bibnamefont {Otani}},\ }\bibfield  {title} {\enquote {\bibinfo {title} {Modulation of effective damping constant using spin Hall effect},}\ }\href {\doibase 10.1063/1.4867649} {\bibfield  {journal} {\bibinfo  {journal} {Applied Physics Letters}\ }\textbf {\bibinfo {volume} {104}},\ \bibinfo {pages} {092408} (\bibinfo {year} {2014})}\BibitemShut {NoStop}%
\bibitem [{\citenamefont {Nan}\ \emph {et~al.}(2015)\citenamefont {Nan}, \citenamefont {Emori}, \citenamefont {Boone}, \citenamefont {Wang}, \citenamefont {Oxholm}, \citenamefont {Jones}, \citenamefont {Howe}, \citenamefont {Brown},\ and\ \citenamefont {Sun}}]{Nan2015a}%
  \BibitemOpen
  \bibfield  {author} {\bibinfo {author} {\bibfnamefont {Tianxiang}\ \bibnamefont {Nan}}, \bibinfo {author} {\bibfnamefont {Satoru}\ \bibnamefont {Emori}}, \bibinfo {author} {\bibfnamefont {Carl~T.}\ \bibnamefont {Boone}}, \bibinfo {author} {\bibfnamefont {Xinjun}\ \bibnamefont {Wang}}, \bibinfo {author} {\bibfnamefont {Trevor~M.}\ \bibnamefont {Oxholm}}, \bibinfo {author} {\bibfnamefont {John~G.}\ \bibnamefont {Jones}}, \bibinfo {author} {\bibfnamefont {Brandon~M.}\ \bibnamefont {Howe}}, \bibinfo {author} {\bibfnamefont {Gail~J.}\ \bibnamefont {Brown}}, \ and\ \bibinfo {author} {\bibfnamefont {Nian~X.}\ \bibnamefont {Sun}},\ }\bibfield  {title} {\enquote {\bibinfo {title} {Comparison of spin-orbit torques and spin pumping across NiFe/Pt and NiFe/Cu/Pt interfaces},}\ }\href {\doibase 10.1103/PhysRevB.91.214416} {\bibfield  {journal} {\bibinfo  {journal} {Physical Review B}\ }\textbf {\bibinfo {volume} {91}},\ \bibinfo {pages} {214416} (\bibinfo {year} {2015})}\BibitemShut {NoStop}%
\bibitem [{\citenamefont {Wang}\ \emph {et~al.}(2019)\citenamefont {Wang}, \citenamefont {Wang}, \citenamefont {Amin}, \citenamefont {Wang}, \citenamefont {Radhakrishnan}, \citenamefont {Davidson}, \citenamefont {Allen}, \citenamefont {Silva}, \citenamefont {Ohldag}, \citenamefont {Balzar}, \citenamefont {Zink}, \citenamefont {Haney}, \citenamefont {Xiao}, \citenamefont {Cahill}, \citenamefont {Lorenz},\ and\ \citenamefont {Fan}}]{Wang2019a}%
  \BibitemOpen
  \bibfield  {author} {\bibinfo {author} {\bibfnamefont {Wenrui}\ \bibnamefont {Wang}}, \bibinfo {author} {\bibfnamefont {Tao}\ \bibnamefont {Wang}}, \bibinfo {author} {\bibfnamefont {Vivek~P.}\ \bibnamefont {Amin}}, \bibinfo {author} {\bibfnamefont {Yang}\ \bibnamefont {Wang}}, \bibinfo {author} {\bibfnamefont {Anil}\ \bibnamefont {Radhakrishnan}}, \bibinfo {author} {\bibfnamefont {Angie}\ \bibnamefont {Davidson}}, \bibinfo {author} {\bibfnamefont {Shane~R.}\ \bibnamefont {Allen}}, \bibinfo {author} {\bibfnamefont {T.~J.}\ \bibnamefont {Silva}}, \bibinfo {author} {\bibfnamefont {Hendrik}\ \bibnamefont {Ohldag}}, \bibinfo {author} {\bibfnamefont {Davor}\ \bibnamefont {Balzar}}, \bibinfo {author} {\bibfnamefont {Barry~L.}\ \bibnamefont {Zink}}, \bibinfo {author} {\bibfnamefont {Paul~M.}\ \bibnamefont {Haney}}, \bibinfo {author} {\bibfnamefont {John~Q.}\ \bibnamefont {Xiao}}, \bibinfo {author} {\bibfnamefont {David~G.}\ \bibnamefont {Cahill}}, \bibinfo {author} {\bibfnamefont {Virginia~O.}\ \bibnamefont
  {Lorenz}}, \ and\ \bibinfo {author} {\bibfnamefont {Xin}\ \bibnamefont {Fan}},\ }\bibfield  {title} {\enquote {\bibinfo {title} {Anomalous spin–orbit torques in magnetic single-layer films},}\ }\href {\doibase 10.1038/s41565-019-0504-0} {\bibfield  {journal} {\bibinfo  {journal} {Nature Nanotechnology}\ }\textbf {\bibinfo {volume} {14}},\ \bibinfo {pages} {819–824} (\bibinfo {year} {2019})}\BibitemShut {NoStop}%
\bibitem [{\citenamefont {Soya}\ \emph {et~al.}(2023)\citenamefont {Soya}, \citenamefont {Yamada}, \citenamefont {Hamaya},\ and\ \citenamefont {Ando}}]{Soya2023}%
  \BibitemOpen
  \bibfield  {author} {\bibinfo {author} {\bibfnamefont {Nozomi}\ \bibnamefont {Soya}}, \bibinfo {author} {\bibfnamefont {Michihiro}\ \bibnamefont {Yamada}}, \bibinfo {author} {\bibfnamefont {Kohei}\ \bibnamefont {Hamaya}}, \ and\ \bibinfo {author} {\bibfnamefont {Kazuya}\ \bibnamefont {Ando}},\ }\bibfield  {title} {\enquote {\bibinfo {title} {Isotropic spin Hall effect in an epitaxial ferromagnet},}\ }\href {\doibase 10.1103/PhysRevLett.131.076702} {\bibfield  {journal} {\bibinfo  {journal} {Physical Review Letters}\ }\textbf {\bibinfo {volume} {131}},\ \bibinfo {pages} {076702} (\bibinfo {year} {2023})}\BibitemShut {NoStop}%
\bibitem [{\citenamefont {Kim}\ \emph {et~al.}(2018)\citenamefont {Kim}, \citenamefont {Kim}, \citenamefont {Chun}, \citenamefont {Moon},\ and\ \citenamefont {Hwang}}]{Kim2018b}%
  \BibitemOpen
  \bibfield  {author} {\bibinfo {author} {\bibfnamefont {Changsoo}\ \bibnamefont {Kim}}, \bibinfo {author} {\bibfnamefont {Dongseuk}\ \bibnamefont {Kim}}, \bibinfo {author} {\bibfnamefont {Byong~Sun}\ \bibnamefont {Chun}}, \bibinfo {author} {\bibfnamefont {Kyoung-Woong}\ \bibnamefont {Moon}}, \ and\ \bibinfo {author} {\bibfnamefont {Chanyong}\ \bibnamefont {Hwang}},\ }\bibfield  {title} {\enquote {\bibinfo {title} {Evaluation method for fieldlike-torque efficiency by modulation of the resonance field},}\ }\href {\doibase 10.1103/PhysRevApplied.9.054035} {\bibfield  {journal} {\bibinfo  {journal} {Physical Review Applied}\ }\textbf {\bibinfo {volume} {9}},\ \bibinfo {pages} {054035} (\bibinfo {year} {2018})}\BibitemShut {NoStop}%
\bibitem [{\citenamefont {Emori}\ \emph {et~al.}(2016)\citenamefont {Emori}, \citenamefont {Nan}, \citenamefont {Belkessam}, \citenamefont {Wang}, \citenamefont {Matyushov}, \citenamefont {Babroski}, \citenamefont {Gao}, \citenamefont {Lin},\ and\ \citenamefont {Sun}}]{Emori2016}%
  \BibitemOpen
  \bibfield  {author} {\bibinfo {author} {\bibfnamefont {Satoru}\ \bibnamefont {Emori}}, \bibinfo {author} {\bibfnamefont {Tianxiang}\ \bibnamefont {Nan}}, \bibinfo {author} {\bibfnamefont {Amine~M.}\ \bibnamefont {Belkessam}}, \bibinfo {author} {\bibfnamefont {Xinjun}\ \bibnamefont {Wang}}, \bibinfo {author} {\bibfnamefont {Alexei~D.}\ \bibnamefont {Matyushov}}, \bibinfo {author} {\bibfnamefont {Christopher~J.}\ \bibnamefont {Babroski}}, \bibinfo {author} {\bibfnamefont {Yuan}\ \bibnamefont {Gao}}, \bibinfo {author} {\bibfnamefont {Hwaider}\ \bibnamefont {Lin}}, \ and\ \bibinfo {author} {\bibfnamefont {Nian~X.}\ \bibnamefont {Sun}},\ }\bibfield  {title} {\enquote {\bibinfo {title} {Interfacial spin-orbit torque without bulk spin-orbit coupling},}\ }\href {\doibase 10.1103/PhysRevB.93.180402} {\bibfield  {journal} {\bibinfo  {journal} {Physical Review B}\ }\textbf {\bibinfo {volume} {93}},\ \bibinfo {pages} {180402} (\bibinfo {year} {2016})}\BibitemShut {NoStop}%
\bibitem [{\citenamefont {Greening}\ \emph {et~al.}(2020)\citenamefont {Greening}, \citenamefont {Smith}, \citenamefont {Lim}, \citenamefont {Jiang}, \citenamefont {Barber}, \citenamefont {Dail}, \citenamefont {Heremans},\ and\ \citenamefont {Emori}}]{Greening2020}%
  \BibitemOpen
  \bibfield  {author} {\bibinfo {author} {\bibfnamefont {Ryan~W.}\ \bibnamefont {Greening}}, \bibinfo {author} {\bibfnamefont {David~A.}\ \bibnamefont {Smith}}, \bibinfo {author} {\bibfnamefont {Youngmin}\ \bibnamefont {Lim}}, \bibinfo {author} {\bibfnamefont {Zijian}\ \bibnamefont {Jiang}}, \bibinfo {author} {\bibfnamefont {Jesse}\ \bibnamefont {Barber}}, \bibinfo {author} {\bibfnamefont {Steven}\ \bibnamefont {Dail}}, \bibinfo {author} {\bibfnamefont {Jean~J.}\ \bibnamefont {Heremans}}, \ and\ \bibinfo {author} {\bibfnamefont {Satoru}\ \bibnamefont {Emori}},\ }\bibfield  {title} {\enquote {\bibinfo {title} {Current-induced spin–orbit field in permalloy interfaced with ultrathin Ti and Cu},}\ }\href {\doibase 10.1063/1.5131665} {\bibfield  {journal} {\bibinfo  {journal} {Applied Physics Letters}\ }\textbf {\bibinfo {volume} {116}},\ \bibinfo {pages} {052402} (\bibinfo {year} {2020})}\BibitemShut {NoStop}%
\bibitem [{\citenamefont {Inyang}\ \emph {et~al.}(2023)\citenamefont {Inyang}, \citenamefont {Swindells}, \citenamefont {Rianto}, \citenamefont {Bouchenoire}, \citenamefont {Morris}, \citenamefont {Merkulov}, \citenamefont {Caruana}, \citenamefont {Kinane}, \citenamefont {Hase},\ and\ \citenamefont {Atkinson}}]{Inyang2023}%
  \BibitemOpen
  \bibfield  {author} {\bibinfo {author} {\bibfnamefont {O.}~\bibnamefont {Inyang}}, \bibinfo {author} {\bibfnamefont {C.}~\bibnamefont {Swindells}}, \bibinfo {author} {\bibfnamefont {D.}~\bibnamefont {Rianto}}, \bibinfo {author} {\bibfnamefont {L.}~\bibnamefont {Bouchenoire}}, \bibinfo {author} {\bibfnamefont {R.~J.H.}\ \bibnamefont {Morris}}, \bibinfo {author} {\bibfnamefont {A.}~\bibnamefont {Merkulov}}, \bibinfo {author} {\bibfnamefont {A.}~\bibnamefont {Caruana}}, \bibinfo {author} {\bibfnamefont {C.}~\bibnamefont {Kinane}}, \bibinfo {author} {\bibfnamefont {T.~P.A.}\ \bibnamefont {Hase}}, \ and\ \bibinfo {author} {\bibfnamefont {D.}~\bibnamefont {Atkinson}},\ }\bibfield  {title} {\enquote {\bibinfo {title} {Non-uniform Gd distribution and magnetization profiles within GdCoFe alloy thin films},}\ }\href {\doibase 10.1063/5.0165423/2911708} {\bibfield  {journal} {\bibinfo  {journal} {Applied Physics Letters}\ }\textbf {\bibinfo {volume} {123}},\ \bibinfo {pages} {122403} (\bibinfo {year}
  {2023})}\BibitemShut {NoStop}%
\bibitem [{Ref()}]{Ref1Dweb}%
  \BibitemOpen
  \href {https://www.nist.gov/ncnr/data-reduction-analysis/reflectometry-software} {\enquote {\bibinfo {title} {Reflectometry software, Ref1D, NIST, available at https://www.nist.gov/ncnr/data-reduction-analysis/reflectometry-software},}\ }\BibitemShut {NoStop}%
\bibitem [{\citenamefont {Fitzsimmons}\ and\ \citenamefont {Majkrzak}(2005)}]{Fitzsimmons2005}%
  \BibitemOpen
  \bibfield  {author} {\bibinfo {author} {\bibfnamefont {M.~R.}\ \bibnamefont {Fitzsimmons}}\ and\ \bibinfo {author} {\bibfnamefont {C.~F.}\ \bibnamefont {Majkrzak}},\ }\bibfield  {title} {\enquote {\bibinfo {title} {Application of polarized neutron reflectometry to studies of artificially structured magnetic materials},}\ }\href {\doibase 10.1007/0-387-23395-4_3} {\bibfield  {journal} {\bibinfo  {journal} {Modern Techniques for Characterizing Magnetic Materials}\ ,\ \bibinfo {pages} {107--155}} (\bibinfo {year} {2005})}\BibitemShut {NoStop}%
\bibitem [{\citenamefont {Schoen}\ \emph {et~al.}(2017{\natexlab{b}})\citenamefont {Schoen}, \citenamefont {Lucassen}, \citenamefont {Nembach}, \citenamefont {Silva}, \citenamefont {Koopmans}, \citenamefont {Back},\ and\ \citenamefont {Shaw}}]{Schoen2017b}%
  \BibitemOpen
  \bibfield  {author} {\bibinfo {author} {\bibfnamefont {Martin A.~W.}\ \bibnamefont {Schoen}}, \bibinfo {author} {\bibfnamefont {Juriaan}\ \bibnamefont {Lucassen}}, \bibinfo {author} {\bibfnamefont {Hans~T.}\ \bibnamefont {Nembach}}, \bibinfo {author} {\bibfnamefont {T.~J.}\ \bibnamefont {Silva}}, \bibinfo {author} {\bibfnamefont {Bert}\ \bibnamefont {Koopmans}}, \bibinfo {author} {\bibfnamefont {Christian~H.}\ \bibnamefont {Back}}, \ and\ \bibinfo {author} {\bibfnamefont {Justin~M.}\ \bibnamefont {Shaw}},\ }\bibfield  {title} {\enquote {\bibinfo {title} {Magnetic properties of ultrathin 3d transition-metal binary alloys. i. spin and orbital moments, anisotropy, and confirmation of slater-pauling behavior},}\ }\href {\doibase 10.1103/PhysRevB.95.134410} {\bibfield  {journal} {\bibinfo  {journal} {Physical Review B}\ }\textbf {\bibinfo {volume} {95}},\ \bibinfo {pages} {134410} (\bibinfo {year} {2017}{\natexlab{b}})}\BibitemShut {NoStop}%
\bibitem [{\citenamefont {Bandiera}\ \emph {et~al.}(2011)\citenamefont {Bandiera}, \citenamefont {Sousa}, \citenamefont {Rodmacq},\ and\ \citenamefont {Dieny}}]{Bandiera2011}%
  \BibitemOpen
  \bibfield  {author} {\bibinfo {author} {\bibfnamefont {S.}~\bibnamefont {Bandiera}}, \bibinfo {author} {\bibfnamefont {R.~R.}\ \bibnamefont {Sousa}}, \bibinfo {author} {\bibfnamefont {B.~B.}\ \bibnamefont {Rodmacq}}, \ and\ \bibinfo {author} {\bibfnamefont {B.}~\bibnamefont {Dieny}},\ }\bibfield  {title} {\enquote {\bibinfo {title} {Asymmetric interfacial perpendicular magnetic anisotropy in Pt/Co/Pt trilayers},}\ }\href {\doibase 10.1109/LMAG.2011.2174032} {\bibfield  {journal} {\bibinfo  {journal} {IEEE Magnetics Letters}\ }\textbf {\bibinfo {volume} {2}} (\bibinfo {year} {2011})}\BibitemShut {NoStop}%
\bibitem [{\citenamefont {Haazen}\ \emph {et~al.}(2013)\citenamefont {Haazen}, \citenamefont {Murè}, \citenamefont {Franken}, \citenamefont {Lavrijsen}, \citenamefont {Swagten},\ and\ \citenamefont {Koopmans}}]{Haazen2013}%
  \BibitemOpen
  \bibfield  {author} {\bibinfo {author} {\bibfnamefont {P~P~J}\ \bibnamefont {Haazen}}, \bibinfo {author} {\bibfnamefont {E}~\bibnamefont {Murè}}, \bibinfo {author} {\bibfnamefont {J~H}\ \bibnamefont {Franken}}, \bibinfo {author} {\bibfnamefont {R}~\bibnamefont {Lavrijsen}}, \bibinfo {author} {\bibfnamefont {H~J~M}\ \bibnamefont {Swagten}}, \ and\ \bibinfo {author} {\bibfnamefont {B}~\bibnamefont {Koopmans}},\ }\bibfield  {title} {\enquote {\bibinfo {title} {Domain wall depinning governed by the spin Hall effect.}}\ }\href {\doibase 10.1038/nmat3553} {\bibfield  {journal} {\bibinfo  {journal} {Nature Materials}\ }\textbf {\bibinfo {volume} {12}},\ \bibinfo {pages} {299--303} (\bibinfo {year} {2013})}\BibitemShut {NoStop}%
\bibitem [{\citenamefont {Je}\ \emph {et~al.}(2013)\citenamefont {Je}, \citenamefont {Kim}, \citenamefont {Yoo}, \citenamefont {Min}, \citenamefont {Lee},\ and\ \citenamefont {Choe}}]{Je2013}%
  \BibitemOpen
  \bibfield  {author} {\bibinfo {author} {\bibfnamefont {Soong-Geun}\ \bibnamefont {Je}}, \bibinfo {author} {\bibfnamefont {Duck-Ho}\ \bibnamefont {Kim}}, \bibinfo {author} {\bibfnamefont {Sang-Cheol}\ \bibnamefont {Yoo}}, \bibinfo {author} {\bibfnamefont {Byoung-Chul}\ \bibnamefont {Min}}, \bibinfo {author} {\bibfnamefont {Kyung-Jin}\ \bibnamefont {Lee}}, \ and\ \bibinfo {author} {\bibfnamefont {Sug-Bong}\ \bibnamefont {Choe}},\ }\bibfield  {title} {\enquote {\bibinfo {title} {Asymmetric magnetic domain-wall motion by the dzyaloshinskii-moriya interaction},}\ }\href {\doibase 10.1103/PhysRevB.88.214401} {\bibfield  {journal} {\bibinfo  {journal} {Physical Review B}\ }\textbf {\bibinfo {volume} {88}},\ \bibinfo {pages} {214401} (\bibinfo {year} {2013})}\BibitemShut {NoStop}%
\bibitem [{\citenamefont {Noma}\ \emph {et~al.}(1999)\citenamefont {Noma}, \citenamefont {Takada},\ and\ \citenamefont {Iida}}]{Noma1999}%
  \BibitemOpen
  \bibfield  {author} {\bibinfo {author} {\bibfnamefont {T}~\bibnamefont {Noma}}, \bibinfo {author} {\bibfnamefont {K}~\bibnamefont {Takada}}, \ and\ \bibinfo {author} {\bibfnamefont {A}~\bibnamefont {Iida}},\ }\bibfield  {title} {\enquote {\bibinfo {title} {Surface-sensitive x-ray fluorescence and diffraction analysis with grazing-exit geometry},}\ }\href {\doibase 10.1002/(SICI)1097-4539(199911/12)28:6} {\bibfield  {journal} {\bibinfo  {journal} {X-ray Spectrometry}\ }\ \textbf {\bibinfo {volume} {28}},\ \bibinfo {pages} {433--439} (\bibinfo {year} {1999})}\BibitemShut {NoStop}%
\bibitem [{\citenamefont {Colombi}\ \emph {et~al.}(2006)\citenamefont {Colombi}, \citenamefont {Zanola}, \citenamefont {Bontempi}, \citenamefont {Roberti}, \citenamefont {Gelfi},\ and\ \citenamefont {Depero}}]{Colombi2006}%
  \BibitemOpen
  \bibfield  {author} {\bibinfo {author} {\bibfnamefont {Paolo}\ \bibnamefont {Colombi}}, \bibinfo {author} {\bibfnamefont {Paolo}\ \bibnamefont {Zanola}}, \bibinfo {author} {\bibfnamefont {Elza}\ \bibnamefont {Bontempi}}, \bibinfo {author} {\bibfnamefont {Roberto}\ \bibnamefont {Roberti}}, \bibinfo {author} {\bibfnamefont {Marcello}\ \bibnamefont {Gelfi}}, \ and\ \bibinfo {author} {\bibfnamefont {Laura~E.}\ \bibnamefont {Depero}},\ }\bibfield  {title} {\enquote {\bibinfo {title} {Glancing-incidence x-ray diffraction for depth profiling of polycrystalline layers},}\ }\href {\doibase 10.1107/S0021889805042779/HTTPS://JOURNALS.IUCR.ORG/SERVICES/TERMSOFUSE.HTML} {\bibfield  {journal} {\bibinfo  {journal} {Journal of Applied Crystallography}\ }\textbf {\bibinfo {volume} {39}},\ \bibinfo {pages} {176--179} (\bibinfo {year} {2006})}\BibitemShut {NoStop}%
\bibitem [{\citenamefont {Filianina}\ \emph {et~al.}(2020)\citenamefont {Filianina}, \citenamefont {Hanke}, \citenamefont {Lee}, \citenamefont {Han}, \citenamefont {Jaiswal}, \citenamefont {Rajan}, \citenamefont {Jakob}, \citenamefont {Mokrousov},\ and\ \citenamefont {Kläui}}]{Filianina2020}%
  \BibitemOpen
  \bibfield  {author} {\bibinfo {author} {\bibfnamefont {Mariia}\ \bibnamefont {Filianina}}, \bibinfo {author} {\bibfnamefont {Jan~Philipp}\ \bibnamefont {Hanke}}, \bibinfo {author} {\bibfnamefont {Kyujoon}\ \bibnamefont {Lee}}, \bibinfo {author} {\bibfnamefont {Dong~Soo}\ \bibnamefont {Han}}, \bibinfo {author} {\bibfnamefont {Samridh}\ \bibnamefont {Jaiswal}}, \bibinfo {author} {\bibfnamefont {Adithya}\ \bibnamefont {Rajan}}, \bibinfo {author} {\bibfnamefont {Gerhard}\ \bibnamefont {Jakob}}, \bibinfo {author} {\bibfnamefont {Yuriy}\ \bibnamefont {Mokrousov}}, \ and\ \bibinfo {author} {\bibfnamefont {Mathias}\ \bibnamefont {Kläui}},\ }\bibfield  {title} {\enquote {\bibinfo {title} {Electric-field control of spin-orbit torques in perpendicularly magnetized w/cofeb/mgo films},}\ }\href {\doibase 10.1103/PHYSREVLETT.124.217701/FIGURES/4/MEDIUM} {\bibfield  {journal} {\bibinfo  {journal} {Physical Review Letters}\ }\textbf {\bibinfo {volume} {124}},\ \bibinfo {pages} {217701} (\bibinfo {year}
  {2020})}\BibitemShut {NoStop}%
\bibitem [{\citenamefont {Wong}\ \emph {et~al.}(2021)\citenamefont {Wong}, \citenamefont {Xu}, \citenamefont {Gan}, \citenamefont {Ang}, \citenamefont {Law}, \citenamefont {Tang}, \citenamefont {Zhang}, \citenamefont {Wong}, \citenamefont {Yu}, \citenamefont {Xu}, \citenamefont {Wee}, \citenamefont {Seet},\ and\ \citenamefont {Lew}}]{Wong2021strainSOT}%
  \BibitemOpen
  \bibfield  {author} {\bibinfo {author} {\bibfnamefont {Grayson Dao~Hwee}\ \bibnamefont {Wong}}, \bibinfo {author} {\bibfnamefont {Zhan}\ \bibnamefont {Xu}}, \bibinfo {author} {\bibfnamefont {Weiliang}\ \bibnamefont {Gan}}, \bibinfo {author} {\bibfnamefont {Calvin Ching~Ian}\ \bibnamefont {Ang}}, \bibinfo {author} {\bibfnamefont {Wai~Cheung}\ \bibnamefont {Law}}, \bibinfo {author} {\bibfnamefont {Jiaxuan}\ \bibnamefont {Tang}}, \bibinfo {author} {\bibfnamefont {Wen}\ \bibnamefont {Zhang}}, \bibinfo {author} {\bibfnamefont {Ping Kwan~Johnny}\ \bibnamefont {Wong}}, \bibinfo {author} {\bibfnamefont {Xiaojiang}\ \bibnamefont {Yu}}, \bibinfo {author} {\bibfnamefont {Feng}\ \bibnamefont {Xu}}, \bibinfo {author} {\bibfnamefont {Andrew~T.S.}\ \bibnamefont {Wee}}, \bibinfo {author} {\bibfnamefont {Chim~Seng}\ \bibnamefont {Seet}}, \ and\ \bibinfo {author} {\bibfnamefont {Wen~Siang}\ \bibnamefont {Lew}},\ }\bibfield  {title} {\enquote {\bibinfo {title} {Strain-mediated spin-orbit torque enhancement in pt/co on
  flexible substrate},}\ }\href {\doibase 10.1021/ACSNANO.0C09404/ASSET/IMAGES/MEDIUM/NN0C09404_M012.GIF} {\bibfield  {journal} {\bibinfo  {journal} {ACS Nano}\ }\textbf {\bibinfo {volume} {15}},\ \bibinfo {pages} {8319--8327} (\bibinfo {year} {2021})}\BibitemShut {NoStop}%
\bibitem [{\citenamefont {Go}\ \emph {et~al.}(2018)\citenamefont {Go}, \citenamefont {Jo}, \citenamefont {Kim},\ and\ \citenamefont {Lee}}]{Go2018}%
  \BibitemOpen
  \bibfield  {author} {\bibinfo {author} {\bibfnamefont {Dongwook}\ \bibnamefont {Go}}, \bibinfo {author} {\bibfnamefont {Daegeun}\ \bibnamefont {Jo}}, \bibinfo {author} {\bibfnamefont {Changyoung}\ \bibnamefont {Kim}}, \ and\ \bibinfo {author} {\bibfnamefont {Hyun-Woo}\ \bibnamefont {Lee}},\ }\bibfield  {title} {\enquote {\bibinfo {title} {Intrinsic spin and orbital Hall effects from orbital texture},}\ }\href {\doibase 10.1103/PhysRevLett.121.086602} {\bibfield  {journal} {\bibinfo  {journal} {Physical Review Letters}\ }\textbf {\bibinfo {volume} {121}},\ \bibinfo {pages} {086602} (\bibinfo {year} {2018})}\BibitemShut {NoStop}%
\bibitem [{\citenamefont {Go}\ and\ \citenamefont {Lee}(2020)}]{Go2020a}%
  \BibitemOpen
  \bibfield  {author} {\bibinfo {author} {\bibfnamefont {Dongwook}\ \bibnamefont {Go}}\ and\ \bibinfo {author} {\bibfnamefont {Hyun~Woo}\ \bibnamefont {Lee}},\ }\bibfield  {title} {\enquote {\bibinfo {title} {Orbital torque: Torque generation by orbital current injection},}\ }\href {\doibase 10.1103/PHYSREVRESEARCH.2.013177/FIGURES/5/MEDIUM} {\bibfield  {journal} {\bibinfo  {journal} {Physical Review Research}\ }\textbf {\bibinfo {volume} {2}},\ \bibinfo {pages} {013177} (\bibinfo {year} {2020})}\BibitemShut {NoStop}%
\bibitem [{\citenamefont {Seki}\ \emph {et~al.}(2021)\citenamefont {Seki}, \citenamefont {Lau}, \citenamefont {Iihama},\ and\ \citenamefont {Takanashi}}]{Seki2021}%
  \BibitemOpen
  \bibfield  {author} {\bibinfo {author} {\bibfnamefont {Takeshi}\ \bibnamefont {Seki}}, \bibinfo {author} {\bibfnamefont {Yong-Chang}\ \bibnamefont {Lau}}, \bibinfo {author} {\bibfnamefont {Satoshi}\ \bibnamefont {Iihama}}, \ and\ \bibinfo {author} {\bibfnamefont {Koki}\ \bibnamefont {Takanashi}},\ }\bibfield  {title} {\enquote {\bibinfo {title} {Spin-orbit torque in a Ni-Fe single layer},}\ }\href {\doibase 10.1103/PhysRevB.104.094430} {\bibfield  {journal} {\bibinfo  {journal} {Physical Review B}\ }\textbf {\bibinfo {volume} {104}},\ \bibinfo {pages} {094430} (\bibinfo {year} {2021})}\BibitemShut {NoStop}%
\bibitem [{\citenamefont {Fu}\ \emph {et~al.}(2022)\citenamefont {Fu}, \citenamefont {Liang}, \citenamefont {Wang}, \citenamefont {Yang}, \citenamefont {Zhou}, \citenamefont {Li}, \citenamefont {Yan}, \citenamefont {Li}, \citenamefont {Li},\ and\ \citenamefont {Liu}}]{Fu2022}%
  \BibitemOpen
  \bibfield  {author} {\bibinfo {author} {\bibfnamefont {Qingwei}\ \bibnamefont {Fu}}, \bibinfo {author} {\bibfnamefont {Like}\ \bibnamefont {Liang}}, \bibinfo {author} {\bibfnamefont {Wenqiang}\ \bibnamefont {Wang}}, \bibinfo {author} {\bibfnamefont {Liupeng}\ \bibnamefont {Yang}}, \bibinfo {author} {\bibfnamefont {Kaiyuan}\ \bibnamefont {Zhou}}, \bibinfo {author} {\bibfnamefont {Zishuang}\ \bibnamefont {Li}}, \bibinfo {author} {\bibfnamefont {Chunjie}\ \bibnamefont {Yan}}, \bibinfo {author} {\bibfnamefont {Liyuan}\ \bibnamefont {Li}}, \bibinfo {author} {\bibfnamefont {Haotian}\ \bibnamefont {Li}}, \ and\ \bibinfo {author} {\bibfnamefont {Ronghua}\ \bibnamefont {Liu}},\ }\bibfield  {title} {\enquote {\bibinfo {title} {Observation of nontrivial spin-orbit torque in single-layer ferromagnetic metals},}\ }\href {\doibase 10.1103/PhysRevB.105.224417} {\bibfield  {journal} {\bibinfo  {journal} {Physical Review B}\ }\textbf {\bibinfo {volume} {105}},\ \bibinfo {pages} {224417} (\bibinfo {year} {2022})}\BibitemShut
  {NoStop}%
\bibitem [{\citenamefont {Fan}\ \emph {et~al.}(2013)\citenamefont {Fan}, \citenamefont {Wu}, \citenamefont {Chen}, \citenamefont {Jerry}, \citenamefont {Zhang},\ and\ \citenamefont {Xiao}}]{Fan2013}%
  \BibitemOpen
  \bibfield  {author} {\bibinfo {author} {\bibfnamefont {Xin}\ \bibnamefont {Fan}}, \bibinfo {author} {\bibfnamefont {Jun}\ \bibnamefont {Wu}}, \bibinfo {author} {\bibfnamefont {Yunpeng}\ \bibnamefont {Chen}}, \bibinfo {author} {\bibfnamefont {Matthew~J}\ \bibnamefont {Jerry}}, \bibinfo {author} {\bibfnamefont {Huaiwu}\ \bibnamefont {Zhang}}, \ and\ \bibinfo {author} {\bibfnamefont {John~Q}\ \bibnamefont {Xiao}},\ }\bibfield  {title} {\enquote {\bibinfo {title} {Observation of the nonlocal spin-orbital effective field.}}\ }\href {\doibase 10.1038/ncomms2709} {\bibfield  {journal} {\bibinfo  {journal} {Nature Communications}\ }\textbf {\bibinfo {volume} {4}},\ \bibinfo {pages} {1799} (\bibinfo {year} {2013})}\BibitemShut {NoStop}%
\bibitem [{TUW()}]{TUWienXrayCalculator}%
  \BibitemOpen
  \href {https://gixa.ati.tuwien.ac.at/tools/refractive.xhtml} {\enquote {\bibinfo {title} {X-ray refractive index calculator, TU Wien, available at https://gixa.ati.tuwien.ac.at/tools/refractive.xhtml},}\ }\BibitemShut {NoStop}%
\end{thebibliography}

%

\end{document}